\documentclass{svjour3}       
\usepackage{silence}
\usepackage{paralist}
\usepackage{booktabs}
\usepackage[linesnumbered,noend, noline]{styles/algorithm2e}
\usepackage{algpseudocode}
\usepackage{url}
\usepackage[gen]{eurosym}
\makeatletter
\g@addto@macro{\UrlBreaks}{\UrlOrds}
\makeatother
\usepackage{breakurl}
\newcommand{\urls}[1]{{\scriptsize\url{#1}}}

\def\it{\textit}

\def\tt{\ft}

\usepackage{wrapfig}

\def\bf{\textbf}

\def\fig {Figure~}

\usepackage{subfig}
\usepackage{balance}  
\usepackage{graphics} 
\usepackage{color}
\usepackage{booktabs}
\usepackage{ccicons}

\usepackage{changepage}
\usepackage{threeparttable}
\usepackage{cite}
\usepackage{url}
\usepackage{fancybox}
\usepackage{multirow}
\usepackage{flushend}
\usepackage{booktabs}
\usepackage{tabularx}
\usepackage{comment}
\usepackage{array}

\usepackage{mdframed}
\graphicspath{{}{images/}{dia/}}
\DeclareGraphicsExtensions{.pdf,.png}

\usepackage{listings}
\usepackage{hyperref}

\newcounter{o}
\newcounter{d}
\newcounter{t}
\usepackage{tikz}
\usepackage{styles/pgf-pie}
\usetikzlibrary{positioning,shadows}
\newif\ifpienumberinlegend
\pgfkeys{/number in legend/.code=
    \expandafter\let\expandafter\ifpienumberinlegend
    \csname if#1\endcsname
    \ifpienumberinlegend

    \def\beforenumber##1\afternumber{}%
    \fi,
    /number in legend/.default=true
}
\definecolor{1c1}{RGB}{188,162,6}
\definecolor{1c2}{RGB}{137,129,80}
\definecolor{1c3}{RGB}{239,167,31}
\definecolor{1c4}{RGB}{88,194,241}
\definecolor{1c5}{RGB}{6,180,188}

\tikzset{mynode/.style={draw=white,solid,circle,fill=green,inner sep=1pt, thick,
text=black}}
\tikzset{arrow line/.style={dashed, line width= 2.5pt, color=#1}}

\def\bf{\textbf}

\def\fig {Figure~}

\def\tbl {Table~}

\def\sec {Section~}

\def\it{\textit}

\def\tt{\mct}
\newcommand{\mct}[1]{{\footnotesize {\texttt {#1}}}}

\usepackage{paralist}
\usepackage{soul}

\normalsize

\usepackage{enumitem}

\newcommand{\nd}{\vspace{1mm}\noindent}
\usepackage{tikz}

\usepackage[figuresright]{rotating}
\usepackage{pdflscape}
\lstdefinestyle{inlinecode}{basicstyle={\ttfamily\scriptsize\bfseries}}

\usepackage{tcolorbox}

\usepackage{paralist}
\usepackage{amssymb}
\usepackage{pifont}

\newcommand{\dq}[1]{\href{https://stackoverflow.com/questions/#1/}{$Q_{#1}$}}
\newcommand{\da}[1]{\href{https://stackoverflow.com/answers/#1/}{$A_{#1}$}}

\usepackage{enumitem}
\hypersetup{
    colorlinks=true,
    linkcolor=blue,
    filecolor=magenta,      
    urlcolor=cyan,
}
\newtcolorbox
{mybox}[2][]{colbacktitle=red!10!white,
colback=blue!10!white,coltitle=black!70!black,
title={#2},fonttitle=\bfseries,#1}
\usepackage{bchart}

\usepackage{xcolor}
\usepackage[numbers]{natbib}
\definecolor{ao(english)}{rgb}{0.0, 0.5, 0.0}

\usepackage{xcolor}
\definecolor{ao(english)}{rgb}{0.0, 0.5, 0.0}

\def\test#1{%
    \ifnum #1 > 0
      #1
    \fi
}

\newcommand{\sixbars}[6]{
{{\color{black}\rule{#1pt}{4pt}} \test{#1}}
{{\color{ao(english)}\rule{#2pt}{4pt}} \test{#2}}
{{\color{magenta}\rule{#3pt}{4pt}} \test{#3}}
{{\color{red}\rule{#4pt}{4pt}} \test{#4}}
{{\color{cyan}\rule{#5pt}{4pt}} \test{#5}}
{{\color{orange}\rule{#6pt}{4pt}} \test{#6}}
}
\newcommand{\rev}[1]{#1}
\begin{document}

\title{Developer Discussion Topics on the Adoption and Barriers of Low Code Software Development Platforms}
\titlerunning{Topics of Developer Discussions of the LCSD Platforms}
\author{Md Abdullah Al Alamin, Gias Uddin, Sanjay Malakar, Sadia Afroz, Tameem Haider, Anindya Iqbal}
\authorrunning{Alamin et al.}
\institute{Md Abdullah Al Alamin (corresponding author) and Gias Uddin \at
              DISA Lab, University of Calgary,
              \email{{mdabdullahal.alamin, gias.uddin}@ucalgary.ca}  \\
              Sanjay Malakar, Sadia Afroz, Tameem Bin Haider, and Anindya Iqbal \at
              Bangladesh University of Engineering \& Technology 
             }

\date{}
\maketitle

\sloppy

\begin{abstract}
Low-code software development (LCSD) is an emerging approach to democratize application development for software practitioners from diverse backgrounds. LCSD platforms promote rapid application development with a drag-and-drop interface and minimal programming by hand. As it is a relatively new paradigm, it is vital to study developers' difficulties when adopting LCSD platforms. Software engineers frequently use the online developer forum Stack Overflow (SO) to seek assistance with technical issues. We observe a growing body of LCSD-related posts in SO. This paper presents an empirical study of around 33K SO posts (questions + accepted answers) containing discussions of 38 popular LCSD platforms. We use Topic Modeling to determine the topics discussed in those posts. Additionally, we examine how these topics are spread across the various phases of the agile software development life cycle (SDLC) and which part of LCSD is the most popular and challenging. Our study offers several interesting findings. First, we find 40 LCSD topics that we group into five categories: Application Customization, Database and File Management, Platform Adoption, Platform Maintenance, and Third-party API Integration. 
Second, while the Application Customization (30\%) and Data Storage (25\%) \rev{topic} categories are the most common, inquiries relating to several other categories  (e.g., the Platform Adoption \rev{topic} category) have gained considerable attention in recent years.
Third, all topic categories are evolving rapidly, especially during the Covid-19 pandemic. Fourth, the How-type questions are prevalent in all \rev{topics}, but the What-type and Why-type (i.e., detail information for clarification) questions are more prevalent in the Platform Adoption and Platform Maintenance category. Fifth, LCSD practitioners find topics related to Platform Query the most popular, while topics related to Message Queue and Library Dependency Management as the most difficult to get accepted answers to. Sixth, the Why-type and What-type questions and Agile Maintenance and Deployment phase are the most challenging among practitioners. The findings of this study have implications for all three LCSD stakeholders: LCSD platform vendors, LCSD developers/practitioners, Researchers, and Educators. Researchers and LCSD platform vendors can collaborate to improve different aspects of LCSD, such as better tutorial-based documentation, testing, and DevOps support.

\end{abstract}

\section{Introduction}\label{sec:intro}
There is a massive shortage of skilled software developers \rev{in} this age of digitalization. According to Gartner, the demand for IT professionals will be multiple times more than supply~\cite{waszkowski2019low-automating, torres2018demand}. To make matters worse, training and hiring new software developers are very expensive in this rapidly evolving world. LCSD aims to address this issue by democratizing software development to domain experts and accelerating the development and deployment process. It tries to bridge the gap between the system requirement and the developer constraints, which is a common reason for long development times in complex business applications.

LCSD is a novel paradigm for developing software applications with minimal hand-coding through visual programming, a graphical user interface, and model-driven design. LCSD embodies End User Software Programming~\cite{Pane-MoreNatureEUSE-Springer2006} by democratizing application development to software practitioners from diverse backgrounds~\cite{di2020democratizing}.  It combines various approaches such as visual modeling, rapid app development, model-driven development, cloud computing, \rev{and} automatic code generation. Low-code development tools enable the development of production-ready apps with less coding by facilitating automatic code generation. Additionally, LCSD platforms also provide more flexibility and agility, faster development time that allows responding quickly to market needs, less bug fixing, less deployment effort, and easier maintenance~\rev{~\cite{sahay2020supporting, di2020democratizing}}. These platforms are used to develop high-performance database-driven mobile and online applications for various purposes. As a result, low-code development is rapidly growing in popularity. According to Forrester, the LCSD platform market is estimated to reach \$21 billion by 2022. By 2024, over 65\% of big companies will utilize LCSD systems to some extent, according to a Gartner report~\cite{wong2019low}.

To date, there are more than 400  LCSD platforms~\rev{~\cite{total_low_code}}, offered by almost all major companies like Google~\cite{googleappmaker} and Salesforce~\cite{salesforce}. 
Naturally,  LCSD has some unique challenges~\cite{sahay2020supporting}. Wrong
choice of  LCSD application/platforms may cause a waste of
time and resources. There is also concern about the security/scalability of
 LCSD applications~\cite{lowcodetesting}. With interests in  LCSD growing, we observe discussions about  LCSD platforms are becoming prevalent in online developer forums like Stack Overflow (SO). SO is a large online technical Q\&A site with around 120 million posts and 12 million registered users~\cite{website:stackoverflow}. Several research has been conducted to
analyze SO posts (e.g., IoT~\cite{iot21}, big data~\cite{bagherzadeh2019going}, blockchain~\cite{wan2019discussed} concurrency~\cite{ahmed2018concurrency}, , microservices~\cite{bandeira2019we}). 
The studies, however, did not analyze discussions about LCSD platforms in SO.

In 2021, we conducted an empirical study~\cite{alamin2021empirical} by analyzing 4,785 posts (3,597 questions + 1,118 accepted answers) from SO that contained discussion about nine LCSD platforms. The study offered, for the first time, an overview of the challenges software developers face while using LCSD platforms. However, to date, there are over 400 LCSD platforms and we observed discussions about many of those platforms in SO. Therefore, it was important that we revisit our empirical study with a larger dataset of discussions about LCSD platforms in SO. In addition, given that the previous empirical study was a conference paper, the analysis was not as in-depth as we could have provided due to \rev{space limitations}. Therefore, a \rev{larger-scale} empirical study of the challenges developers face to adopt and use the LCSD platforms was warranted. Such insights can complement our previous empirical study~\cite{alamin2021empirical} as well as the existing  LCSD literature -- which so far has mainly used surveys or controlled studies to understand the needs of low-code practitioners~\cite{lowcodeapp,kourouklidis2020towards,alonso2020towards,lowcodetesting}.

Specifically, in this paper, we present an empirical study of 33.7K SO posts relating to the top 38 LCSD platforms (according to Gartner~\cite{gartner}) at the time of our analysis to ascertain the interest and challenges of LCSD practitioners. \rev{We answer five research questions by analyzing the dataset.}

\nd\bf{RQ1. What topics do LCSD practitioners discuss?}
Given that LCSD is a novel paradigm, it is vital to study the types of topics discussed by LCSD practitioners on a technical Q\&A platform such as SO. As a result, we use the topic modelling method LDA~\cite{blei2003latent} on our 33.7K post dataset. We find a total of 40  LCSD topics grouped into five categories: Application Customization (30\% Questions, 11 Topics),  Data Storage (25\% Questions, 9 Topics), Platform Adoption (20\% Questions, 9 Topics),  Platform Maintenance (14\% Questions, 6 Topics), and Third-Party Integration (12\% Questions, 5 Topics). Around 34\% of questions are particular to the many supported capabilities of LCSD platforms, while the remaining 66\% are regarding development activities, namely application customization. This is because the LCSD platform's features are naturally oriented around a graphical user interface (GUI) in a drag-and-drop environment. As a result, any customization of such features that are not native to the LCSD platforms becomes difficult.

\nd\bf{RQ2. How do the LCSD topics evolve over time?}
We elaborate \rev{on} our findings from RQ1 by examining how the observed LCSD topics evolved in SO over time. We conduct an in-depth analysis of LCSD-related discussions from 2008 to \rev{mid-2021} in SO. We discover that since 2012, discussion about LCSD has piqued community interest, which has increased significantly throughout the pandemic, i.e., since 2020. In recent years, Platform Adoption-related discussions have acquired more traction than original application customization or database query-related discussions. Future research and LCSD platform vendors should support emerging topics such as Library Dependency Management, External Web Request Processing, Platform Infrastructure API, \rev{and} Data Migration.

\nd\bf{RQ3. What types of questions are asked across the observed topic categories?}
From RQ1, we find some of the unique challenges for LCSD practitioners regarding Customization, Data Storage on the completely managed cloud platforms. This motivates us to explore further to understand more of those challenges. For instance, we want to understand if practitioners mostly ask about different solution approaches (i.e., How-type) or further explanation clarification type (Why/What-type). Following previous studies\cite{iot21, abdellatif2020challenges}, we manually annotated a statistically significant number of posts (e.g., 471 Questions) into four categories. We find that How-type (57\%) is the most common form of inquiry across all five \rev{topic categories}, followed by What-type (18\%), Why-type (14\%), and Other-type (12\%) questions. Most of the How-type questions are application implementation-related, and most of the What-type and Why-type Questions are server configuration and troubleshooting related. According to our findings, proper documentation and tutorials might significantly reduce these challenges.

\nd\bf{RQ4. How are the \rev{observed topic categories} discussed across SDLC phases?}
Our findings from the previous research questions examined the practitioners' challenges on LCSD platforms and their evolution. The acceptance of this emerging technology depends largely on effective adoption into the various stages of a software development life cycle (SDLC). So, following our previous study~\cite{alamin2021empirical} we manually annotate statistically significant samples (e.g., 471 Questions) into six agile SDLC stages. We find that the Implementation (65\%) is the most prominent phase in terms of the number of questions, followed by Application Design (17\%) and Requirement Analysis \& Planning (9.1\%).

\nd\bf{RQ5.What LCSD topics are the most difficult to answer?}
LCSD practitioners face many different challenges to understand different features of the cloud platform, server configuration. LCSD vendors aim to provide support from requirement gathering to deployment and maintenance, but practitioners still struggle with customization, data management, and cloud configuration. We find that, while the topic of application customization and the Implementation-SDLC are the most prevalent, Platform Adoption \rev{topic} category and the Deployment-SDLC and Maintenance-SDLC as the most popular and hardest \rev{to get accepted answers.}

This paper extends our previous paper~\cite{alamin2021empirical} along two major dimensions: the data used and the results reported. We offer details about the extensions below.
\begin{enumerate}
\item\bf{Data (see \sec\ref{sec:methodology}).} The dataset in this study is significantly larger and more diverse than our previous paper as follows.
\begin{itemize}
    \item \bf{Size.} The size of the SO dataset in this paper is almost seven times bigger than the dataset used in our previous paper. This study analyses 33766 posts (26763 Questions + 11355 Accepted Answers). Our prior paper examined 4,785 posts (3597 Q + 1188 A).
    \item\bf{Time.} This study analyzes\rev{LCSD-related} discussions in SO between July 2008 to May 2021, while the previous study analyzed the discussions between July 2008 to May 2020.
    \item\bf{LCSD Platforms.} This study analyzes 64 \rev{LCSD-related} tags which contain 38 LCSD platforms, while the previous study analyzed 19 SO tags related \rev{to} 9 LCSD platforms.
\end{itemize}
\item\bf{Empirical Study (see \sec\ref{sec:results}).} This paper considerably enhances our understanding of LCSD platforms over our previous paper \cite{alamin2021empirical} as follows.
\begin{itemize}
    \item \bf{Research Questions (RQ).} We have answered five research questions (RQ2, RQ3) in this paper compared to three RQs in our previous paper (RQ1, RQ4, RQ5). The two new RQs offer insights on the type of LCSD questions asked and the evolution of the LCSD topics. Our revision of the previous three RQs provided several new results as follows.
    \item \bf{LCSD Topics.} In this study, we found 40 \rev{topics} organized into five high-level categories. We found 13 \rev{topics} organized into four high-level categories in our previous paper. While we found all the previous 13 topics, we also found 27 new LCSD topics. This study found  Platform Maintenance as a new high-level topic category (see \sec\ref{sec:rq_topic}). 
    \item \bf{Finer-Grained Analysis}. Due to our use of more data, we find better results from our topic modeling. For example, some topics from our previous studies are broken down into more informative/coherent topics. For example, Client-Server Communication and IO from Platform Adoption \rev{topic category} became \rev{topics} Web Service Communication and Message Queue under Asynchronous Service to Service Communication sub-sub-category in this study as those topics contained more coherent discussions. Similarly, we have expanded our understanding of software development lifecycle phases (SDLC) around the new 40 topics (see \sec\ref{rq:rq_sdlc}).
        \item \bf{Topic Evolution.} Our new RQ2 analyzes the evolution of the observed LCSD topics in \sec\ref{rq:evolve}. We further \rev{discuss} the prevalence and evolution of the topics across the top 10 LCSD platforms in our dataset (see \sec\ref{sec:evolutionTopLCSDPlatforms}).
    \item \bf{Question Type.} Our new RQ3 offers insights into the type of questions asked across the observed LCSD topics (see \sec\ref{rq:type}). 
    \item \bf{Popularity vs Difficulty}. In addition to analyzing the popularity and difficulty of all 40 topics in \sec\ref{rq:pop_diff}, we also offer the following new insights. \begin{inparaenum}
    \item Following a recent study~\cite{iot21}, we report the popularity and the difficulty using two fused metrics (see \sec\ref{rq:pop_diff}).
    \item We report the popularity and difficulty of the LCSD question types and SDLC phases (see \sec\ref{sec:implications})
    \end{inparaenum}

\end{itemize}
\item \bf{Related Work.} We have expanded our literature review with a comparison of key metrics around our observed LCSD topics against those previously reported for other domains while using the SO data (see \sec\ref{sec:rel-se-res-topic}).
\end{enumerate}

Our study findings can enhance our understanding of the developers' struggle while using LCSD platforms. The findings would help the research community and platform vendors better focus on the specific LCSD areas. The practitioners can prepare for difficult areas. LCSD platforms can design more effective and usable tools. All stakeholders can collaborate to provide enhanced documentation assistance. The  LCSD vendors can support increased customization of the  LCSD middleware and UI to make the provided functionalities more usable.


\nd\bf{Replication Package}: The code and data are shared in \url{https://github.com/al-alamin/LCSD_challenge_EMSE}


\section{Background} \label{sec:background}
\rev{This section aims to provide a high-level overview of LCSD development, as well as some of the relevant technologies and research that have shaped this industry. We hope that this will serve as a resource for future researchers (particularly those interested in the underlying technologies)) and practitioners to learn and contribute more to this emerging new field.}

\nd\bf{Low-code Software Application.} To cater to the demand of the competitive market, business organizations often need to quickly develop and deliver customer-facing applications.  LCSD platform allows the quick translation of the business requirement into a usable software application. It also enables citizen developers of varying levels of software development experience to develop applications using visual tools to design the user interface in a drag-and-drop manner and deploy them easily~\cite{lowcodewiki}.  LCSD is inspired by the model-driven software principle where abstract representations of the knowledge and activities drive the development, rather than focusing on algorithmic computation~\cite{sahay2020supporting}.  LCSD platforms aim to abstract away the complexity of testing, deployment, and maintenance that we observe in traditional software development. Some of the most popular low-code platforms are Appian~\cite{appian}, Google App Maker~\cite{googleappmaker}, Microsoft Powerapps~\cite{powerapps}, and Salesforce Lightning~\cite{salesforce}.

\nd\bf{Technologies that Shaped LCSD.} \rev{Model-driven} Software Engineering (MDSE) field proposes \rev{the} adoption of domain-specific modeling practices \cite{brambilla2017modelmdse}. Low-code platforms adopt model-driven engineering (MDE) principles as their core that \rev{has} been applied in several engineering disciplines for the purpose of automation, analysis, user interface design~\cite{botterweck2006model, pleuss2013model, brambilla2017model} and abstraction possibilities enabled by the adoption of modelling and meta modeling \cite{basciani2014mdeforge}. Besides, End-User Development (EUD) is a set of methods, techniques, and tools that allow users of software systems, who are mainly non-professional software developers, at some point to create, modify or extend a software \rev{artifact} \cite{paterno2013end, fischer2004meta}. EUD for GUIs can be \rev{a} good example of \rev{its} usage~\cite{costabile2007visual}. Scratch \cite{resnick2009scratch}, Bloqqi \cite{fors2016design}, EUD-MARS \cite{akiki2020eud}, App Inventor \cite{wolber2011app}, AppSheet \cite{googleAppSheet} are such ``low-code/no-code'' application development tools that offer visual drag-and-drop facilities. \rev{ Similarly, there are several other research areas within the domains of HCI~\cite{sinha2010human} and Software engineering, such as Visual Programming~\cite{burnett1995visual}, Programming by example~\cite{halbert1984programming}, End users programming~\cite{myers2006invited}, domain specific language~\cite{mernik2005and, van2000domain}, trigger action programming~\cite{ur2014practical} that aim to enhance the technologies underlying low-code software development.} Thus, gaining a better knowledge of the problems associated with low-code platforms through developer discussion would be extremely beneficial for further improving these studies.

\begin{figure}[t]
\centering
\includegraphics[scale=.50]{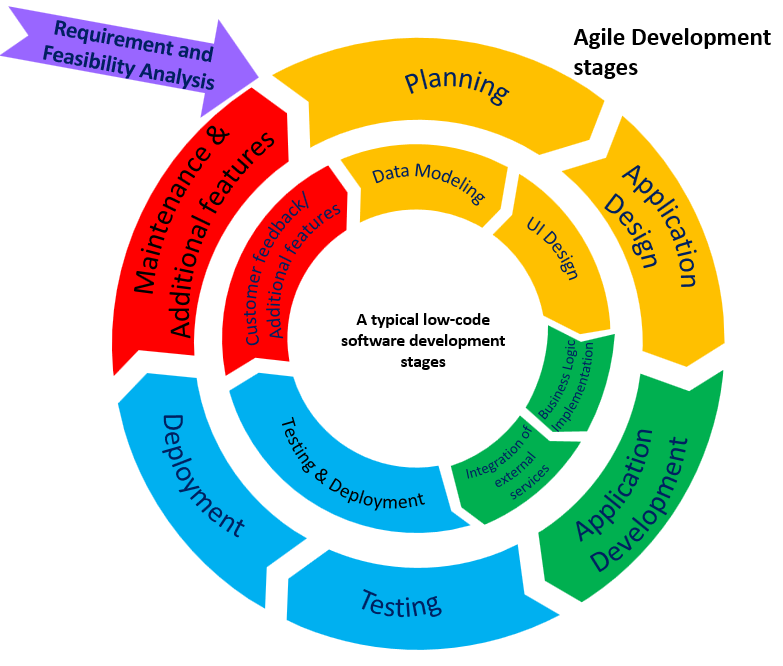}
\caption{Agile methodologies in traditional vs  LCSD development}
\label{fig:low-code-agile}
\vspace{-5mm}
\end{figure}

\nd\bf{Development Phases of an  LCSD Application.} A typical  LCSD application can be built in two ways~\cite{sahay2020supporting}: \begin{inparaenum}
\item ``UI to Data Design'', where developers create UI and then connect the UI to necessary data sources, or \item ``Data to UI'' where the design of the data model is followed by the design of the user interfaces. \end{inparaenum} In both approaches, application logic is implemented, and then \rev{third-party} services and APIs are integrated. APIs are interfaces to reusable software libraries~\cite{Robillard-APIProperty-IEEETSE2012}.  
A major motivation behind  LCSD is to build applications, get reviews from the users, and incorporate those changes quickly~\cite{waszkowski2019low-automating}. Some software development approaches are quite popular and supported by different LCSD platforms, such as Iterative software development~\cite{basil1975iterative} which is based on the iterative development of the application. In this way, every step is cyclically repeated one after another. In practice, this is very helpful because it allows developing and improving the application gradually. Another approach can be Rapid application development (RAD)~\cite{beynon1999rapid} is a software development methodology that promotes the rapid release of a software prototype. It is an agile approach and aims to utilize user feedback from the prototype to \rev{deliver a better product}. Another popular methodology is the agile development methodology~\cite{beck2001manifesto} which is a collection of approaches and \rev{practices} that promote the evolution of software development through collaboration among cross-functional teams.

Different LCSD teams may adopt different SDLC approaches. However, we focus mostly on Agile methodology for this study because Agile and  LCSD can go hand in hand because the fundamental principle and objective are customer satisfaction and continuous incremental delivery. Traditional software development teams widely use agile, which also provides the generalizability for other methodologies. So, in this study, we map agile software development life cycle phases with LCSD methodologies. The inner circle of Figure~\ref{fig:low-code-agile} shows the important development phases of an  LCSD application, as outlined in~\cite{sahay2020supporting}. The outer circle of \fig\ref{fig:low-code-agile} shows the phases in a traditional agile software development environment. As  LCSD platforms take care of many application development challenges, some of the agile application development phases have shorter time/execution spans in  LCSD than traditional software development.

\section{Study Data Collection and Topic Modeling} \label{sec:methodology}
In this Section, we discuss our data collection process to find  LCSD related posts (Section~\ref{sub-sec:data_collection}). We then discuss the details about our pre-processing and topic modeling steps on the selected posts (Section~\ref{sub-sec:topic_modeling}).

\begin{sidewaysfigure}
\vspace{12cm}
\includegraphics[scale=.45]{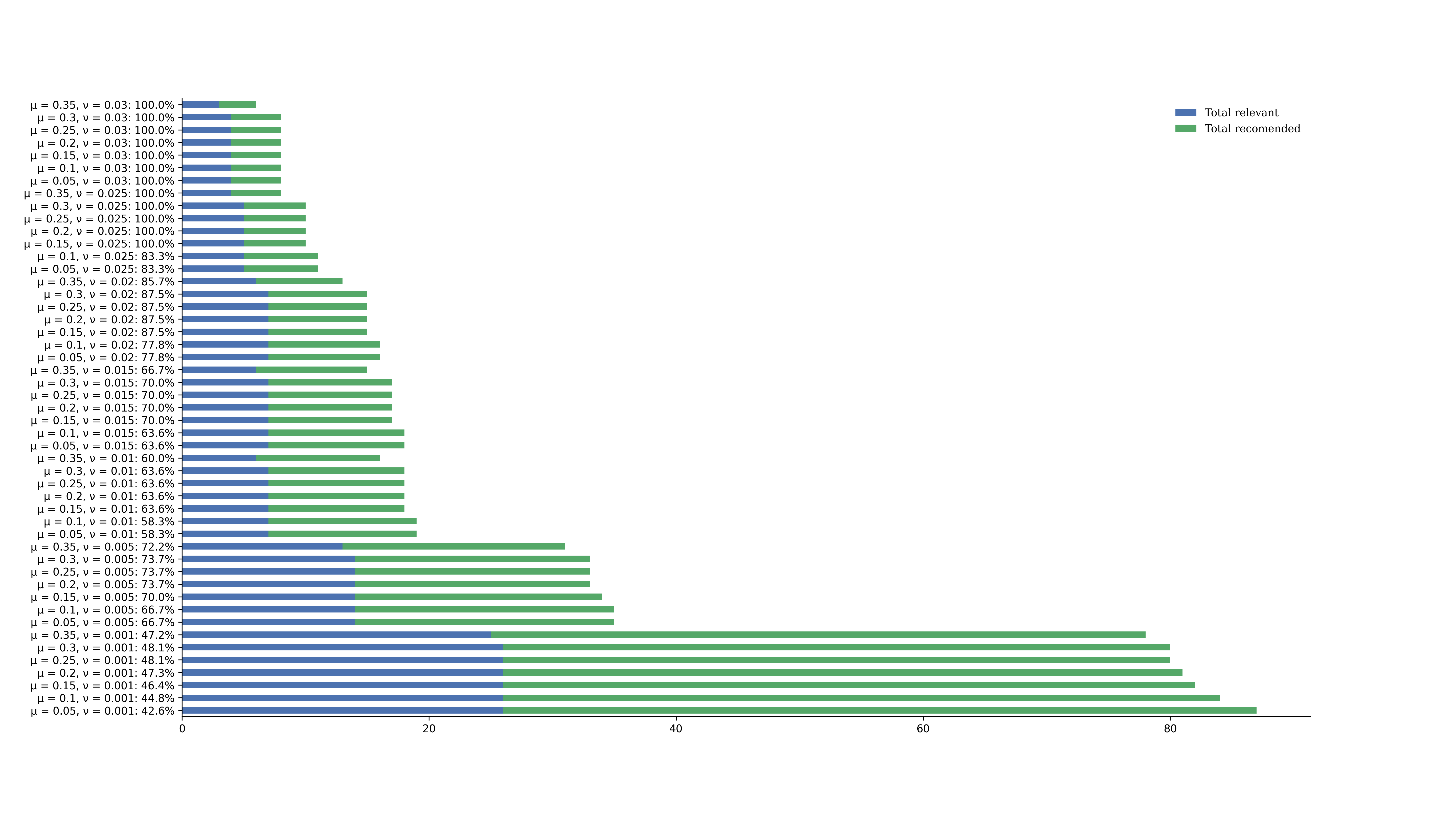}
\caption{Distribution of LCSD related  recommended vs relevant tags  based on different $u$ and $V$ values from our initial taglist.}
\label{fig:significance_vs_relevance}
\end{sidewaysfigure}

\subsection{Data Collection} \label{sub-sec:data_collection}

We collect  LCSD related SO posts in three steps: \begin{inparaenum}[(1)]
\item Download SO data dump,
\item Identify  LCSD related tag list, and
\item Extract  LCSD related posts from the data dump based on our selected tag list.
\end{inparaenum} We describe the steps below.

\nd\textbf{Step 1: Download SO data dump.} For this study, we used the most popular Q\&A site, Stack Overflow (SO), where developers from diverse background \rev{discuss} about various software and hardware related issues~\cite{website:stackoverflow}. For this study, We downloaded SO data dump~\cite{SOdump} of May 2021 which was the latest dataset available during the starting of this study. We used the contents of ``Post.xml'' file, which contained information about each post like the post's unique ID, type (Question or Answer), title, body, associated tags, creation date, view-count, etc. Our data dump included discussion of 12 years from July 2008 to July 2021 and contained around 53,086,327 posts. Out of them, 21,286,478 (i.e., 40.1\%) are questions, 31,799,849 (i.e., 59.9\%) are answers, and 51.5\% questions had accepted answers. Around 12 million users from all over the world participated in the discussions.

Each SO post contains 19 attributes, and some of the relevant are: \begin{inparaenum}[(1)]
        \item Post's body with code snippets,
        \item Post's Id, creation and modification time,
        \item Post's view count, favorite count, score,
        \item User Id of the creator,
        \item Accepted answer Id and a list of 0 to 5 tags.
    \end{inparaenum}

\nd\textbf{Step 2: Identify low-code tags.}
We need to identify the tags that are related to LCSD in order to extract low-code related posts from SO discussions. To find relevant tags, we followed a similar procedure used in prior work~\cite{alamin2021empirical, iot21,  abdellatif2020challenges, ahmed2018concurrency, wan2019discussed, linares2013exploratory}. At Step 1, we identify the initial low-code related tags and call them $T_{init}$. At Step 2, we finalize our low-code tag list following related work~\cite{bagherzadeh2019going, yang2016security}. Our final tag list $T_{final}$ contains 64 tags from the top 38 LCSD platforms. We discuss each step in details below.

(1) Identifying Initial low-code tags.
The SO posts do not have tags like ``low-code'' or ``lowcode''. Instead, we find that low-code developers use an LCSD platform name as a tag, e.g., ``appmaker'' for Google Appmaker~\cite{googleappmaker}. 
Hence, to find relevant tags, first, we compile a list of top  LCSD platforms by analyzing a list of platforms that are considered as the market leaders in Gartner~\cite{vincent2019magic}, Forrester~\cite{rymer2019forrester}, related research work~\cite{sahay2020supporting}, and other online resources like PC magazine~\cite{pcmag}. Our compiled list contained 137 LCSD platforms, including all of our previous nine platforms from previous study~\cite{alamin2021empirical}. Then for each of the LCSD platforms, we manually searched for the SO tags \rev{in SO. For example, we search for Oracle Apex via SO search engine and find a list of SO posts. We build a potential list of tags related with this platform based on manual inspection, such as ``oracle" and ``oracle-apex". Then, manually examine the metadata associated with each of these tags \footnote{https://meta.stackexchange.com/tags}. For example, ``oracle-apex'' tag's metadata says ``Oracle Application Express (Oracle APEX) is a rapid Web application development tool that lets you share data and create applications. Using only a Web browser and limited programming experience, you can develop and deploy applications that are fast and secure.'' and ``oracle'' tag's metadata says ``Oracle Database is a Multi-Model Database Management System created by Oracle Corporation. Do NOT use this tag for other products owned by Oracle, such as Java and MySQL.''. Therefore, we select the ``oracle-apex'' tag for Oracle Apex platform.   Not all LCSD platforms have associated SO tags; thus, they were excluded. For example, OneBlink~\cite{oneblink} low-code platform there is no associated SO tags and thus we exclude this from our list.}  In order to better understand the evolution of this domain, we excluded discontinued LCSD platforms. For example, In Jan 2020, Google announced that they would no longer release new features for Google App Maker and discontinue it by 2021~\cite{google-disc} and so we excluded this platform from our list. Finally, we found 38 relevant SO tags from 38 platforms. The fifth and the first author participated in this step, and the complete list of the platforms and tags are available in our replication package.

So, our initial list contains 38 LCSD platforms such as: Zoho Creator~\cite{zohocreator}, Salesforce~\cite{salesforce}, Quickbase~\cite{quickbase}, Outsystems~\cite{quickbase}, Mendix~\cite{mendix}, Vinyl~\cite{vinyl}, Appian~\cite{appian}, and Microsoft Powerapps~\cite{powerapps}. We thus focus on the discussions of the above 38 LCSD platforms in SO. We find one tag per LCSD platform as the name of the platform (e.g., ``powerapps'' for Microsoft Powerapps platform). Thus, We refer to these 38 tags as $T_{init}$. 

(2) Finalizing low-code related tags.
Intuitively, there might be more variations to tags of 38  LCSD platforms other than those in $T_{init}$. We use heuristics from previous related works~\cite{alamin2021empirical, bagherzadeh2019going, yang2016security} to find other relevant tags. First, we denote our entire SO dump data as $Q_{all}$. Second, we extract all the questions $Q$ that contain any tag from $T_{init}$. Third, we create a candidate tag list $T_{candidate}$ using all the tags found in questions $Q$. Fourth, we select significantly relevant tags from  $T_{candidate}$ for our  LCSD discussions. Following related works~\cite{iot21, bagherzadeh2019going, yang2016security}, we compute significance and relevance for each tag $t$ in $T_{candidate}$ with respect to $Q$ (our extracted questions that has $T_{init}$ tag) and $Q_{all}$ (i.e., our data dump) as follows,
{ \[
( Significance) \ \ S_{tag} \ =\ \ \frac{\#\ of\ ques.\ with\ the\ tag\ t\ in\ Q}{\ \ \#\ of\ ques.\ with\ the\ tag\ t\ in\ Q_{all}}
\]

\[
( Relevance) \ \ R_{tag} \ =\ \ \frac{\#\ of\ questions\ with\ tag\ t\ in\ Q}{\ \ \#\ of\ questions\ in\ Q}
\]} A tag t is significantly relevant to  LCSD if the $S_{tag}$ and  $R_{tag}$ are higher than a threshold value. 
We experimented with a wide range of values of $S_{tag}$ = \{0.05, 0.10, 0.15, 0.20, 0.25, 0.30, 0.35\} and  $R_{tag}$ = \{0.001, 0.005, 0.010, 0.015, 0.020, 0.025, 0.03\}. Figure~\ref{fig:significance_vs_relevance} shows the total number of recommended vs relevant tags from our 49 experiments. It shows that as we increase $S_{tag}$ and  $R_{tag}$ the total number of recommend tags decreases. For example, we find that for $S_{tag}$=.05 and  $R_{tag}$ = 0.001 the total number of recommended tags is 61 which is highest. \rev{However, not all of the suggested tags are LCSD-related. For instance, according to our significance and relevance analysis, tags such as ``oracle-xe'', ``ems'', ``aura-framework'', ``power-automate'' etc are frequently correlated with other LCSD platform tags, although they do not contain low-code-related discussions.} After manually analysing these 61 tags we find that 26 tags are relevant to LCSD-related discussions. So, for the lowest $S_{tag}$ = 0.3 and $R_{tag}$ = 0.001 we find 26 additional LCSD-related tags. These values are consistent with related work~\cite{iot21, alamin2021empirical, bagherzadeh2019going, ahmed2018concurrency}. The final tag list $T_{final}$ contains 64 significantly relevant tags. So, after combining with out initial taglist, i.e., $T_{init}$, our final tag list $T_{final}$ contains 64 significantly relevant LCSD-related tags \rev{which are: 
\begin{description}
\item\{ apex-code, 
lotus-notes, 
domino-designer-eclipse, 
visualforce, 
salesforce-chatter, 
apex, 
salesforce-service-cloud, 
simple-salesforce, 
salesforce-ios-sdk, 
apex-trigger, 
oracle-apex-5, 
salesforce-lightning, 
salesforce-communities, 
oracle-apex-5.1, 
servicenow-rest-api, 
powerapps-formula, 
salesforce-marketing-cloud, 
powerapps-selected-items, 
powerapps-modeldriven, 
powerapps-collection, 
powerapps-canvas, 
oracle-apex-18.2, 
lwc, 
salesforce-development, 
oracle-apex-19.1, 
oracle-apex-19.2, 
outsystems, 
appian, 
quickbase, 
powerapps, 
oracle-apex, 
salesforce, 
zoho, 
mendix, 
servicenow, 
pega, 
retool, 
vinyl, 
kissflow, 
bizagi, 
neutrinos-platform, 
rad, 
joget, 
filemaker, 
boomi, 
opentext, 
tibco, 
webmethods, 
conductor, 
temenos-quantum, 
shoutem, 
oracle-cloud-infrastructure, 
amazon-honeycode, 
convertigo, 
lotus-domino, 
genero, 
genesis, 
gramex, 
processmaker, 
orocrm, 
slingr, 
unqork, 
uniface, 
structr\}
\end{description}
}

\nd\textbf{Step 3: Extracting low-code related posts.}
An SO question can have at most five tags, and we consider a question as low-code related question if at least one of its tag is in our chosen tag list $T_{final}$. Based on our $T_{final}$ tag set, we found a total of 27,880 questions from our data dump. SO has a score-based system (up-vote and down-vote) to ensure the questions are in proper language with necessary information (code samples and error messages), not repeated, off-topic \rev{ or incorrectly tagged}. Here is an example for a question with score ``-4'' where a practitioner is making an API related query in Powerapps(\dq{61147923})\footnote{$Q_i$ and $A_i$ denote a question Q or answer A in SO with an ID $i$} platform. However, it is not clear what the practitioner is asking as the question is poorly written and without any clear example. So, in order to ensure good quality discussions, we excluded questions that had a negative score. Following previous research~\cite{iot21, bagherzadeh2019going, rosen2016mobile, barua2014developers}, we also excluded unaccepted answers and only considered accepted answers for our dataset. Hence, our final dataset $B$ contained 37,766 posts containing 67.4\% Questions (i.e., 26,763) and 32.6\% Accepted Answers (i.e., 11,010).  

\rev{To ensure that our final taglist $T_{final}$ comprises discussions relating to low-code software development, we randomly select 96 questions from our dataset that are statistically significant with a 95\% confidence level and 10 confidence interval. First and third authors contributed to this manual analysis, and after manual analysis, we found that 98\% of questions from our selected taglist contain low-code platform-related discussion, with only two questions containing discussion that is not particularly related to low-code platforms. For instance, question \dq{59402662} includes the tag ``appian'', yet the question body describes only about a MySQL database performance-related issue on the Azure platform.  Similarly, the question \dq{19289762} contains the ``apex-code'' tag, but exclusively discusses AWS cloud authentication signature-related issues in its problem description.
}

\subsection{Topic Modeling} \label{sub-sec:topic_modeling}
We produce  LCSD topics from our extracted posts in three steps: \begin{inparaenum}[(1)]
\item Preprocess the posts, 
\item Find optimal number of topics, and
\item Generate topics.
\end{inparaenum} We discuss the steps below.

\nd\textbf{Step 1. Preprocess the posts.} For each post text, we remove noise following related works~\cite{abdellatif2020challenges,bagherzadeh2019going,barua2014developers}. First, we remove the code snippets from the body, which is inside \textless code\textgreater \textless /code\textgreater\ tag, HTML tags such as (\textless p\textgreater \textless /p\textgreater, \textless a\textgreater \textless /a\textgreater, \textless li\textgreater \textless /li\textgreater\ etc), and URLs. Then we remove the stop words such as ``the'', ``is'', ``are'', punctuation marks, numbers, non-alphabetical characters using the stop word list from MALLET~\cite{mccallum2002mallet}, NLTK~\cite{loper2002nltk}, and our custom low-code specific (i.e.,  LCSD platform names) stop word list. \rev{We remove the platform's name from the dataset since, based on our experiments with LDA topic modeling for this study and our past work~\cite{alamin2021empirical}, the resultant topics sometimes tend to cluster around LCSD platforms rather than the technical challenges discussed. Thus, we remove the LCSD platform names from our dataset.}  After this, we use porter stemmer~\cite{ramasubramanian2013effective} to get the stemmed representations of the words e.g., ``wait'', ``waits'', ``waiting'', and ``waited'' - all of which are stemmed to base form ``wait''.

\nd\textbf{Step 2. Finding the optimal number of topics.}  After the prepossessing, we use Latent Dirichlet Allocation~\cite{blei2003latent} and the MALLET tool~\cite{mccallum2002mallet} to find out the  LCSD-related topics in SO discussions. We follow similar studies in Software engineering research using topic modeling~\cite{arun2010finding, asuncion2010software, yang2016security, bagherzadeh2019going,abdellatif2020challenges}. Our goal is to find the optimal number of topics $K$ for our dataset $B$ so that the \textit{coherence} score, i.e., encapsulation of underlying topics, is high. We use Gensim package~\cite{rehurek2010software} to determine the coherence score following previous works~\cite{uddin2017automatic, roder2015exploring}. We experiment with different values of $K$ that range from \{5, 10, 15, 20, 25, 30, 35, 40, 45, 50, 55, 60, 65, 70\} and for each value, we run  MALLET LDA on our dataset for 1000 iterations~\cite{bagherzadeh2019going}. Then we observe how the coherence score is changing with respect to $K$. We pick the topic model with the highest coherence score. Choosing the right value of $K$ is important because, for smaller values of $K$, multiple real-world concepts merge, and for a large value of $K$, a topic breaks down. \rev{For example, in our result, the highest coherence score 0.50  for $K$ = 45 and $K$ = 40. The first, third, fourth, and fifth authors participate in the analysis  and we choose $K$ = 45 as it captures our underlying topics better.} MALLET also uses two hyper-parameters, $\alpha$ and $\beta$, to distribute words and posts across the topics. Following the previous works~\cite{alamin2021empirical, iot21, bagherzadeh2019going, ahmed2018concurrency, bajaj2014mining, rosen2016mobile}, we use the standard values $50/K$ and 0.01 for hyper-parameters $\alpha$ and $\beta$ in our experiment.  

\nd\textbf{Step 3. Generating topics.} Topic modeling is a method of extracting a set of topics by analysing a collection of documents without any predefined taxonomy. Each document has a probability distribution of topics, and every topic has a probability distribution of a set of related words~\cite{barua2014developers}. We produced 45 topics using the above LDA configuration on our dataset $B$. Each topic model offers a list of top $N$ words and a list of $M$ posts associated with the topic. In our settings, a topic consists of 30 most frequently co-related words, which represent a concept. Each post had a correlation score between 0 to 1, and following the previous work~\cite{alamin2021empirical, iot21,  wan2019discussed}, we assign a document with a topic that it correlates most.

\section{Empirical Study}


\label{sec:results}
We report the results of an empirical study by answering the following five research questions (RQ) based on our analysis of LCSD topics in our dataset.
\begin{enumerate}[leftmargin=30pt, label=\bf{RQ\arabic{*}.}]
  \item What topics do LCSD practitioners discuss? (\sec\ref{sec:rq_topic})
  \item How do the LCSD topics evolve over time in SO? (\sec\ref{rq:evolve})
  \item What types of questions are asked across the \rev{observed topic categories?} (\sec\ref{rq:type})
  \item How are the \rev{observed topic categories} discussed across SDLC phases? (\sec\ref{rq:rq_sdlc})
  \item What LCSD topics are the most difficult to answer? (\sec\ref{rq:pop_diff})
\end{enumerate}
The first two research questions (RQ1, RQ2) provide insights about what topics practitioners discuss in SO and how these topics have evolved over time. The third and fourth research questions (RQ3, RQ4) explore the types of questions in these topics and they affected different SDLC phases. At the end, we discuss the popularity and difficulty of the LCSD topics in the last research question (RQ5).


\subsection{What topics are discussed about 
LCSD in Stack Overflow? (RQ1)} \label{sec:rq_topic}
\subsubsection{Motivation}
The increased popularity of LCSD as a flexible and straightforward approach helps \rev{develop} practical business applications. The challenges of LCSD are yet to be studied as this is a new approach to software development. SO is an established source of knowledge repository to systematically study the real-world challenges that the practitioners face. An understanding of the LCSD topics in SO developer discussions will
help LCSD platform providers and researchers to have a better understanding of
the underlying prevalent issues, which can then help guide efforts to improve the quality of LCSD platforms.

\subsubsection{Approach}
\label{sub-sec:lda-labeling}
We applied LDA topic modeling to our LCSD-related discussion in SO. We get 45 low-code related topics from our LDA topic modeling, as discussed in \sec\ref{sec:methodology}. We use card sorting~\cite{fincher2005making} to label these topics  following previous works ~\cite{bagherzadeh2019going, ahmed2018concurrency, yang2016security, rosen2016mobile, abdellatif2020challenges}.
In open card sorting, there is no predefined list of labels. Following related works  \cite{alamin2021empirical,iot21,bagherzadeh2019going,abdellatif2020challenges}, we label each topic by analyzing the top 30 words for the topic and a random sample of at least 20 questions that are assigned to the topic. Four of the authors participated in the labeling process in group sessions (first, third to fifth). Each author assigns a label to each topic and discusses it until there is an agreement. The authors reached an agreement after around 10 iterations of meetings over Skype and email and labeled the 45 topics from the LDA output. 

After this initial labeling, we merged a few topics because they contained similar discussions with different vocabularies. For example, we merged topic 36 and 43 into Dynamic form controller because both topics contained discussions related to forms with a predefined list of values, dynamically changing the fields (or options) of forms values based on users' actions or previous selections. Similarly, we merged topic 2 and 19 to DB Setup \& Migration. In the end, we obtained 40 distinct LCSD-related topics.

After the labeling of the topics, we revisited the labels \rev{in} an attempt to find any clusters/groups among the topics. For example, Date \& Time Manipulation, Formatted Data Parsing, and Pattern Matching topics are related, and thus, they are grouped under the General Programming category. We repeated this process multiple times to find increasingly higher categories. For example, we found another category called Dynamic Content which contained two topics Dynamic Data Binding and Dynamic Data Filtering. We then put these two categories under called Business Logic Implementation. This higher abstraction helped us to place other topics related to implementing business logic under this category. Following a similar strategy, we put this Business logic implementation under the Customization category, which discussed customizing applications. For example, under Customization, there were a category called UI which contained Dynamic Layout, and Script category, which contained topics such as Dynamic Page Layout, Dialog Box Manipulation, Window Style Manipulation, and Dynamic Form Controller. The entire process of creating this hierarchy of topic categories took multiple iterations and revisions. We created a coding guideline for creating the taxonomy of topics to ensure consistency and reproducibility. We share the coding guide with our replication package.

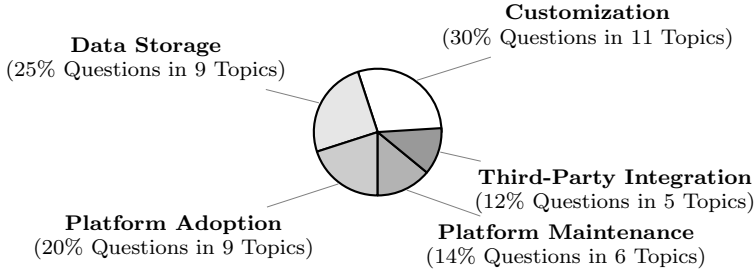
\begin{figure}[t]
	\centering\begin{tikzpicture}[scale=0.28]-
    \pie[
        /tikz/every pin/.style={align=center},
        text=pin, number in legend,
        explode=0.0,
        color={black!0, black!10, black!20, black!30, black!40,  black!50},
        ]
        {
            30/\bf{Customization} \\ (30\% Questions in 11 Topics),
            25/\bf{Data Storage} \\ (25\% Questions in 9 Topics),
            20/\bf{Platform Adoption}\\ (20\% Questions in 9 Topics),
            14/\bf{Platform Maintenance}\\ (14\% Questions in 6 Topics),
            12/\bf{Third-Party Integration}\\ (12\% Questions in 5 Topics)
        }
    \end{tikzpicture}
	\caption{Distribution of Questions and Topics per Topic Category}
	\vspace{-5mm}
	\label{fig:distribution_of_questions_topic_categories_pie_chart}
\end{figure}

\begin{figure}[t]
\centering
\includegraphics[scale=0.65]{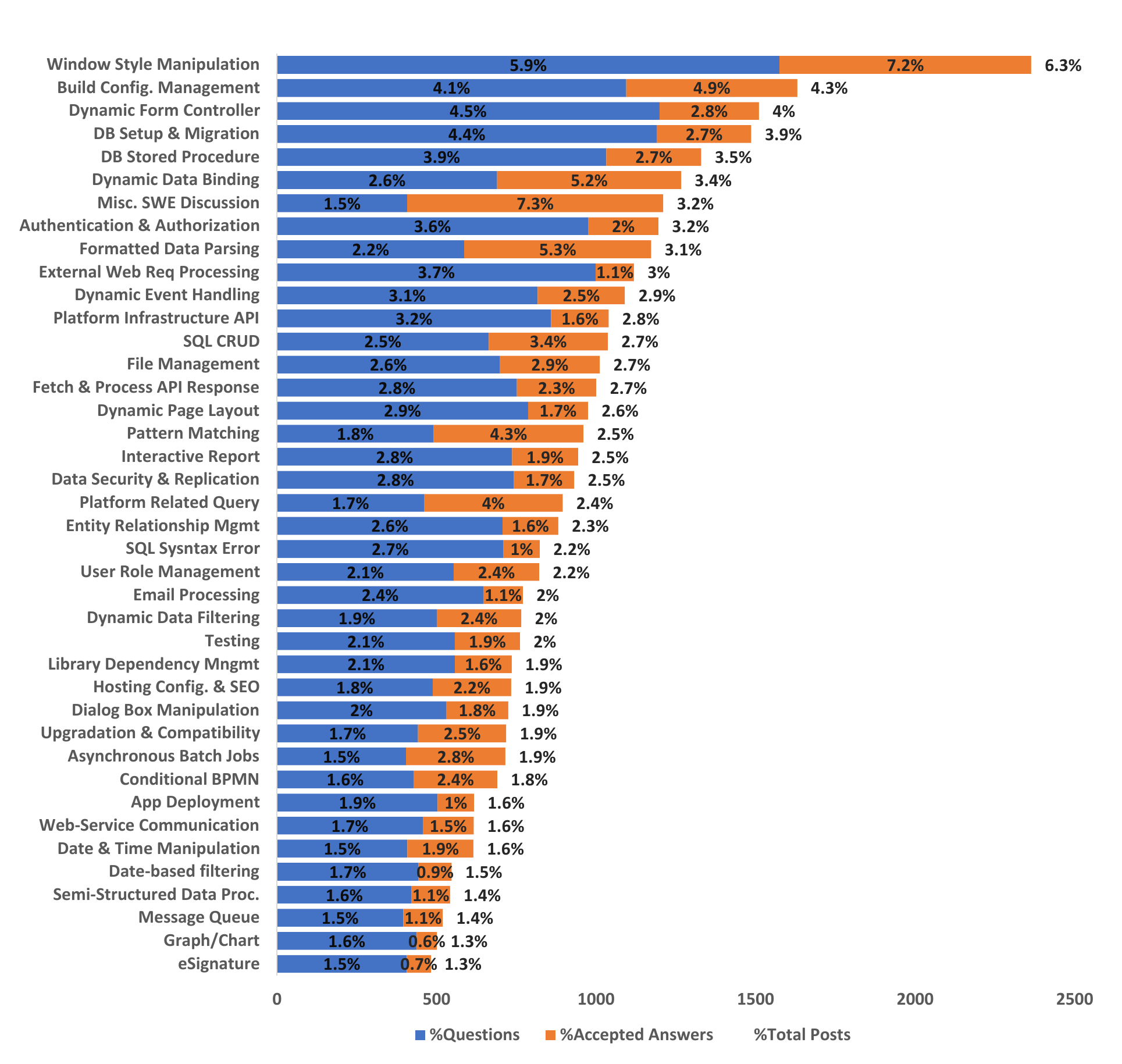}
\caption{Distribution of questions by low-code topics by total number of posts (Number of Questions + Number of Accepted Answers)}
\label{fig:distribution_of_topic_posts_bar_chart}
\end{figure}
 
\begin{figure}[t]
\centering
\includegraphics[scale=0.50]{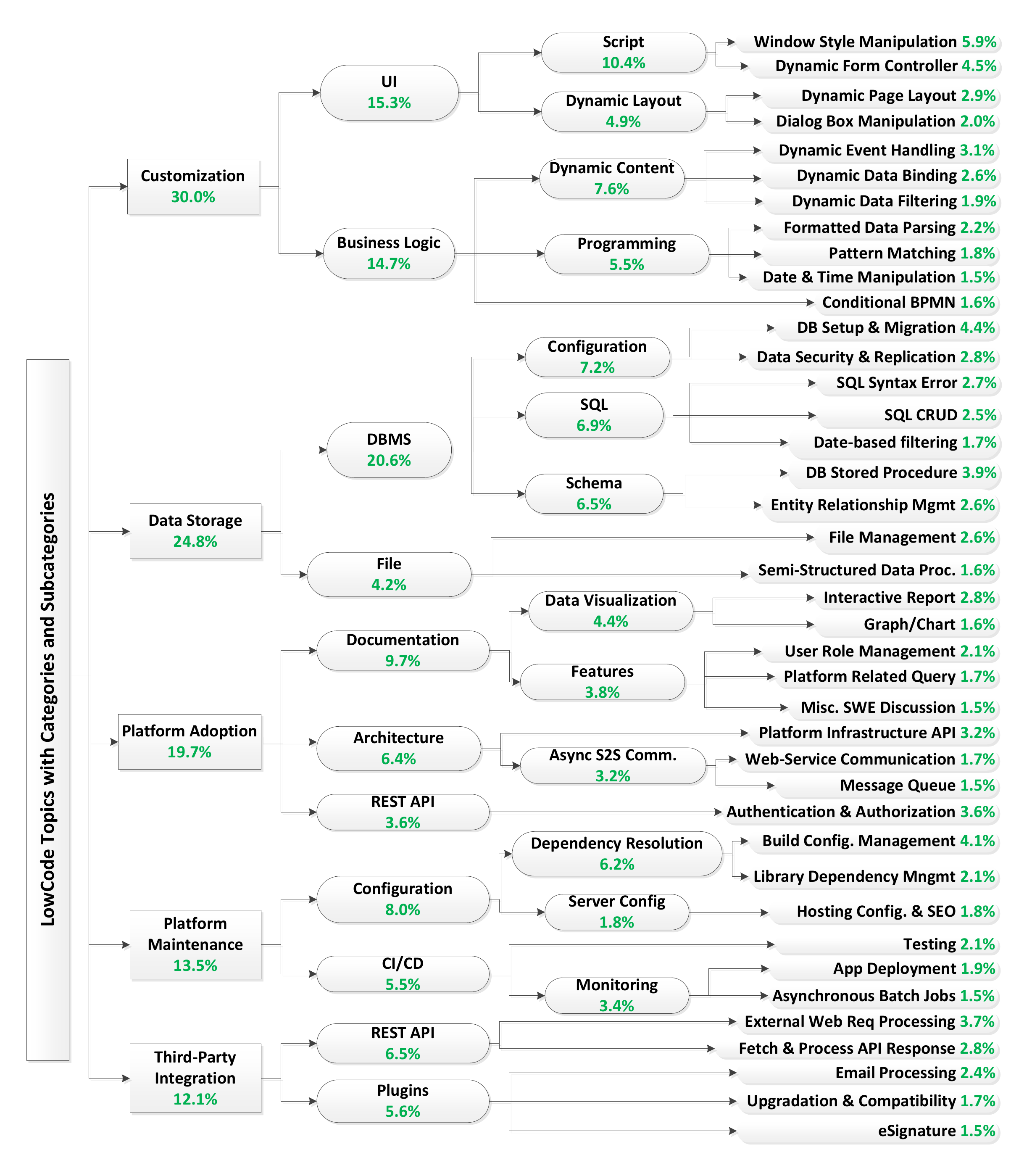}
\caption{ A taxonomy of LCSD topics with sub-categories.}
\label{fig:taxonomy_TM}
\end{figure}



\subsubsection{Results}
After labeling and merging, we find 40 LCSD-related topics. Then after grouping these topics into higher categories, we find five high-level categories: \begin{inparaenum}[(1)]
\item Customization, 
\item Data Storage, 
\item Platform Adoption, 
\item Platform Maintenance, and 
\item Third-Party Integration
\end{inparaenum}. Figure \ref{fig:distribution_of_questions_topic_categories_pie_chart} shows the distribution of topics and questions into these five categories. Among these categories, Customization has the highest coverage of questions and topics (30\% Questions in 11 Topics), followed by Data Storage (25\% Questions in 9 Topics), Platform Adoption (20\% Questions in 9 Topics), Platform Maintenance (14\% Questions in 6 Topics), Third-Party Integration (12\% Questions in 5 Topics).

Figure~\ref{fig:distribution_of_topic_posts_bar_chart} shows the 40 LCSD topics sorted by numbers of posts. A post means an LCSD-related question or an accepted answer in our case. As discussed in Section \ref{sub-sec:data_collection}, our dataset has total 37,773 posts containing 26,763 questions and 11,010 accepted answers. The topic with the highest number of posts is placed on top of the list. On average, each topic has 944 posts (question + accepted answer). The topic Window Style Manipulation has the highest number of posts (6.3\%) with Questions 5.9\% of total questions and 7.2\% total accepted answers. On average, each topic has around 669 questions.

Figure \ref{fig:taxonomy_TM} provides a taxonomy of 40 LCSD related topics into five categories. The topics are organized in descending order of the number of questions. For example, the Customization category has the highest number of questions, followed by Data Storage. Each category may have some sub-categories of topics. For example, the Customization category has two sub-categories: UI and Business Logic. The topics under each sub-category can further be grouped into multiple sub-sub-categories. For example, the UI sub-category has 4 topics grouped into Script and Dynamic Layout sub-sub-categories. Each sub-category, sub-sub-categories, and topics are also presented in descending order of the number of questions.

We discuss the five categories and the 40 topics in detail below.

\nd\bf{\ul{Customization Topic Category.}}
Customization is the largest topic category in terms of the number of topics and percentage of questions. Out of the 40 topics, 11 topics belong to the Customization category, with around 30\% of questions \rev{in} our dataset. These topics contain discussions about implementing business logic, customizing UI, input and form data validation, general programming-related query to implement some features, etc. This category has two sub-categories: \begin{inparaenum}[(1)]
        \item UI contains discussion about customizing the UI, dynamically changing window components, interactive dialog boxes, and
        \item Business Logic contains discussion about different programming customization-related queries, dynamically binding UI elements to backend data.
    \end{inparaenum}

    \nd\bf{$\bullet$ UI Sub-Category} contains 15\% questions and four topics divided in two sub-sub-categories: \begin{inparaenum}[(1)]
        \item Script contains discussion about manipulation of text widgets, formatting components, and
        \item Dynamic Layout is about hiding and moving components, showing popups.
    \end{inparaenum}
    
    The \bf{Script} sub-sub-category contains 10.4\% questions and has two topics: \begin{inparaenum}[(1)]
        \item Topic \textit{Window Style Manipulation (5.9\%)} concerns about manipulating the style of the HTML documents such as adding/removing margins/padding (e.g., \dq{36503030}), adding links, manipulating navigation bar and embedded views (e.g., \dq{30453620}). 
        \item  Topic \textit{Dynamic Form Controller (4.5\%)} are about dynamic form, i.e., forms with predefined list of values (e.g., \dq{64373454}), multi select content (e.g., \dq{39318510}), changing forms option based on previous selection (e.g., \dq{43725028}).
    \end{inparaenum}
    The \bf{Dynamic Layout} sub-sub-category covers 4.9\% questions and has two topics: \begin{inparaenum}[(1)]
        \item  Topic \textit{Dynamic Page Layout (2.9\%)} contains discussion about UI (i.e. page) customization (e.g., \dq{65964413}), pagination in \dq{4536018}, hiding or moving element based on some user action or an event (e.g., \dq{13231072}).,
        \item  Topic \textit{Dialog Box Manipulation (2.0\%)} is about manipulating dialog box (e.g., pop up/ modals) such as hiding them in \dq{49804455}, close them in \dq{55513527}, show popup, refresh web-page (e.g., \dq{60606986}, \dq{21701437}).
    \end{inparaenum}
    
    \nd\bf{$\bullet$ Business Logic Sub-Category} contains 14.7\% questions and 7 topics in two sub-sub-categories: \begin{inparaenum}[(1)]
        \item Programming is about discussion related to different programming-related questions and data access, and
        \item Dynamic Content is about discussions related to dynamically querying data from different data sources, dynamically changing the web-page content.
    \end{inparaenum}
    The Business Logic sub-category contains one topic \textit{Conditional BPMN} that does not belong to any sub-sub-category. Topic Conditional BPMN (1.6\%) contains LCSD platform's application customization related discussions on Business Process Model and Notation (BPMN) (e.g. \dq{38265464}) and conditional logic features (e.g., \dq{66335289}, \dq{65838553}).

    The \bf{Dynamic Content} sub-sub-category contains 7.6\% of questions and has three topics: \begin{inparaenum}[(1)]
        \item Topic \textit{Dynamic Event Handling (3.1\%)}  discusses about different JavaScript related issues such as JavaScript feature not working (e.g., ``JavaScript promise is not working'' in \dq{65550370}), browser compatibility, JS event initialization issue (e.g., \dq{64507615}) etc.
        \item  Topic \textit{Dynamic Data Binding (2.6\%)} is about discussions related to the design of forms with predefined values (e.g., \dq{45051098}), the implementation of multi-select, customized drop-down values, form validation (e.g., \dq{51115573}), changing content of one field based on the value of other field in \dq{47652192}.
        \item  Topic \textit{Dynamic Data Filtering (1.9\%)} contains business logic customization related discussion based on advanced filtering criteria and querying multiple tables. (e.g., ``Find Records Based on the Contents of Records in a Related Table?'' in \dq{20665253} and ``find the latest record grouped by name in layout'' in \dq{38128584}).
    \end{inparaenum}
    
    The \bf{Programming} sub-sub-category contains 5.5\% of questions and has three topics: \begin{inparaenum}[(1)]
        \item Topic \textit{Formatted Data Parsing (2.2\%)} is about programming related discussion on parsing formatted data, i.e., JSON (e.g., \dq{50184058}, \dq{44803257}), XML (e.g., \dq{13785513}), array of objects (e.g., \dq{66744874}) etc.
        \item  Topic \textit{Pattern Matching (1.8\%)} topic concerns programming related discussions about searching and modifying \rev{strings} by pattern matching using regular expression (e.g., ``How do I search for an Exact phrase'' in \dq{51258323}, ``Regex pattern to replace html in a given text string'' in \dq{48251198}).
        \item  Topic \textit{Date \& Time Manipulation (1.5\%)} contains programming discussions about date-time manipulation like conversion of formatted string from data-time in \dq{51714301}, calculation of difference between date-time (e.g., \dq{59230493}) and timezone, time conversion (e.g., ``How to convert a Date value to unix timestamp?'' in \dq{60601201}).
    \end{inparaenum}

\nd\bf{\ul{Data Storage Topic Category.}} Data Storage Category is the second largest topic category with a total of 9 topics and around 25\% of the questions of our dataset. This topic category contains discussions on database management and file storage. It contains two sub-categories: \begin{inparaenum}[(1)]
        \item DBMS is about discussion related to database setup, migration, DB query,
        \item File concerns storing and retrieving files (i.e., images, CSV files, etc.).
    \end{inparaenum}
    
    \nd\bf{$\bullet$ DBMS Sub-Category} contains around around 20.6\% questions with seven topics under three sub-sub-categories: \begin{inparaenum}[(1)]
        \item Configuration contains discussions about database setup, database connection, DB data security,
        \item SQL contains discussion about SQL query,
        \item Schema is about database schema design (i.e., Primary key, foreign key design), different stored procedure.
    \end{inparaenum}
    
    \bf{Configuration} sub-sub-category contains 7.2\% questions and two topics: \begin{inparaenum}[(1)]
        \item  Topic \textit{DB Setup \& Migration (4.4\%)} topic is about connecting applications to different vendor databases (i.e., MySQL, \rev{Postgres}, Oracle etc.) (e.g., ``is ODBC Firebird connection possible?'' in \dq{28251836}), DB users (e.g., \dq{58815776}), issues about different database versions, data migration to LCSD platform (e.g., ``How to add External data source into MySQL?'' in \dq{28251836} or \dq{22626970}).
        \item  Topic \textit{Data Security \& Replication (2.8\%)} topic concerns about discussion related to data security (e.g., encryption and decryption in \dq{1567252}), accessing stored of database file (e.g., \dq{5730482}), data backup or replication (e.g., \dq{10997987}) etc.
    \end{inparaenum}
    \bf{SQL} sub-sub-category contains 6.9\% questions and three topics: \begin{inparaenum}[(1)]
        \item Topic \textit{SQL Syntax Error (2.7\%)} discusses about errors in syntax in different SQL query and stored procedure. For example, there are questions such as ``I'm getting error while creating a procedure in pl/sql'' in \dq{63990287}, ``SQL parsing fails, not sure what the issue is?'' in \dq{8298486}.
        \item Topic \textit{SQL CRUD (2.5\%)} is about database Create, read, update and delete (i.e., CRUD) related queries (e.g., \dq{22712624}), and advanced queries too, such as inner join, nested join, aggregate (e.g., ``SQL Query: JOIN Three tables then Not showing the results after joining the 3rd table'' in \dq{22712624}). This topic also contains discussion about Object query language, which is a high-level wrapper over SQL in \dq{64812548}.
        \item Topic \textit{Date-based filtering (1.7\%)} contains database query related discussion specially for date-time based filtering (e.g., \dq{52389335}), i.e., monthly/quarterly query, time-based data grouping etc. For example, there are questions such as How to count total amount of value by day (e.g., \dq{65142062}). 
    \end{inparaenum}
    \bf{Schema} sub-sub-category contains 6.5\% questions and two topics: \begin{inparaenum}[(1)]
        \item Topic \textit{DB Stored Procedure (3.9\%)} is about database schema and advanced database related discussion on stored procedure (e.g.,  support of triggers in LCSD platform, \dq{11799577}, \dq{37810803}).
        \item Topic \textit{Entity Relationship Mgmt (2.6\%)} concerns about discussion on advanced database schema design (e.g., ``How to automatically insert foreign key into table after submit in [LCSD platform]'', \dq{66968187}) and database discussion to automatically update database (e.g., ``Auto increment item in Oracle APEX'' in \dq{61961348}).
    \end{inparaenum}
    
    \nd\bf{$\bullet$ File Sub-Category} contains 4.2\% questions with two topics: \begin{inparaenum}[(1)]
        \item Topic \textit{File Management (2.6\%)} contains discussion of file management, i.e., storing and processing files, renaming it in \dq{56414466}, converting files from one format to another in \dq{45796962}, handling image files (e.g, \dq{34203211}).
        \item Topic \textit{Semi-Structured Data Proc. (1.6\%)} is about different programming related discussion on processing, modifying and storing semi-structured data files, i.e., XML, CSV files. For example, there are questions such as Fetch CSV file columns dynamically Using [platform] package in \dq{66310875}.
    \end{inparaenum}

\nd\bf{\ul{Platform Adoption Topic Category.}} A total of nine topics belong to the Platform Adoption category with around 20\% questions. The nine topics belong to three sub-categories: \begin{inparaenum}[(1)]
        \item Documentation contains LCSD platform's feature-related discussions and how to use those features, 
        \item Architecture concerns about what type of software development architecture (e.g., client-server communication) is supported by the LCSD platforms, 
        \item REST API contains LCSD platform's RESTful APIs. 
    \end{inparaenum}
    
    \nd\bf{$\bullet$ Documentation Sub-Category} contains 9.7\% questions and five topics. Four of the topics fall under two sub-sub categories: \begin{inparaenum}[(1)]
        \item Data Visualization contains discussion related to interactive reports and graphs,
        \item Features is about LCSD platform provides features such as user's role management, support on SDLC management.
    \end{inparaenum}
    Topic \textit{Misc. SWE Discussion (1.5\%)} concerns about discussions related to general software engineering such as Unix Threading (e.g., \dq{30530873}), Object-oriented programming (e.g., \dq{314241}), auto scaling, ambiguous documentation in \dq{10348746}.
    
    \bf{Data Visualization} sub-sub-category contains 4.4\% questions and has two topics: \begin{inparaenum}[(1)]
        \item Topic \textit{Interactive Report (2.8\%)} is about data visualization and interactive data reports. It contains developers' discussions about how they can use different platform features for customized reports. For example, ``using jquery hide column heading when no data in column in interactive report in [platform]'' in \dq{53294659}.
        \item Topic  \textit{Graph/Chart (1.6\%)} discusses about platform's support and documentation request to draw different graphs (e.g., \dq{41257691}) and charts using stored data. For example, ``How to overlay a line plot over a bar graph in [platform]?'' in \dq{28727869}
    \end{inparaenum}
    \bf{Features} sub-sub-category contains 3.8\% questions and has two topics: \begin{inparaenum}[(1)]
        \item Topic \textit{User Role Management (2.1\%)} contains discussion about different user role management features (i.e., administrators and regulators) provided by the LCSD platforms. It discusses about user's profile management (e.g., \dq{66853056}), user group and access-level (e.g., \dq{35457840}).
        \item Topic \textit{Platform Related Query (1.7\%)} contains general discussion about LCSD platforms such as comparison of features between different platforms (e.g., ``How is [platform A] better than [platform B] in BPM?'' in \dq{39127918}), Agile and RAD development support (e.g., \dq{2512396}), performance of a specific feature of a platform in \dq{6068882}.
    \end{inparaenum}

    \nd\bf{$\bullet$ Architecture Sub-Category} contains 6.4\% questions and two topics and one sub-sub category: \begin{inparaenum}[(1)]
        \item Async S2S Comm contains discussion related to distributed applications with service to service communication.
    \end{inparaenum}
    Topic \textit{Platform Infrastructure API (3.2\%)} contains cloud-based REST API from the LCSD platforms to configure and utilize different platform features, connect to other services or data sources via connectors or cloud REST APIs. For example, the questions are about how [platform] apps portal integration with [DB] On-premise in \dq{63688934}, ``Change Shape OCI instance with Ansible'' in \dq{60511836}.
    
    \bf{Async S2S Comm} sub-sub-category contains 3.2\% questions and has two topics: \begin{inparaenum}[(1)]
        \item Topic \textit{Web-Service Communication (1.7\%)} contains discussions about micro-service architecture, service to service communication via web service description language (e.g., \dq{16278661}, \dq{2567466}), HTTP REST message in \dq{58689313}, Windows Communication Foundation (e.g., \dq{36849686}).
        \item Topic \textit{Message Queue (1.5\%)} is about discussion about different asynchronous \rev{service-to-service} data exchange mechanisms such as using a message queue. It generally contains discussions about micro-service design patterns and producer and consumer mechanisms (e.g., \dq{41640881}) for data exchange. For example, ``How to know who is connected to a [Platform] EMS Queue'' in \dq{66999418}, \dq{56334001}.
    \end{inparaenum}
        
    \bf{REST API} sub-category contains 3.6\% questions and has one topics: \begin{inparaenum}[(1)]
        \item Topic \textit{Authentication \& Authorization (3.6\%)} contains discussion about LCSD platforms support on different standard authentication and authorization protocol such as OAuth2 (e.g., \dq{30475542}), SAML (e.g., \dq{23624206}), access token (e.g., ``access token in android [Platform] sdk'' in \dq{32943204}).
    \end{inparaenum}

\nd\bf{\ul{Platform Maintenance Topic Category.}} We find 6 topics and 13.5\% questions in Platform Maintenance category. It has two sub-categories: \begin{inparaenum}[(1)]
        \item Configuration contains discussion on LCSD platforms library and build configuration, 
        \item CI/CD. is about discussion related to DevOps tasks such \rev{as} continuous integration and continuous delivery, testing etc.
    \end{inparaenum}

    \nd\bf{$\bullet$ Configuration Sub-Category} contains 8.0\% questions and three topics under two sub-sub categories:  \begin{inparaenum}[(1)]
        \item Dependency Resolution is about LCSD platforms server's library dependency management, 
        \item Server Configuration is about LCSD platform's servers configuration and hosting settings such as SSL configuration.
    \end{inparaenum}
    
    \bf{Dependency Resolution}  sub-sub-category contains 6.2\% questions and has two topics: \begin{inparaenum}[(1)]
        \item Topic \textit{Build Config. Management (4.1\%)} contains discussion about system build configuration and external library management-related issues in \dq{48727452}. This topic also contains discussion about compilation failure (e.g., \dq{29243987}), library dependency, build path not configured properly (e.g., \dq{57015131}) etc.
        \item Topic \textit{Library Dependency Mngmt (2.1\%)} is about the library and dependencies of the system (e.g. \dq{23471590}, \dq{62872836}), server configuration, different library version compatibility issues in \dq{60050869}.
    \end{inparaenum}
    \bf{Server Config.}  sub-sub-category contains 1.8\% questions and has one topic: \begin{inparaenum}[(1)]
        \item Topic \textit{Hosting Config. \& SEO (1.8\%)}  is about discussions about LCSD platforms support on server configuration, i.e., configuring SSL certificate (e.g., ``[platform] client ignoring expired certificate'' in \dq{55044903}), LCSD platform's support on making the application accessible and index-able (e.g., \dq{34860991}).
    \end{inparaenum}
    
    \nd\bf{$\bullet$ CI/CD Sub-Category} contains 5.5\% questions and three topics. Two of the topics fall under one sub-sub category:\begin{inparaenum}[(1)]
        \item Monitoring is about the discussion on monitoring the deployed applications and scheduled job status.
    \end{inparaenum}
    Topic \textit{Testing (2.1\%)} contains discussions about LCSD platforms support on testing and test coverage. For example, ``How to know overall code coverage of multiple test classes?'' in \dq{67724447}, How to write a test class as in \dq{50586482}.
    
    \bf{Monitoring}  sub-sub-category contains 3.4\% questions and has two topics: \begin{inparaenum}[(1)]
        \item Topic \textit{App Deployment (1.9\%)} discusses about the LCSD platform's CI/CD features such as incrementally updating the application code in \dq{39045129}, deployment packages as in \dq{4813597}, monitoring the changes in the application code (e.g., \dq{61938011}).
        \item Topic \textit{Asynchronous Batch Jobs (1.5\%)} contains discussions about LCSD platforms' support for monitoring and scheduling asynchronous batch jobs and scheduled tasks. For example, ``How to get your failing batch records?'' in \dq{11068830}, ``How can I schedule apex to run every 30 seconds?'' in \dq{17143633}.
    \end{inparaenum}

\nd\bf{\ul{Third-Party Integration Topic Category}} is smallest topic category based on number of questions (12.1\% questions). It has five Topics. Four of its topics fall under two sub-categories: \begin{inparaenum}[(1)]
        \item REST API contains discussion related to RestFul API communication with third-party services,
        \item Plugins is about discussion about external plugins and APIs that \rev{are} supported by the LCSD platforms.
    \end{inparaenum}

    \nd\bf{$\bullet$ REST API Sub-Category}  contains 6.5\% questions and two topics: \begin{inparaenum}[(1)]
        \item Topic \textit{External Web Req Processing (3.7\%)} contains discussion about integrating 3rd party REST APIs, processing and parsing external requests such as ``Connect to [Platform] REST API with [Service] data integration'' in \dq{51865601}, \dq{46033973}.
        \item Topic \textit{Fetch \& Process API Response (2.8\%)} contains discussions about making HTTP request to remote servers (e.g., ``REST http post method - what does -d mean in a curl?'' in \dq{65877037}), analyzing and processing the response, handling web security issues (e.g., CORS policies in \dq{60270574}).
    \end{inparaenum}

    \nd\bf{$\bullet$ Plugins Sub-Category} contains 5.6\% questions and three topics:  \begin{inparaenum}[(1)]
        \item Topic \textit{Email Processing (2.4\%)} discusses about processing automating emailing in \dq{65626477}, sending formatted HTML email (e.g, ``Send HTML email using [platform]'' in \dq{41546887}), forwarding emails in \dq{18234790} etc.
        \item Topic \textit{Upgradation \& Compatibility (1.7\%)} contains discussion about application version migration as in \dq{16894245}, upgradation and compatibility issues of different plugins used in low-code applications (e.g., \dq{18231293}, \dq{49017642}).
        \item Topic \textit{eSignature (1.5\%)} contains discussion about different issues and customization for electronic signature of documents, i.e., docusign about collecting user's agreement/permission for sales or account opening. For example, there are questions such as ``Auto Add Document to DocuSign [Platform] Using Custom Button'' in \dq{34804072}, \dq{27512874}.
    \end{inparaenum}

\begin{tcolorbox}[flushleft upper,boxrule=1pt,arc=0pt,left=0pt,right=0pt,top=0pt,bottom=0pt,colback=white,after=\ignorespacesafterend\par\noindent]
\noindent\textbf{RQ1. What topics are discussed about LCSD in SO?}
We found 40 Topics organized into five high-level categories. The Customization category (30\%) has the highest number of questions, followed by Data Storage (25\%), Platform Adoption (20\%), Platform Maintenance (14\%), and Third-Party Integration (12\%). Window Style Manipulation from the Customization category has the highest number of questions (5.9\%) followed by build Configuration (4.1\%) Management from the Platform Maintenance category. \rev{Our studies reveal that low-code practitioners struggle with RESTful API Integration, configuration and maintenance of the platforms. We also \rev{observed} that proper documentation could have mitigated these challenges to a great \rev{extent}.}
\end{tcolorbox} 

\subsection{How do the LCSD topics evolve over time? (RQ2)}\label{rq:evolve}

\subsubsection{Motivation}
Our analysis of RQ1 finds that LCSD topics are diverse. For example, the Customization \rev{topic category} contains discussions about developing and customizing the application, and Platform Adoption and Platform Maintenance topic contains discussions related to different features provided by the LCSD platform providers. The platforms for LCSD continue to evolve, as do the underlying \rev{topics} and question types. We study the evolution of these topics and question types to understand better the evolution and adoption of LCSD development and its community. This analysis will provide valuable insights into the LCSD community and help identify if any topic needs special attention.

\subsubsection{Approach}\label{sub:sec:evolution_approach}
Following related studies~\cite{iot21}, we study the absolute and relative impacts of each of our observed five LCSD topic categories as follows.

\textbf{Topic Absolute Impact.} We apply LDA topic for our corpus $C$ and get $K$ topics ($t_1$, $t_2$,........,$t_k$). The absolute impact metric for a topic $t_k$ in a month ($m$) is defined as:
\begin{equation}
Impact_{absolute}(t_k;m) = \sum_{p_i=1}^{D(m)} \theta (p_i; t_k)   
\label{eq:absoluteeq}
\end{equation}
Here the $D(m)$ is the total number of SO posts in the month $m$ and $\theta (p_i; t_k)$ denotes the possibility for a post ($p_i$) belonging to a topic $t_k$.

From our topic modeling, we found 40 topics that were categorized into five high \rev{topic categories}, i.e., \textit{Customization, Data Storage, Platform Adoption, Platform Maintenance, Third-Party Integration}. Now, we further refine the equation for absolute impact for LCSD topics to find absolute impact metrics for a \rev{topic category} ($TC_j$) for a month $m$ as follows:
\begin{equation}
Impact_{absolute}(TC_j;m) = \sum_{t_k}^{TC_j}Impact_{absolute}(t_k; m), 0 < j < TC
\label{eq:absoluteeqcat}
\end{equation}

\textbf{Topic Relative Impact.} Related impact metric signifies the proportion of posts for a particular LCSD topic $t_k$ relative to all the posts in our corpus $C$ for a particular month $m$. Following related studies~\cite{iot21}, we compute the related impact metrics for LCSD topics. We use the following equation to compute the metric for a topic $t_k$ in a month $m$ as follows:

\begin{equation}
Impact_{relative}(t_k,m) = \frac{1}{|D(m)|}\sum_{p_i=1}^\theta(p_i; t_k), 1\le i\le C 
\label{relativeeq}
\end{equation}
Here $D(m)$ denotes the total number of posts that belongs to a topic $t_k$ for a particular month $m$. Here $\delta$ denotes the probability of a particular post $p_i$ for our Corpus $C$ belonging to a particular topic $t_k$.

Similar to the absolute impact, we refine the equation to compute the relative impact on LCSD \rev{topic categories} as follows:
\begin{equation}
Impact_{relative}(TC;m) = \sum_{t_k}^{TC}Impact_{relative}(t_k;m)
\label{relativeeqsw}
\end{equation}

Here $TC$ donates one of our five \rev{topic categories} and topics that belong to each topic category.

\subsubsection{Result}



\begin{figure}[t]
\centering
\includegraphics[scale=0.50]{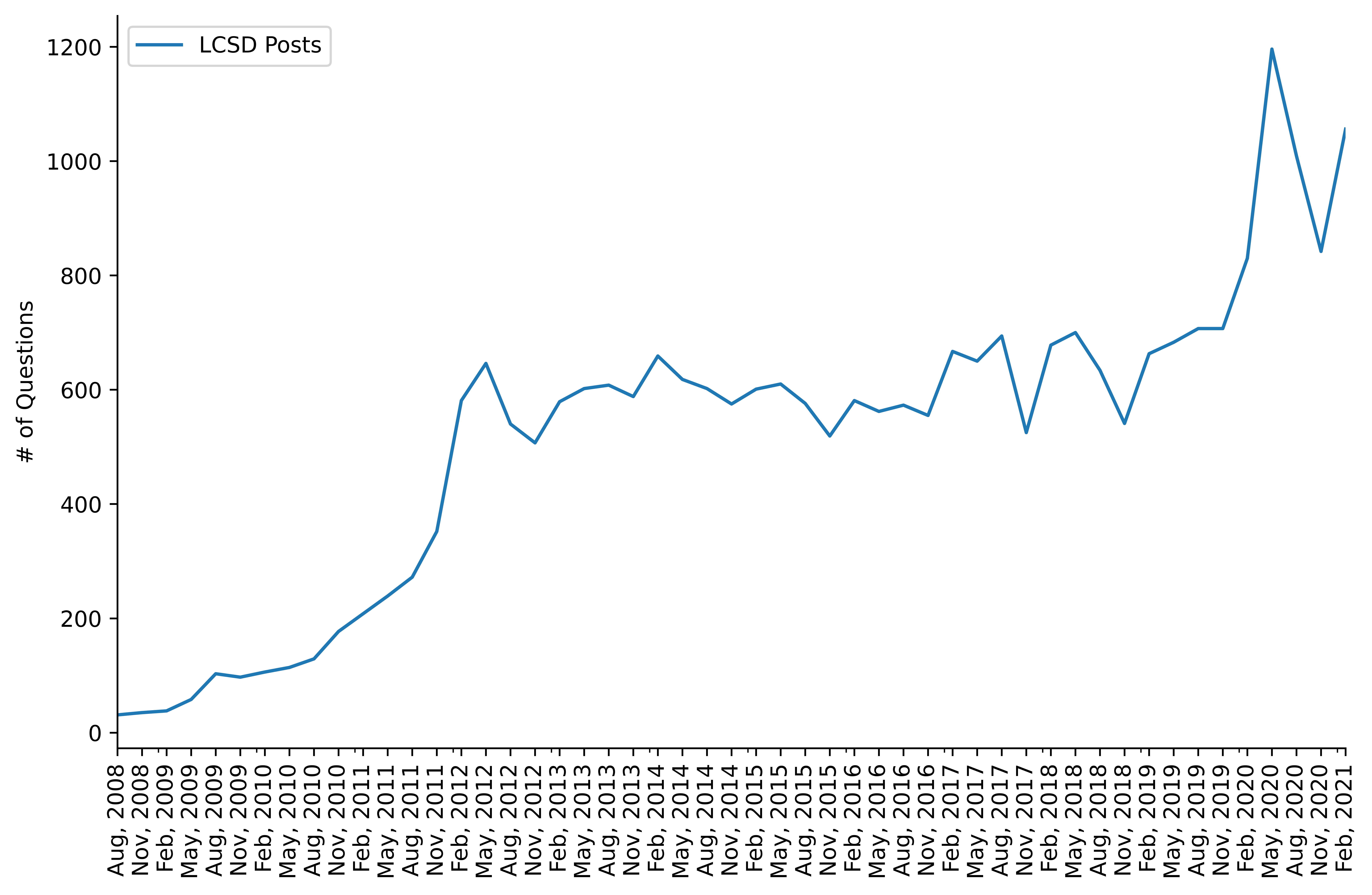}
\caption{The evolution of overall LCSD-related discussions over time.}
\label{fig:all_questions_absolute_impact}
\end{figure}

Figure~\ref{fig:all_questions_absolute_impact} depicts the progression of overall LCSD-related conversation using the absolution impact equation from our extracted dataset between 2008 to 2021. Additionally, it demonstrates that the peaks of LCSD-related talks occurred in mid-2020 (i.e., around 400 questions per month). It also reveals that LCSD-related discussion has been gaining popularity since 2012. 
In the section below, we provide a more extensive explanation for the minor spikes in the Figure~\ref{fig:all_questions_absolute_impact}.


\begin{figure}[t]
\centering
\includegraphics[scale=0.50]{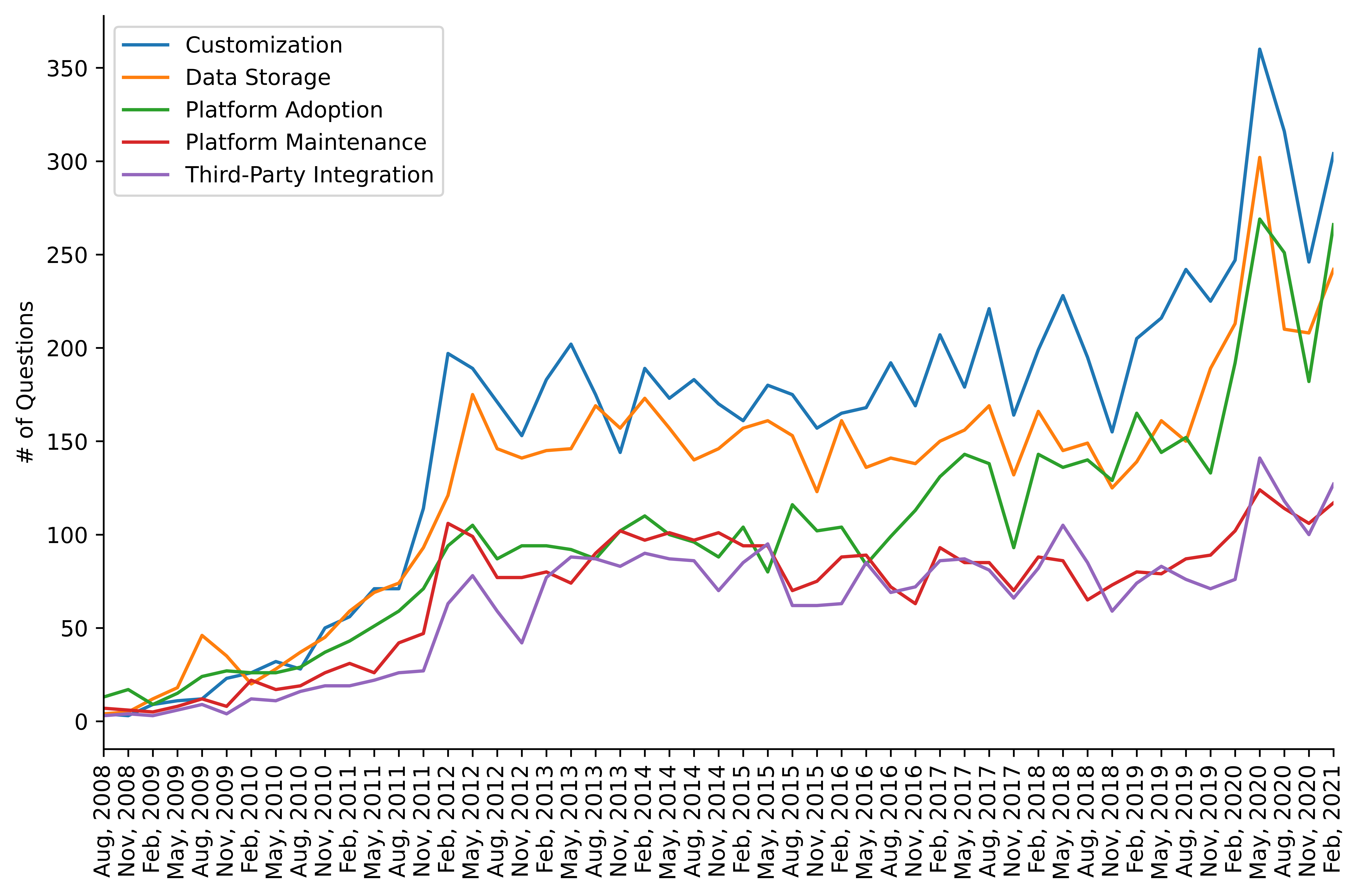}
\caption{The absolute impact for LCSD topic categories over time.}
\label{fig:topic_absolute_impact}
\end{figure}

\begin{figure}[t]
\centering
\includegraphics[scale=0.50]{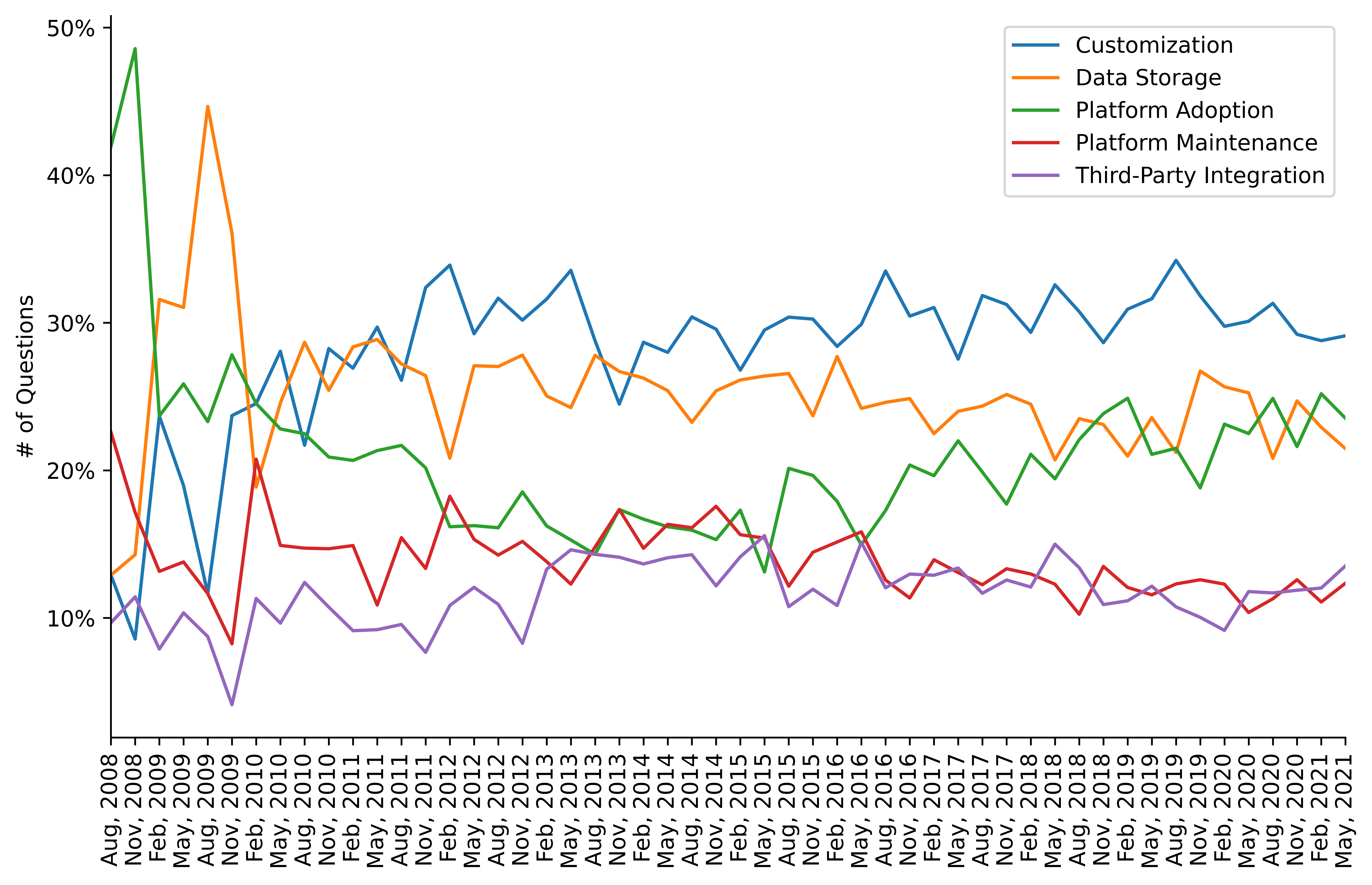}
\caption{The relative impact for LCSD topic categories over time.}
\label{fig:topic_relative impact}
\end{figure}

We observe that in the early days (i.e., 2009), Platform Adoption along with Data Storage \rev{topic category} has more questions compared to Customization. Customization topic category starts to get a dominant position from mid (i.e., August) of 2011 over Platform Adoption, and it remains the dominant topic till the end of 2021. The number of questions in the Customization topic category gradually increased over time, from August 2011 (23) to March 2012 (81) to May 2020 (99). Data Storage topic category briefly exceeds Customization Category during August 2013, but it mostly holds a dominant position other times compared to Platform Adoption topic category. On the other hand, Platform Maintenance and Third-Party Integration exhibits very similar evolution over the whole time.

We further look into the Figure~\ref{fig:topic_absolute_impact} and see there are mainly two significant upticks in the number of posts about LCSD. The first one is between August 2011 to May 2012, \rev{when} there is a sharp increase in the number of questions for almost all topic categories, especially for Customization and Data Storage Category. By this Salesforce\cite{salesforce} LCSD platform-related discussion becomes quite popular in SO, and around that time, it ranks very high as a CRM platform. The second increase in posts is between February 2020 and August 2020. During this time of the Covid19 pandemic, many businesses began to operate online and there is a significant uptick in the number of questions in the Customization category, followed by Data Storage and Platform Adoption
Moreover, there is an uptick in the number of questions on building a simple app to collect and report data, especially the Salesforce platform. There is an increase in the Platform Adoption topic category between mid-2016 to mid-2017. During this time Oracle Apex platform released version 5.0, and there is an uptick of questions regarding different features such as interactive grid in \dq{43316233}, drag and drop layout in \dq{45818292}. Now we provide a detailed analysis of each of the five topic categories.

\textbf{Customization}
This is the biggest topic category with 11 topics. From 2008 to mid of 2011, all of these topics evolve homogeneously. From the mid of 2011 to the first Quartile of 2012, Dynamic Page Layout topic becomes dominant. ``How to get fields in page layout'' in \dq{7256190}, issues with page layout in different LCSD platforms (e.g., \dq{7421985}). From the end of 2012 to 2017, Window Style Manipulation topic remains most dominant. ``Passing values from child window to parent window which is on another domain?'' in \dq{16463602}, view related issues \dq{15715645}. From the end of 2017 to the end, our dataset Dynamic Form Controller topic remains the most dominant.

\textbf{Data Storage Category}
From mid-2015, Database Setup \& Migration topic becomes the most dominant topic in this category and has some high spikes during the pandemic and mid of 2017. For instance, there are queries like ``Using Jenkins for OCI database migration'' in \dq{62217796} and  ``Almost all the cloud service providers have 99.95\% of data available on the cloud. What will happen if the whole region sinks in an earthquake?'' in \dq{62102679}. Since 2017 DevOps and database ``Domino Xpage database building automation or continuous integration using Jenkins with maven.'' in \dq{43092239}. From mid-2011 to mid-2014, DB Stored Procedure topic remains dominant. ``Oracle APEX: Call stored procedure from javascript'' in \dq{20501834}.

\textbf{Platform Adoption Category}
From 2008 to mid-2011, Platform Adoption related topics were the most dominant (e.g., ``Suggested platform/tools for rapid game development and game prototyping'' in \dq{312357}). Between mid-2011 to mid-2017, Authentication \& Authorization topic becomes dominant (e.g., ``Can I implement my own authentication process in force.com or it is against terms of service?'' \dq{13059568}, \dq{13034866}). Since the end of 2017, Platform Infrastructure API remains the most dominant. So, practitioners ask queries like  ``VirtualBox VM changes MAC address after imported to Oracle Cloud Infrastructure'' in \dq{61501108} and ``How to send a classic report as mail body in oracle Apex 19.2'' in \dq{59693984}, report layout (\dq{59833909}, \dq{59752159}).

\textbf{Platform Maintenance Topic Category}
From 2008 to mid-2019, the Build Configuration Management topic remains the most dominant topic. It has some high spikes in the number of questions during the beginning of 2012 and the first quartile of 2014. Build error \dq{21720165}, \dq{21326163}, build projects automatically \dq{21758244}. From mid-2019, Library Dependency Management topic-related questions became popular (e.g., library-related issues (e.g., \dq{62825046}, \dq{61100705}), library not found\dq{61911916}).

\textbf{Third-Party Integration Topic Category.}
The five topics from this category evolve simultaneously. From the beginning of 2015, the External Web Request Processing topic has become more dominant than other topics with a slight margin. External Web Request Processing and Fetch \& Process API response, E-signature topics become dominant during the pandemic with queries such as platform support on e-signature \dq{62417381} and etc.

In Figure~\ref{fig:topic_relative impact}, we now provide more insight into the evolution of LCSD topic categories. It confirms the findings presented in Figure~\ref{fig:topic_absolute_impact} and adds some previously unknown insights. For instance, in the last quartile of 2009, it is apparent that Data Storage is the most popular Topic Category. According to the absolute impact metric, all five themes are increasing monotonically. The relative impact measure, on the other hand, indicates that the Customization, Platform Maintenance, and Third-Party Integration Topic group evolves in a nearly identical manner. However, this Figure demonstrates that, beginning in 2016, Platform Adoption-related conversation increased and eventually surpassed Data Storage-related discussion. This in-depth examination of evolution is significant because it demonstrates that, while Data Storage Topics are the second-largest Topic category, Platform Adoption-related queries are evolving rapidly and require further attention from platform vendors.

\begin{tcolorbox}[flushleft upper,boxrule=1pt,arc=0pt,left=0pt,right=0pt,top=0pt,bottom=0pt,colback=white,after=\ignorespacesafterend\par\noindent]
\noindent\textbf{RQ2. How does the LCSD-related discussion evolve?}
Since 2012, LCSD-related talks in SO have grown in popularity, and this trend has accelerated since 2020.
Initially, the Customization and Data Storage Topic Categories dominated, but in recent years, Platform Adoption-related inquiries have grown in popularity.
\end{tcolorbox}
\subsection{What types of questions are asked across the \rev{observed topic categories?} (RQ3)}\label{rq:type}
\subsubsection{Motivation}
\rev{This research question aims to provide a deeper understanding of LCSD-related topics based on the types of questions asked about the LCSD platforms in SO. For example, “what” types of questions denote that developers are not sure about some specific characteristics of LCSD platforms, while “how” types of questions denote that they do not know how \rev{to} solve a problem using an LCSD platform. Intuitively, the prevalence of “what” types of questions would denote that the LCSD platforms need to better inform the services they offer, while the prevalence of “how” type of questions would denote that the LCSD platforms need to have better documentation so that developers can learn easily on how to use those. Initially, in 2011 Treude et al.~\cite{treude2011programmers} investigated different types of questions on stack overflow. Later Rosen et al.~\cite{Rosen-MobileDeveloperSO-ESE2015} conducts an empirical study like ours on Mobile developers’ discussion in stack overflow with these four types of \rev{questions}. Later, very similar studies on chatbot development~\cite{abdellatif2020challenges} and IoT developers’ discussion on Stack overflow~\cite{iot21} also explore this research question to provide more insights about specific \rev{domains} and complement the findings of topic modeling.}


\subsubsection{Approach}
\label{rq_type_appraoch}
In order to understand what-type of questions are discussed in SO by the LCSD practitioners, we take a statistically significant sample from our extracted dataset and then manually analyze each question and label them into one of four types: How-type, Why-type, What-type, Others-type following \rev{related studies~\cite{iot21, abdellatif2020challenges, Rosen-MobileDeveloperSO-ESE2015}.} So, our approach is divided into two steps: Step 1. We generate a statistically significant sample size, Step 2. we manually analyze and label them.

\nd \textbf{Step 1. Generate Sample.} As discussed in Section \ref{sub-sec:data_collection} our final dataset has 26,763 questions. A statistically significant sample with a 95\% confidence level and five confidence intervals would be at least 379 random questions, and a 10 confidence interval would have a sample size of 96 questions. A random sample represents a representative for the entire dataset, and thus this could miss questions from the subset of questions that may belong to smaller topic categories. For example, as discussed in RQ1, we have 40 topics organized into five categories. As random sampling is not uniform across the topic categories, it might miss important questions from smaller topic categories such as Third-Party Integration. Therefore, following previous empirical studies \cite{iot21,abdellatif2020challenges}, we draw a statistically significant random sample from each of the five topic categories. Specifically, we draw the distribution of questions in our sample from each of the 5 topic categories with a 95\% confidence level and ten confidence intervals. The sample is drawn as follows: 95 questions from the Customization category (total question 8014), 95 questions from the Data Storage category (total question 6610), 94 questions from Platform Adoption category (total question 5285), 94 questions from Platform Maintenance category (total question 3607), 93 questions from Third-Party Integration category (total question 3247). In summary, we sampled a total of 471 questions.

\nd \textbf{Step 2. Label Question Types.} We analyze and label each question from our samples into the following four categories. The categories and the coding guides follow previous research \cite{iot21, abdellatif2020challenges, rosen2016mobile}
\begin{itemize}
    \item \textbf{How-type} post contains a discussion about the implementation details of a technical task~\cite{iot21}. The questions primarily focus on the steps required to solve certain issues or complete certain tasks (e.g., ``How to create a submit button template in Oracle APEX?'' in \dq{1730566
}).

    \item \textbf{Why-type} post is about troubleshooting and attempting to determine the cause/reason for a behavior. These questions help practitioners understand the problem-solving or debugging approach, e.g., in \dq{25176669}, a user is trying to find out why an SSL server certificate is rejected.
       
     \item \textbf{What-type} question asks for more information about a particular architecture/event. The practitioners ask for more information that helps them to make informed decisions. For example, in \dq{11608661} a practitioner is asking for detailed information about the Oracle Apex platform's secure cookies.
     
    \item \textbf{Other-type} question do not fall into any of the above three categories, e.g., ``Initiating Salesforce API in Google App Script'' in \dq{66317111}
    
\end{itemize}

Three authors (first, third and fourth) participated together in the labeling process. We assessed our level of agreement using Cohen kappa\cite{mchugh2012interrater}. The disagreement and annotation difficulties were resolved by discussing with the first author. In general, the authors achieved a substantial agreement ($k$ \textgreater~0.6) on the 471 questions classified. Our coding \rev{guidelines} and the final annotated dataset are available in our replication package.



\subsubsection{Result}
Table \ref{tab:qtypes} shows the percentage of type of questions across our five LCSD topic categories.
Similar to related studies \cite{iot21, abdellatif2020challenges, rosen2016mobile}, during our labeling process, we observed that some of the questions can have multiple labels, e.g., What-type and How-type. For example, ``How can I get jQuery post to work with a [platform] WebToLead'' in \dq{2339550} discusses making Ajax request using jQuery where the practitioner is getting an error response. At the same time, the practitioner is further querying some detailed information on how Ajax requests. 
Therefore, the sum of percentages of question types reported in the result section is more than 471 classified posts. We now discuss each question type with examples. 

\begin{figure}[t]
	\centering\begin{tikzpicture}[scale=0.28]-
    \pie[
        /tikz/every pin/.style={align=center},
        text=pin, number in legend,
        explode=0.0,
        color={black!0, black!10, black!20, black!30, black!40},
        ]
        {
            55.6/\bf{How-type} \\ (55.6\%),
            17.9/\bf{What-type} \\ (17.9\%),
            14/\bf{Why-type} \\ (14\%),
            12.5/\bf{Other-type}\\ (12.5\%)
        }
    \end{tikzpicture}
	\caption{Distribution of different question Types}
	\label{fig:pie_chart_question_type}
\end{figure}
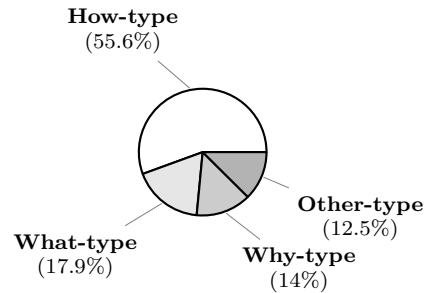
\begin{table}[t]
  \centering
    \caption{Types of questions across the LCSD topic categories}
    \begin{tabular}{lrrrr}\toprule
    \textbf{Topic Category} & \multicolumn{1}{l}{\textbf{How}} & \multicolumn{1}{l}{\textbf{What}} & \multicolumn{1}{l}{\textbf{Why}} & \multicolumn{1}{l}{\textbf{Other}} \\
    \midrule
    Customization  & 51.0\% &  18.4\% &  17.3\% &  13.3\% \\
    Data Storage & 59.8\% &  13.4\% &  17.5\% &  9.3\% \\
    Platform Adoption & 57.3\% &  19.8\% &  9.4\% &  13.5\% \\
    Platform Maintenance & 49.0\% &  20.4\% &  17.3\% &  13.3\% \\
    Third-Party Integration & 61.2\% &  17.3\% &  8.2\% &  13.3\% \\
    \midrule
    \textbf{Overall} & 55.6\% &  17.9\% &  14.0\% &  12.5\%  \\
    \bottomrule
    \end{tabular}%
  \label{tab:qtypes}%
\end{table}%

\textbf{How-type.} Around 57\% of our annotated questions fall under this type. This type of question is most prevalent in the topic categories Third-Party Integration (61\%) followed by  Data Storage (60\%),  Platform Adoption (57\%), Customization (51\%), Platform Maintenance (49\%). This high prevalence is not surprising, given that SO is a Q\&A platform and the LCSD practitioners ask many questions about how to implement certain features or debug an issue. Additionally this also signifies that LCSD practitioners are asking a lot questions while integrating with third-party library (e.g., \dq{62825046}) and plugins (e.g., \dq{61455233}) and managing the data with a database management system (e.g., \dq{38111768}) or file storage (e.g., \dq{63284305}). To explain further we find questions regarding implementing encryption, e.g., \dq{2220076}, or Making HTTP POST request, e.g., \dq{32736416}, or debugging a script, e.g., \dq{45619586}, or implement a feature, e.g., \dq{28990848} etc.

\textbf{What-type.} This is the second biggest question type with 18\% of all annotated questions. This type of question is the most dominant in the topic categories Platform Maintenance (20.5\%), Platform Adoption (20\%), and Customization (18\%). This type of question can be associated with How-type questions, where the practitioners require further information to implement certain features. For instance, in this question, a practitioner is querying about ``How to implement circular cascade select-lists'' in \dq{60676786}. The questions in this category signify that practitioners fail to find relevant information from the official documentation (e.g., \dq{9377042}) sources. Therefore, as this type of question is prevalent in Platform Maintenance and Platform Adoption category, LCSD platform providers might focus more on improving their resources. As an example, we find questions on JavaScript events not working correctly, e.g., \dq{51564154}, roll back changes, e.g., \dq{11156810}, designing workflow, e.g., \dq{11156810}.

\textbf{Why-type.} This is the third most prevalent question type category, with 14\% of all annotated questions. This type of question is the most prevalent in the topic categories Customization, Data Storage, and Platform Maintenance with around 17\% questions. These questions are mostly related to troubleshooting like when LCSD practitioners implement particular features or deploy an application. For instance, e.g., ``Why does this error happen in [Platform]?'' in \dq{48818859}, ``Why isn't the document going into edit mode'' in \dq{51660117}, ``Not able to launch Android Hybrid app using [Platform] Mobile SDK'' in \dq{20417235}, ``Java code running twice'' in \dq{17147921}.

\textbf{Other-type.} Around 14\% of our annotated questions fall under this type. The questions are almost uniformly distributed across the five topic categories. The questions contain general problems, e.g.,  ``UTF-8 character in attachment name'' in \dq{22808965} or ``Domino Server is installed on Unix or Windows?'' in \dq{10796638}. Some of the questions in this type also contain multiple/ambiguous questions (e.g., \dq{27896327}). For example, How to test an application?, which library is better? etc.

\begin{tcolorbox}[flushleft upper,boxrule=1pt,arc=0pt,left=0pt,right=0pt,top=0pt,bottom=0pt,colback=white,after=\ignorespacesafterend\par\noindent]
\noindent\textbf{RQ3. What types of questions are asked across the observed topic categories?}
How-type (57\%) questions are the most prevalent across all five topic categories, followed by What-type (18\%), Why-type (14\%), and Other-type (12\%) questions. Practitioners in the Customization and Platform Maintenance topic categories are more interested in troubleshooting, i.e. (Why-type, What-type). Practitioners generally ask more implementation questions (i.e., How-type) in the Third-Party Integration Category. Practitioners in the Data Storage topic category are interested in designing databases (i.e., How-type) and troubleshooting (i.e., What-type, Why-type). \rev{This indicates the necessity for a more robust community for troubleshooting and debugging issues.}

\end{tcolorbox} 
\subsection{How are the \rev{observed topic categories} discussed across SDLC phases? (RQ4)} \label{rq:rq_sdlc}

\subsubsection{Motivation}
As we observed the prevalence and evolution of diverse LCSD topics in SO, we also find that the topics contain different types of questions. This diversity may indicate that the topics correspond to the different SDLC phases that are used to \rev{develop} low code software development (see Section \ref{sec:background} for an overview of the LCSD phases). For example, intuitively What-type of questions may be due to the clarification of a low code software requirements during its design phases, which questions/topics related to troubleshooting of issues may be asked during the development, deployment, and maintenance phase. Therefore, an understanding of the SDLC phases in the LCSD questions in SO may offer us an idea about the prevalence of those SDLC phases across our observed LCSD topics in SO. This understanding may help the LCSD practitioners to determine how SO can be used during low code software development across the various SDLC phases.

\subsubsection{Approach}
In order to understand the distribution of LCSD topics across agile SDLC phases, we collect a statistically significant sample of questions from our extracted dataset $D$ into one of the six Agile software development methodology~\cite{beck2001manifesto} phases:  \begin{inparaenum}[(1)]
\item Requirement Analysis \& Planning, 
\item Application Design,
\item Implementation, 
\item Testing,
\item Deployment, and
\item Maintenance. \end{inparaenum} First, we generate a statistically significant sample size.  We use the same set of randomly selected (i.e., 471) posts that we produced during RQ3 (see \sec\ref{rq:type}). \rev{So, we take a statistically significant stratified random sample for each topic category in our dataset with 95\% confidence level and 10 confidence interval to ensure that we have a representative sample from each topic category~\cite{iot21, abdellatif2020challenges}.} We manually annotate each question post with one/more SDLC phases.


We followed the same annotation strategy to label SDLC phased as we did for RQ3 (see \sec\ref{rq_type_appraoch}). Each question was labeled by at least two authors (second and third/fourth) after extensive group discussion on formalizing annotation guidelines and participating in joint sessions. We find our level of agreement using Cohen kappa \cite{mchugh2012interrater, alamin2021empirical}. The authors generally achieved a substantial agreement ($k$ $>$ 0.70). For example, a new practitioner is tasked with finding the right LCSD platform during the planning stage of his/her LCSD application. The practitioner queries, ``Are there any serious pitfalls to Outsystems Agile Platform?'' (\dq{3016015}). We thus assign the SDLC phase as ``Requirement Analysis \& Planning''. Another question asks, ``Google  App  Maker  app  not  working  after  deploy'' (\dq{42506938}). We label the SDLC phase as ``Deployment''. \rev{For some questions, it involved significant manual assessment to assign appropriate SDLC phase, e.g., Requirement Analysis \& Planning phase vs Application Design and Application Design vs Implementation phase. As such, we developed a detailed annotation/coding guide to help us with the manual assessment. This annotation guide was constantly updated during our study to ensure that the guide remained useful with all relevant instructions. For example, one of the questions that helped refine our annotation guide is the question noted by the respected reviewer, i.e., ``Can AppMaker be utilized with SQL Server?" in \dq{55220499}. The user in this question wants to know if Google App Maker and SQL Server databases can be connected. This question was categorized as Application design phase. Based on this, according to our annotation guideline, this question can be labelled as Requirement Analysis \& Planning phase too. However, after discussion, the first and third authors agreed to label it as Application Design phase because from the problem description, it seems the question mainly focuses on connecting the application to a custom data storage. As this question focuses on data source design, which is often explored during the Application Design phase, we concluded that it should be labeled as such. The labeling of each question to determine the precise SDLC phases was conducted by several co-authors in joint discussion sessions spanning over 80 person-hours.}



\begin{figure}[t]
	\centering
	\resizebox{4.8in}{!}{
	\begin{tikzpicture}
    \pie[
        explode=0.0, text=pin, number in legend, sum = auto, 
        color={black!0,black!10, black!20,black!30,black!40,black!60},
        ]
        {
            65/\LARGE{Implementation} 65\%,
            17/\LARGE{Application Design} 17\%,
            9.1/\LARGE{Requirement Analysis \& Planning} 9.1\%,
            3.3/\LARGE{Deployment} 3.3\%,
            2.8/\LARGE{Maintenance} 2.8\%,
            2.7/\LARGE{Testing} 2.7\%
        }
    \end{tikzpicture}
    }
	\caption{Distribution of questions (Q) per SDLC phase}
	\label{fig:distribution_of_SDLC_pie_chart}
\end{figure}
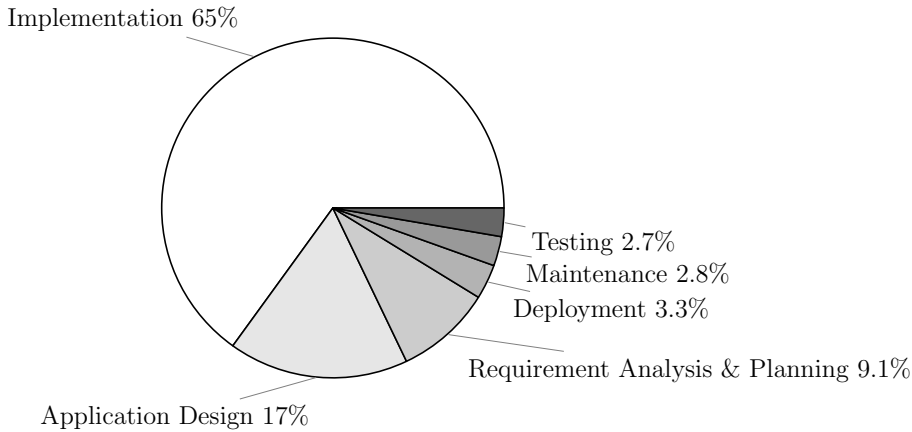

\subsubsection{Results}


Figure \ref{fig:distribution_of_SDLC_pie_chart} shows the distribution of our LCSD questions into six agile SDLC phase. We find that the Implementation phase has 65\% of our 471 annotated questions, followed by Application Design (17\%), Requirement Analysis \& Planning (9.1\%). It is not surprising that the Implementation phase has so many questions because SO is a technical Q\&A platform and practitioners use it mostly to find issues when trying to implement some feature. Though the percentage of questions is not very high (e.g., between 2-3\%), we also find practitioners ask questions regarding Testing and Deployment phases too (e.g., ``Automated Testing for Oracle [Platform] Web Application'' in \dq{1764497}). This analysis highlights that LCSD practitioners ask questions regarding the feasibility analysis of a feature to make design decisions to implement the feature to deployment.
We provide an overview of the types of questions asked during these six SDLC phases.

\bf{Requirement Analysis \& Planning (43, 9.1\%).} 
Requirement analysis is the first and most important stage of software development because the application largely depends on this. Requirement analysis is a process to develop software according to the users' needs. In agile software development methodology, features are implemented incrementally, and requirement and feasibility analysis are crucial in implementing a new feature. During this phase, the operational factors are considered, and the feasibility, time-frame, potential complexity, and reliability. Requirement management tools are typically included with LCSD systems, allowing developers to collect data, modify checklists, and import user stories into sprint plans. Throughout this stage, developers tend to ask questions regarding the platform's features (e.g., ``Does Mendix generates a source code in any particular language, which can be edited and reused?'' in \dq{53043346}), learning curve (e.g., \dq{55304547}, \dq{45631057}), and the LCSD platform's support for faster application development (e.g., \dq{28983651}), general deployment/maintenance support (e.g., \dq{50460088}) in order to select the best platform for their needs. 
For example, in this popular question, a new practitioner is asking for some drawbacks on some potential pitfalls for a particular LCSD platform, e.g., "Are there any serious pitfalls to [Platform] Agile Platform?" (\dq{3016015}). A developer from that platform provider suggests using the platform to build an application and decide for himself as it is hard to define what someone might consider a pitfall. In another question, a practitioner is asking if it is possible to integrate Selenium with an LCSD platform (e.g., \dq{52010004})

\bf{Application Design (80, 17\%).} 
The design specification is created in this step based on the application's needs. The application architecture (e.g., \dq{53820097}), modularity, and extensibility are all reviewed and approved by all critical stakeholders. 
The LCSD developers face challenges regarding data storage design, drag and drop UI design, connecting on-premise data-sources with the LCSD platform (e.g., ``Can AppMaker be used with SQL Server'' (\dq{55220499})), data migration to LCSD platform (\dq{46421271}), following best practices (e.g, ``Salesforce Best Practice To Minimize Data Storage Size'' in \dq{14073151}), designing a responsive web page (e.g., (\dq{52744026})).

\bf{Implementation (306, 65\%).} 
The actual application development begins at this phase. LCSD developers confront a variety of obstacles when they try to customize the application (i.e., personalize UI (e.g, \dq{6454308}), implement business logic (e.g, \dq{40472354})), integrate third-party plugins(e.g, \dq{46538734}), debug (e.g, \dq{35898112}) and test the implemented functionality. For example, LCSD practitioners ask customization questions such as How can they change the timezone in a platform in \dq{47731051}, customizing UI in \dq{40159662}. Many of these challenges arise from incomplete or incorrect documentation. In \dq{34510911}, an LCSD developer asks for sample code to convert a web page to a PDF. The official documentation is not sufficient enough for entry-level practitioners.

\bf{Testing (13, 2.7\%).} LCSD testing differs from standard software testing in some fundamental ways. In LCSD development, many of the features are implemented using graphical interfaces, and they are provided and tested by the LCSD platform providers. As a result, unit testing is less important compared to traditional software development. In LCSD approach practitioners face difficulties to lack of documentation of testing approach in LCSD platform (e.g, ``How to bypass login for unit-testing [Platform]?'' in \dq{54432666}), test coverage (e.g, \dq{54899980}, \dq{57755398}), automated testing (e.g, ``[Platform] 20.1 automated testing'' \dq{63594106}), testing browser compatibility (e.g, \dq{}), troubleshooting errors while running tests (e.g, \dq{47254010}) etc.

\bf{Deployment (16, 3.3\%).} At this phase, the feature of the application  needs to be deployed for the targeted users. One of the goals of LCSD development is to handle many of the complexities of the \textit{deployment and maintenance} phase. Many LCSD platform providers provide advanced Application Life-Cycle Management tools to deploy and maintain the staging (i.e., testing) and the production server (e.g., \dq{65124133}). However, LCSD practitioners still face many challenges regarding deployment configuration issues (\dq{46369742}), Domain name configuration (e.g., DNS configuration (e.g, \dq{65678735}), SSL Configuration (e.g, \dq{67186273})), accessibility issues such as with public URL (\dq{44136328}, \dq{53884162})) etc. 
For example, in this post, a practitioner is having deployment issues (e.g., ``[Platform] app not working after deployment'' (\dq{42506938})). A community member provides a detailed description of how to accomplish this in the answer, highlighting the lack of \textit{Official Documentation} for such a critical use-case. There are a few questions concerning delivering an app with a custom URL or domain name (for example, "How to make friendly custom URL for deployed app" in \dq{47194231}). It was challenging in this scenario because the platform did not have native support.

\bf{Maintenance (13, 2.8\%).}
At this phase, the LCSD application is deployed and requires ongoing maintenance. Sometimes new software development life cycle is agile (i.e., incremental) because new issues are reported that were previously undiscovered and request new features from the users. LCSD practitioners face some problems at this phase, such as event monitoring (e.g, \dq{64322219}), collaboration and developers role management (e.g, ``Role based hierarchy in report access'' in \dq{10436719} or \dq{52762374}), and application reuse (e.g, \dq{64276891}), application version, i.e.,  ``Do I have the latest version of an [Platform] component?'' in \dq{45209796} or \dq{52762374}, etc.

\begin{table}[t]
  \centering
   \caption{Distribution (frequency) of LCSD topics per SDLC phase. Each colored bar denotes a phase 
   (Black = Requirement Analysis, Green = Application Design, Magenta = Implementation, Red = Testing, Blue = Deployment, Orange = Maintenance)}
    \resizebox{\columnwidth}{!}{%
    \begin{tabular}{lr}
    \toprule{}
    \textbf{Topics} & \textbf{Development Phases Found in \#Questions}\\
    \midrule
        \textbf{Customization (95)} & \sixbars{5}{17}{71}{1}{0}{1}  \\
        \textbf{Data Storage (95)} & \sixbars{6}{16}{70}{0}{2}{1}  \\
        \textbf{Platform Adoption (94)} & \sixbars{17}{21}{51}{1}{0}{4}  \\
        \textbf{Platform Maintenance (94)} & \sixbars{11}{9}{46}{10}{12}{6}  \\
        \textbf{Third-Party Integration (93)} & \sixbars{4}{17}{68}{1}{2}{1}  \\ 
    \bottomrule
    \end{tabular}%
   }
  \label{tab:topicSDLC}%
\end{table}%

\nd\bf{\ul{Topic Categories in different SDLC phases.}}
We find that for all five topic categories, LCSD practitioners need some community support from planning to debugging to deployment (e.g., ``How does one deploy after building on [platform]'' in \dq{3952481}). We report how LCSD topics and different types of questions are distributed across six SDLC phases. Table~\ref{tab:topicSDLC} shows the distribution of SDLC phases for each topic category. Our analysis shows that for the Customization topic Category, most questions are asked during the Implementation (75\%) and Design (18\%) phases. The most dominant SDLC phase, i.e., the Implementation phase, is most prevalent in Customization (75\%), Data Storage (74\%), and Third-Party Integration (73\%). Requirement Analysis phase is dominant in Platform Adoption (18\%) and Platform Maintenance (12\%) topic categories where practitioners ask questions like ``Disadvantages of the [platform]'' in \dq{1664503}. Similarly, question in Platform Maintenance topic category is also prevalent in Testing (11\%), deployment (13\%), and Maintenance (6\%) SDLC stage.

\begin{table}[t]
  \centering
    \caption{Types of questions across the Software development life cycle phases}
    \begin{tabular}{lrrrr}\toprule
    \textbf{SDLC phase} & \multicolumn{1}{l}{\textbf{How}} & \multicolumn{1}{l}{\textbf{What}} & \multicolumn{1}{l}{\textbf{Why}} & \multicolumn{1}{l}{\textbf{Other}} \\
    \midrule
        Requirement Analysis \& Planning(9\%) & 28\% & 35\% & 7\% & 30\% \\
        Application Design(17\%) & 75\% & 16\% & 4\% & 10\% \\
        Implementation(65\%) & 59\% & 17\% & 16\% & 11\% \\
        Testing(3\%) & 62\% & 15\% & 15\% & 8\% \\
        Deployment(3\%) & 31\% & 25\% & 38\% & 12\% \\
        Maintenance(3\%) & 46\% & 15\% & 31\% & 15\% \\

    \bottomrule
    \end{tabular}%
  \label{tab:ques_types}%
\end{table}%

\nd\bf{\ul{Types of questions in different SDLC phases.}}
We report the distribution of question types across SDLC phases in Table~\ref{tab:ques_types}. It shows that for Requirement Analysis \& Planning phase, most questions (35\%) belong to What-type. This insight signifies that at this phase, practitioners are making \rev{inquiries} about feature details (e.g., \dq{9577099}). In the Application Design, Implementation, and testing phase, most of the questions belong to How-type, i.e., practitioners are querying about how they can implement a particular feature (e.g., \dq{13933003}) or test it (e.g., \dq{9594709}). At the Deployment phase most prominent is Why-type (38\%) followed by How-type(31\%). We can see a similar pattern for the Maintenance phase, where the most significant question type is How-type (46\%) followed by Why-type (31\%). We see this pattern because, at the Deployment and Maintenance phase, most of the questions belong to some server configuration error (e.g., \dq{4497228}) and the practitioners' inquiry about how they can set up specific server settings (e.g., \dq{8148247}). Similarly, we find \rev{that} What-type questions are more prevalent during Requirement Analysis and Deployment phases.

\begin{tcolorbox}[flushleft upper,boxrule=1pt,arc=0pt,left=0pt,right=0pt,top=0pt,bottom=0pt,colback=white,after=\ignorespacesafterend\par\noindent]
\noindent\textbf{RQ4. How are the \rev{observed topic categories} discussed across SDLC phases?}
Among six agile SDLC phases, the Implementation phase is the most prevalent (65\% questions), followed by Application Design (17\%), Requirement Analysis \& Planning (9.1\%), Deployment (3.3\%), Maintenance (2.8\%) and Testing (2.7\%).
The Implementation Phase is most prevalent in all of the five topic categories and four question types. During Requirement Analysis, Testing, and Deployment phases, Platform Adoption and Platform Maintenance topic categories are more dominant. The How-type question is most popular in the Application Design phase, the what-type question is prevalent in the Requirement Analysis and Planning phase, and the why-type question is prevalent in the Deployment and Requirement Analysis phases.
\end{tcolorbox}

\subsection{What LCSD topics are the most difficult to get an accepted answer? (RQ5)} \label{rq:pop_diff}

\subsubsection{Motivation}
After reviewing LCSD-related topics and discussions in the agile SDLC stages, we discovered that LCSD practitioners encounter generic software development problems and particular challenges specific to LCSD platforms (e.g., Platform Adoption, Platform Maintenance). Some posts come up repeatedly, and some have a lot of community participation (i.e., answers, comments, up-votes). As a result, not all topics and SDLC \rev{phases} are equally difficult \rev{to get a solution}. A thorough examination of the complexity and popularity of the practitioners' conversation might yield valuable information about how to prioritize research and community support. For example, LCSD platform providers and academics can take the required measures to make the architecture, design, features, and tools of LCSD platforms more useable for practitioners, particularly newbies.

\subsubsection{Approach}
We compute the difficulty \rev{of getting an accepted answer} for a group of questions using two metrics for each question in that group \begin{inparaenum}[(1)] 
\item Percentage of questions without an accepted answer,
\item Average median time needed to get an accepted answer.
\end{inparaenum}
In the same way, we use the following three popularity metrics \rev{to calculate popularity of that topic in the SO community}: \begin{inparaenum}[(1)]
\item Average number of views, 
\item Average number of favorites (i.e., for each question number of users marked as favorite), 
\item Average score. 
\end{inparaenum}

The five metrics are standard features of a SO question, and many other related studies \cite{iot21, alamin2021empirical, bagherzadeh2019going, abdellatif2020challenges, ahmed2018concurrency} have used them to analyze the popularity and difficulty \rev{of getting a solution for } a question. In SO, one question can have multiple answers, and The user who posted the question has the option of marking it as accepted. Hence, the accepted answer is considered correct or sound quality. So, the absence of an accepted answer may indicate the user did not find a helpful, appropriate answer. The quality of the question (i.e., problem description) might be one reason for not getting an acceptable answer. However, the SO community collaboratively edits and improves the posts. Therefore, the lack of an accepted answer most likely indicates that the SO community finds those questions challenging to answer. The success and usefulness of a crowd-sourced platform such as SO depends on the community members to quickly provide relevant, helpful correct information. In SO, the median time to get an answer is around 21 minutes only \cite{iot21}, but \rev{a complicated or domain-specific question may necessitate additional time to receive an accepted answer.} 

It can be non-trivial to assess the popularity and difficulty \rev{of getting an accepted answer for } the topics using multiple metrics. We thus compute two fused metrics following related works~\cite{iot21}. We describe the two fused metrics below.

\bf{Fused Popularity Metrics.} First, we compute the popularity metrics for each of the 40 LCSD topics.  However, the average view counts can be in the range of hundreds, average scores, and average favorite count between 0-3. Therefore, following related study~\cite{iot21} we normalize the values of the metrics by dividing the metrics by the average of the metric values of all the groups (e.g., for topics $K$ = 40). Thus, we create three new normalized popularity metrics for each topic. For example the normalized metrics for a group $i$ for all the  $K$ groups can be $ViewN_{i}$, $FavoriteN_{i}$, $ScoreN_{i}$ (e.g., for LCSD topics $K$ = 40). Finally, We calculate the fused popularity $FusedP_{i}$ of a group $i$ by taking the average of the three normalized metric values.
\begin{eqnarray}
ViewN_{i} = \frac{View_{i}}{\frac{\sum_{j=1}^{K}View_j} {K}} \\
FavoriteN_{i} = \frac{Favorite_{i}}{\frac{\sum_{j=1}^{K}Favorite_j} {K}} \\
ScoreN_{i} = \frac{Score_{i}}{\frac{\sum_{j=1}^{K}Score_j} {K}}
\end{eqnarray}
\begin{equation}\label{eq:fusedP}
FusedP_{i} = \frac{ViewN_{i} + FavoriteN_{i} + ScoreN_{i}}{3}
\end{equation}

\bf{Fused Difficulty Metrics.} Similar to popularity metrics, we first compute the difficulty metrics for each topic. Then we normalize the metric values by dividing them by the average of the metric value across all groups (e.g.,  40 for LCSD topics). Thus we, create two new normalized metrics for a given topic $i$. Finally, We calculate the fused difficulty metric $FusedD_{i}$ of topic $i$ by taking the average of the normalized metric values.
\begin{eqnarray}
PctQuesWOAccAnsN_{i} = \frac{PctQWoAcceptedAnswer_{i}}{\frac{\sum_{j=1}^{K}PctQWoAcceptedAnswer_{j}}{K}} \\
MedHrsToGetAccAnsN_{i} = \frac{MedHrsToGetAccAns_{i}}{\frac{\sum_{j=1}^{K}MedHrsToGetAccAns_j}{K}}
\end{eqnarray}   
\begin{equation}\label{eq:fusedD}
FusedD_{i} = \frac{PctQuesWOAccAnsN_{i} + MedHrsToGetAccAnsN_{i}}{2}
\end{equation}

In addition to this, we also aim to determine the correlation between the difficulty and
the popularity of the topics. We use the Kendall Tau correlation measure~\cite{Kendall-TauMetric-Biometrica1938} to find the correlation between topic popularity and topic difficulty. Unlike Mann-Whitney correlation\cite{kruskal1957historical}, it is not susceptible to outliers in the data. We can not provide the evolution of popularity and difficulty for these topics because SO does not provide the data across a time series for all metrics such as view count, score, etc. However, as\rev{LCSD-related} topics are showing increasing trends in recent times, our analysis is valid for recent times.

\subsubsection{Results}
\begin{figure}[t]
\centering
\includegraphics[scale=0.380]{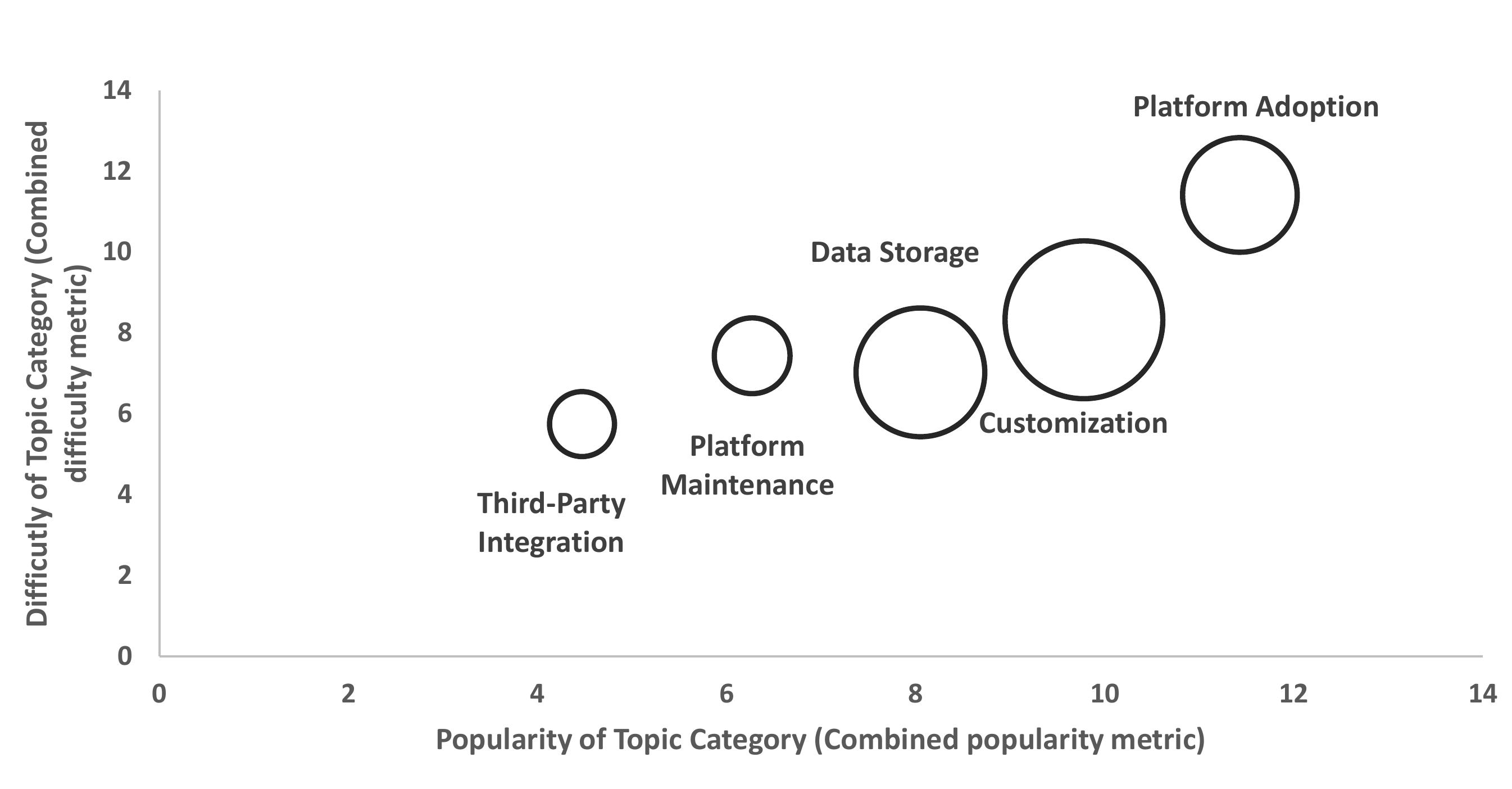}
\caption{The popularity vs. difficulty \rev{of getting an accepted answer for} LCSD Topic categories.}
\label{fig:bubble_diff_pop_per_category}
\end{figure}

\begin{table}[t]
  \centering
   \caption{Popularity \rev{for getting an accepted answer for } LCSD topics}
   \resizebox{\columnwidth}{!}{%
    \begin{tabular}{llr|rrr}\toprule
    \textbf{Topic} & \textbf{Category} & {\textbf{FusedP}} & {\textbf{\#View}} & {\textbf{\#Favorite}} & {\textbf{\#Score}} \\
    \midrule
    Platform Related Query & Platform Adoption & 3.45 &2229.9 &0.9 &2.6 \\
    Message Queue &Platform Adoption & 1.49 &1671.7 &0.3 &1.1 \\
    Dynamic Page Layout & Customization & 1.31 &2447.2 &0.2 &0.7 \\
    Build Config. Management &Platform Maintenance &1.22 &1522.1 &0.2 &1 \\
    Pattern Matching &Customization &1.17 &1539.5 &0.2 &0.9 \\
    SQL CRUD &Data Storage &1.15 &1615.7 &0.2 &0.8 \\
    Web-Service Communication &Platform Adoption &1.15 &1454.9 &0.2 &0.9 \\
    Misc. SWE Discussion &Platform Adoption &1.13 &1193.4 &0.2 &1 \\
    Interactive Report &Platform Adoption &1.12 &1680.8 &0.2 &0.7 \\
    Fetch \& Process API Response &Third-Party Integration &1.11 &1277.8 &0.2 &0.9 \\
    Data Security \& Replication &Data Storage &1.1 &1411.7 &0.2 &0.8 \\
    Hosting Config. \& SEO &Platform Maintenance &1.09 &1377.9 &0.2 &0.8 \\
    Asynchronous Batch Jobs &Platform Maintenance &1.06 &1245.7 &0.2 &0.8 \\
    Authentication \& Authorization &Platform Adoption &1.05 &1057.6 &0.2 &0.9 \\
    Library Dependency Mngmt &Platform Maintenance &1.05 &1230.6 &0.2 &0.8 \\
    Email Processing &Third-Party Integration &1.05 &1600 &0.2 &0.6 \\
    Formatted Data Parsing &Customization &1 &1607.8 &0.1 &0.9 \\
    File Management &Data Storage &0.98 &1302.6 &0.2 &0.6 \\
    DB Setup \& Migration &Data Storage &0.97 &1282.9 &0.2 &0.6 \\
    Testing &Platform Maintenance &0.96 &1810.5 &0.1 &0.7 \\
    Date \& Time Manipulation &Customization &0.94 &1720.5 &0.1 &0.7 \\
    DB Stored Procedure &Data Storage &0.94 &1715.5 &0.1 &0.7 \\
    Dynamic Data Binding &Customization &0.92 &1810.5 &0.1 &0.6 \\
    External Web Req Processing &Third-Party Integration &0.91 &831.8 &0.2 &0.7 \\
    App Deployment &Platform Maintenance &0.89 &757.3 &0.2 &0.7 \\
    Semi-Structured Data Proc. &Data Storage &0.88 &1096.5 &0.2 &0.5 \\
    Upgradation \& Compatibility &Third-Party Integration &0.88 &559.1 &0.2 &0.8 \\
    Dialog Box Manipulation &Customization &0.79 &1479.4 &0.1 &0.5 \\
    Dynamic Event Handling &Customization &0.77 &1245.1 &0.1 &0.6 \\
    Dynamic Form Controller &Customization &0.76 &1382.1 &0.1 &0.5 \\
    Conditional BPMN &Customization &0.75 &1335.1 &0.1 &0.5 \\
    SQL Sysntax Error &Data Storage &0.75 &1343 &0.1 &0.5 \\
    Graph/Chart &Platform Adoption &0.75 &1156.3 &0.1 &0.6 \\
    User Role Management &Platform Adoption &0.73 &1058.3 &0.1 &0.6 \\
    Window Style Manipulation &Customization &0.71 &1000 &0.1 &0.6 \\
    Entity Relationship Mgmt & Data Storage &0.71 &1178 &0.1 &0.5 \\
    Dynamic Data Filtering &Customization &0.66 &1148.1 &0.1 &0.4 \\
    Date-based filtering &Data Storage &0.57 &781.9 &0.1 &0.4 \\
    Platform Infrastructure API &Platform Adoption &0.56 &577.2 &0.1 &0.5 \\
    eSignature &Third-Party Integration &0.52 &574.9 &0.1 &0.4 \\
    \bottomrule
    \end{tabular}%
    }
   \label{tab:topicPopularity}%
\end{table}%

\begin{table}[htbp]
  \centering
   \caption{Difficulty \rev{for getting an accepted answer for} LCSD topics}
   \resizebox{\columnwidth}{!}{%
    \begin{tabular}{llr|rr}\toprule
    \textbf{Topic} & \textbf{Category} & {\textbf{FusedD}} & {\textbf{Med. Hrs To Acc.}} & {\textbf{Ques. W/O Acc.}} \\
    \midrule

    Message Queue &Platform Adoption &1.86 &21.4 &61 \\
    Library Dependency Mngmt &Platform Maintenance &1.78 &18.8 &70 \\
    Web-Service Communication &Platform Adoption &1.76 &19.8 &60 \\
    Authentication \& Authorization &Platform Adoption &1.67 &17.4 &68 \\
    Platform Infrastructure API &Platform Adoption &1.62 &16.3 &70 \\
    Fetch \& Process API Response &Third-Party Integration &1.6 &16.8 &64 \\
    External Web Req Processing &Third-Party Integration &1.37 &13.1 &64 \\
    App Deployment &Platform Maintenance &1.35 &12.1 &70 \\
    Hosting Config. \& SEO &Platform Maintenance &1.35 &12.6 &65 \\
    eSignature &Third-Party Integration &1.32 &11.4 &71 \\
    File Management &Data Storage &1.24 &11.4 &62 \\
    Asynchronous Batch Jobs &Platform Maintenance &1.19 &10.8 &60 \\
    Dynamic Page Layout &Customization &1.15 &10.1 &61 \\
    User Role Management &Platform Adoption &1.03 &8.3 &60 \\
    Graph/Chart &Platform Adoption &0.99 &6.6 &68 \\
    Platform Related Query &Platform Adoption &0.99 &8.4 &54 \\
    DB Stored Procedure &Data Storage &0.97 &7.7 &57 \\
    DB Setup \& Migration &Data Storage &0.95 &6.9 &60 \\
    Dynamic Form Controller &Customization &0.92 &6.3 &61 \\
    Testing &Platform Maintenance &0.92 &6.7 &59 \\
    Conditional BPMN &Customization &0.85 &5.7 &58 \\
    Build Config. Management &Platform Maintenance &0.85 &6.4 &53 \\
    Dynamic Event Handling &Customization &0.84 &4.2 &68 \\
    Semi-Structured Data Proc. &Data Storage &0.82 &4.6 &63 \\
    Email Processing &Third-Party Integration &0.8 &4.7 &59 \\
    Dialog Box Manipulation &Customization &0.79 &4.8 &57 \\
    Interactive Report &Platform Adoption &0.79 &5 &56 \\
    Dynamic Data Binding &Customization &0.77 &5.1 &53 \\
    Dynamic Data Filtering &Customization &0.71 &4.8 &48 \\
    Misc. SWE Discussion &Platform Adoption &0.71 &4.8 &48 \\
    Data Security \& Replication &Data Storage &0.66 &3.5 &52 \\
    Upgradation \& Compatibility &Third-Party Integration &0.66 &4.1 &48 \\
    Date-based filtering &Data Storage &0.65 &2 &61 \\
    SQL Syntax Error &Data Storage &0.62 &1.7 &60 \\
    Entity Relationship Mgmt &Data Storage &0.62 &2.2 &57 \\
    Formatted Data Parsing &Customization &0.61 &3.1 &49 \\
    Date \& Time Manipulation &Customization &0.59 &2.5 &51 \\
    Window Style Manipulation &Customization &0.58 &2.3 &51 \\
    Pattern Matching &Customization &0.52 &1.8 &48 \\
    SQL CRUD &Data Storage &0.50 &1.7 &46 \\
    \bottomrule
    \end{tabular}%
    }
  \label{tab:topicDifficulty}%
\end{table}%

\begin{table}[h]
\centering
\caption{Correlation between the topic popularity and difficulty}
\begin{tabular}{lrrr}\toprule
coefficient/p-value & \bf{View} & \bf{Favorites} & \bf{Score}\\ \midrule
\bf{\% Ques. w/o acc. ans} &   -0.33/0.01 & 0.02/0.88 & -0.17/0.15 \\
\bf{Med. Hrs to acc. ans} &    -0.05/0.65 & 0.30/0.02 & 0.22/0.05  \\
\bottomrule
\end{tabular}%
\label{tab:pop_diff_correlation}%
\end{table}


In Figure~\ref{fig:bubble_diff_pop_per_category} we present an overview of the five high-level topic categories and their popularity and difficulty \rev{to get an accepted answer}. In the Figure, the bubble size represents the number of questions in that category. The Figure shows that Platform Adoption is the most popular and challenging topic category \rev{to get an accepted answer}, followed by Customization, Data Storage, Platform Maintenance, and Third-Party Integration. We can also see that three topic categories, Platform Maintenance, Data Storage, and Customization, are almost similar in terms of difficulty \rev{to get a solution}. From our analysis, we find that practitioners find the Third-Party Integration topic category relatively less difficult because many questions in this category are also relevant to traditional software development (e.g., integrating Google Maps in \dq{63457325} and \dq{1258834})  and thus easier to get community support. Similarly, we find that questions in the Platform Adoption topic category are quite specific to particular LCSD platforms and thus sometimes have less community support to find an acceptable answer quickly. 

\bf{Topic Popularity.} For each of the 40 topics, Table \ref{tab:topicPopularity} shows three popularity metrics: Average number of 1. Views, 2. Favorites, 3. Scores. It also contains the combined popularity metrics (i.e., FusedP) that \rev{are} based on the above three metrics and using the Equation \ref{eq:fusedP}. In the Table, the topics are presented in descending order based on the FusedP popularity metric.

Platform Related Query topic from the Platform Adoption Category has the highest FusedP score. It also has the highest average favorite count (e.g., 0.90) and highest average score (e.g., 2.60). 1.7\% of total questions. This topic contains discussion about LCSD platforms features of different platforms, software development methodologies such as Agile and RAD development. 
The topic Message Queue under Platform Adoption category has the second highest FusedP value. This topic is about different asynchronous \rev{service-to-service} data exchange \rev{mechanisms} such as using a message queue. It generally contains discussions about popular micro-service design patterns. 
The topic Dynamic Page Layout under Customization categories is the third most popular \rev{topic} and it has the highest average view count (e.g., 2447.2). The posts under this topic discuss about UI (i.e. page) customization, hiding or moving \rev{elements} based on some user action or an event (e.g., disable a button for dynamic action in \dq{8640964}. The eSignature topic from Third-Party Integration is the least popular with only 1.15\% of total questions, a fused value of 0.52. It has the lowest favorite and score count. This contains discussion about different issues and customization for electronic signature of documents, i.e., docusign about collecting user's agreement/permission for sales or account opening. This topic is not that much popular \rev{and easy to get an accepted  answer} because this requirement is not generalized and not all the low-code application requires this.

\bf{Topic Difficulty.} In Table~\ref{tab:topicDifficulty} we present the two difficulty metrics: for all the questions in a topic 1. Percentage of questions without accepted \rev{answers}, 2. Median hours to get accepted answer. Similar to topic popularity, we also report the combined topic difficulty metrics (e.g., FusedD) using the Equation \ref{eq:fusedD} and the above two difficulty metrics. The topics in Table \ref{tab:topicDifficulty} are presented in descending order based on the FusedD value.

Topic Message Queue under Platform Adoption category is the most difficult topic \rev{ to get an accepted answer} in terms of FusedD value. Most median hours to get accepted \rev{answers} (21). This topic contains discussion about general micro-service architecture (i.e., producer and consumer) and well as LCSD \rev{platform-specific} support for these architectures. This is why this topic is also second most popular topic.
Library Dependency Mngmt topic from Platform Maintenance is the second most difficult topic \rev{ to get an accepted answer}. Around 70\% of its questions do not have any accepted answers. This topic concerns different troubleshooting issues about library and decencies of the system, server configuration, different library version compatibility issues. 
Web-Service Communication topic from Platform Adoption is the third most difficult topic. It has a long median wait time (around 20 hours) to get an accepted answer. This topic contains discussions about \rev{service-to-service} communication via web service description language, HTTP REST message, and Windows Communication Foundation.

The topics that contain discussion about general software development (not specific to LCSD platforms) are the least difficult topics \rev{ to get an accepted answer}. For example, topic SQL CRUD under Data Storage category is the least difficult topic in terms of FusedD value (e.g., 0.5). This contains database CRUD related queries, and advanced queries too, such as inner join, nested join, aggregate. This also contains discussion about Object query language, which is a \rev{high-level} wrapper over SQL. Topic SQL CRUD and SQL Syntax Error from the Data Storage category are two of the least difficult topics in terms \rev{of} median hours to get accepted answers. Topic Pattern Matching and SQL CRUD are two of the least difficult topics in terms of questions without accepted answers.

Alternatively, topics that are specific to LCSD platforms are the most difficult topics. Four out of five most difficult topic belongs to Platform Adoption Categories. These questions can be popular as well as difficult. For example, \rev{LCSD-related} Third-Party Integration related topic eSignature is the least popular topic  from Table \ref{tab:topicPopularity}, is the most difficult topic in terms of questions without accepted answers (71\%). Topic Platform Related Query is in the mid-range \rev{in terms of} difficulty but most popular \rev{ to get an accepted answer}.

\bf{Correlation between Topic Difficulty and Popularity.} 
Here we want to explore if there is any positive or negative relationship between topic popularity and difficulty. For example, Message Queue is the most difficult and, at the same time second most popular topic \rev{ to get an accepted answer} in terms of FusedD and FusedP metrics.
Platform Related Query is the most popular but mid-range difficult \rev{topic}. 

Table~\ref{tab:pop_diff_correlation} shows six correlation measures between topic difficulty and popularity in Table~\ref{tab:topicPopularity} and~\ref{tab:topicDifficulty}. Three out of six correlation coefficients are negative, and the other three are positive and they are not statistically significant with a 95\% confidence level. Therefore, we can not say the most popular topic is the least difficult \rev{ to get an accepted answer} and vice versa. Nonetheless, LCSD platform provides could use this insight to take necessary steps. Most popular topics should have \rev{an} easy to access-able answer (i.e., least difficult).


\begin{tcolorbox}[flushleft upper,boxrule=1pt,arc=0pt,left=0pt,right=0pt,top=0pt,bottom=0pt,colback=white,after=\ignorespacesafterend\par\noindent]
\noindent\textbf{RQ5. What LCSD topics are the most difficult to answer?}
Platform Adoption is the most popular and challenging topic category, followed by Customization, Data Storage, Platform Maintenance, and Third-Party Integration. We also find that LCSD practitioners find Software Deployment and Maintenance phase most popular and difficult and Testing phase to be least difficult \rev{to an accepted answer. This indicates that LCSD platform providers should provide additional support to enable low-code practitioners understand and utilize the platform's features.}
\end{tcolorbox}

\section{Discussions} \label{sec:discussion}
During our analysis, we observed that several LCSD platforms are more popular across the topics than other platforms. We analyze our findings of LCSD topics across the top 10 most prevalent LCSD platforms in the dataset (\sec\ref{sec:evolutionTopLCSDPlatforms}). 
Finally, we discuss the implications of our study findings in \sec\ref{sec:implications}.




\rev{\subsection{Issues with not accepted answers or posts with negative score.}
\label{sub-sec:accepted_answer_negative_score}
In this paper, for topic modeling we used questions and accepted answers only. We did not consider the posts with negative score too because of the following observations. \begin{inparaenum}[(1)]
        \item Many other similar empirical studies on Topic modeling on SO posts also considered the questions and accepted answers only, e.g., IoT developers discussions in SO~\cite{iot21}, big data related discussion~\cite{bagherzadeh2019going}, concurrency related topics~\cite{ahmed2018concurrency}, mobile app development~\cite{Rosen-MobileDeveloperSO-ESE2015}.
        \item A significant number of studies~\cite{asaduzzaman2013answering, ren2019discovering, ponzanelli2014improving, yang2016query} report quality of questions and unaccepted answers in SO are questionable and therefore it is quite \rev{a} standard practice for SE researchers to consider the accepted answers in SO only. For example, In \dq{7504057} (Fig~~\ref{fig:posts_negative_unanswered}) a user asks question about python code/package to connect and retrieve data from Salesforce. The accepted answer \da{7504244} provides a relevant python code snippet the unaccepted answer \da{34055640} provide resource link for a command line tool which may be relevant but exactly not what the user asked for.
        \item Negative scored questions are typically incorrectly tagged (e.g., \dq{4862071}, \dq{21377026}, \dq{37371712}), duplicates (e.g., \dq{12282151}, \dq{48121405}), lack a detailed problem description (e.g., \dq{25691340}, \dq{1974480}, \dq{50666660}), lack correct formatting (e.g., \dq{32208310}). For instance, in \dq{51635004} (Fig~\ref{fig:posts_negative_unanswered}) a user inquires about an error encountered when attempting to contact a Zoho API. However, crucial important information such as an issue code or error message is lacking from the question description. In \dq{4862071}, an inexperienced user inadvertently tagged a question about the Oracle Apex platform with the Salesforce tag.
    \end{inparaenum}
We, therefore, choose not to include questions with a negative score or unaccepted answers. We also provide potentially missing out some insights for this choice in the threats to validity section (Section~\ref{subsec:validity}).
\begin{figure}[htbp]
\subfloat[A question with negative score]{\includegraphics[height=2.5in, width=2.1in]{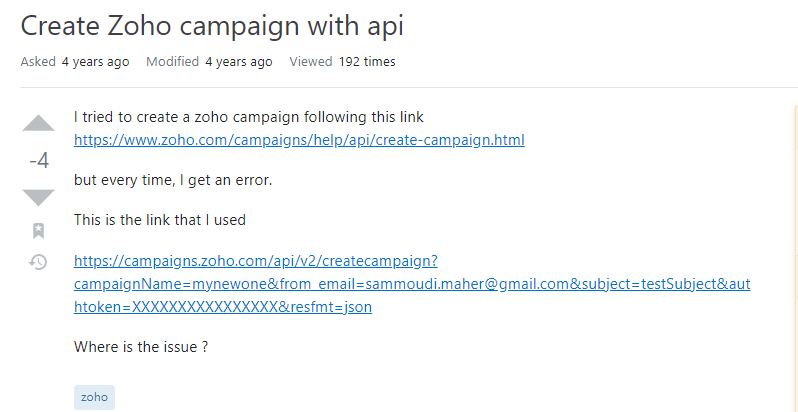}}
\subfloat[A question with irreverent unaccepted answer]{\includegraphics[height=2.5in, width=2.7in]{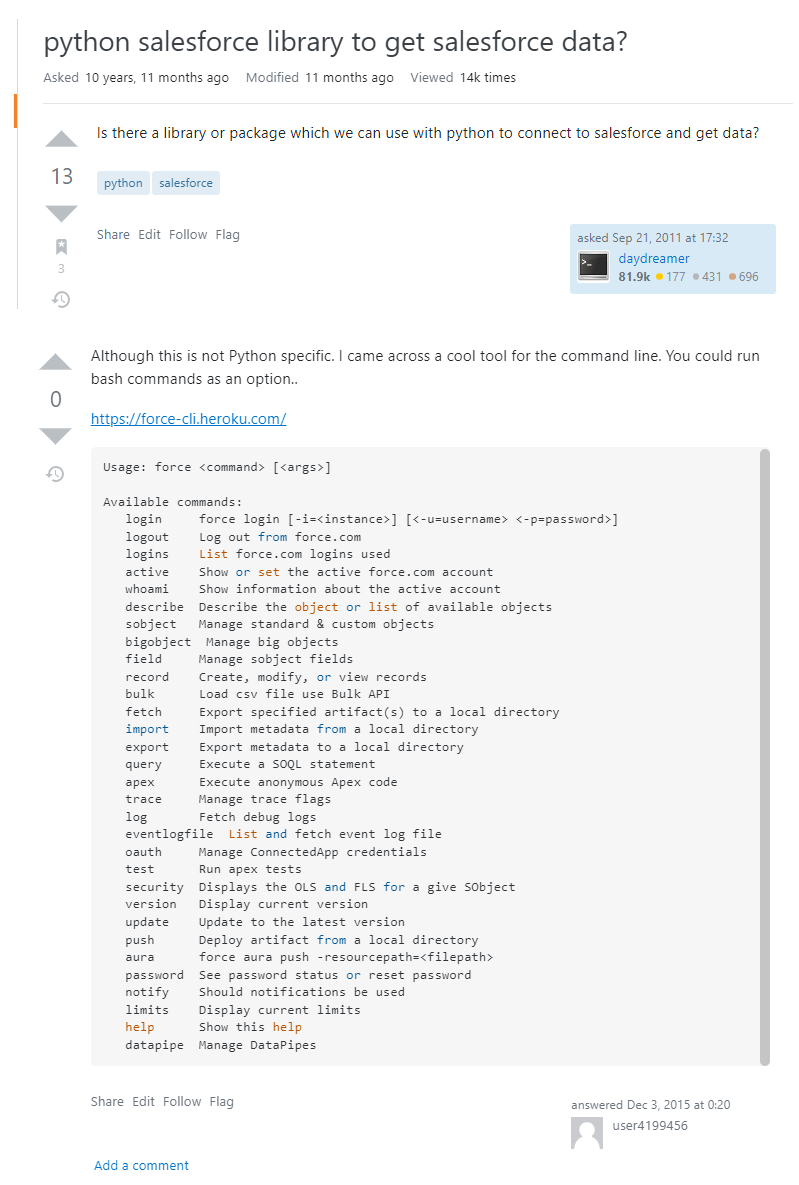}}
\caption{SO questions with a negative score or unaccepted SO answer.}
\label{fig:posts_negative_unanswered}
\end{figure}
}


\rev{\subsection{Discontinued low-code platforms and future trends.}
From our analysis on Section~\ref{rq:evolve} we see the evolution of LCSD platforms, especially from 2012. According to our data, we can see the discontinuation of some low-code platforms but they are usually soon replaced by new low-code/no-code \rev{services}. For example, In Jan 2020, Google announced the discontinuation of Google App Maker~\cite{googleappmaker} by 2021~\cite{google-disc}. But, shortly thereafter, Google announced a ``no-code'' platform called ``AppSheet''~\cite{googleAppSheet} and promoted their fully managed serverless platform called AppEngine~\cite{appengine} to create web application promoting low-code approach. Microsoft and Amazon are also competing for superior low-code/no-code platforms with the emergence of new low-code service platforms such as Microsoft Power FX~\cite{microsoftpowerfx}, Amazon Honeycode~\cite{honeycode}, AWS Amplify Studio~\cite{amplifystudio}. The low-code approach is attracting increasing interest from traditional businesses, particularly during the pandemic~\cite{pandemic-low-code}.
}

\rev{\subsection{LDA parameter Analysis.}
In this study, we applied LDA topic modelling, which employs Dirichlet distribution, to identify practitioners' discussions on low-code. As described in details in Section~\ref{sub-sec:topic_modeling}, we followed the industry standard to configure the parameters and hyperparameters and also followed the industry recommendation to manually annotate the topics as described in Section~\ref{sec:rq_topic} in order to avoid sub-optimal solutions~\cite{de2014labeling}. Following similar studies~\cite{iot21, abdellatif2020challenges, han2020programmers} we use the use the coherence score of of each model for different values of $K$. However, since LDA itself is probabilistic in nature~\cite{agrawal2018wrong} and can produce different results different runs on the same low-code dataset. In order to mintage this problem, we run our LDA model three times and compare the optimal number of topics. Fig.~\ref{fig:coherence_scores} shows the result of different coherence score for different values of $K$. Moreover, we can see after reaching highest coherence values for $K$ = 45 the overall coherence score decreases as the value of $K$ increases.
\begin{figure}[h]
\centering
\includegraphics[scale=0.60]{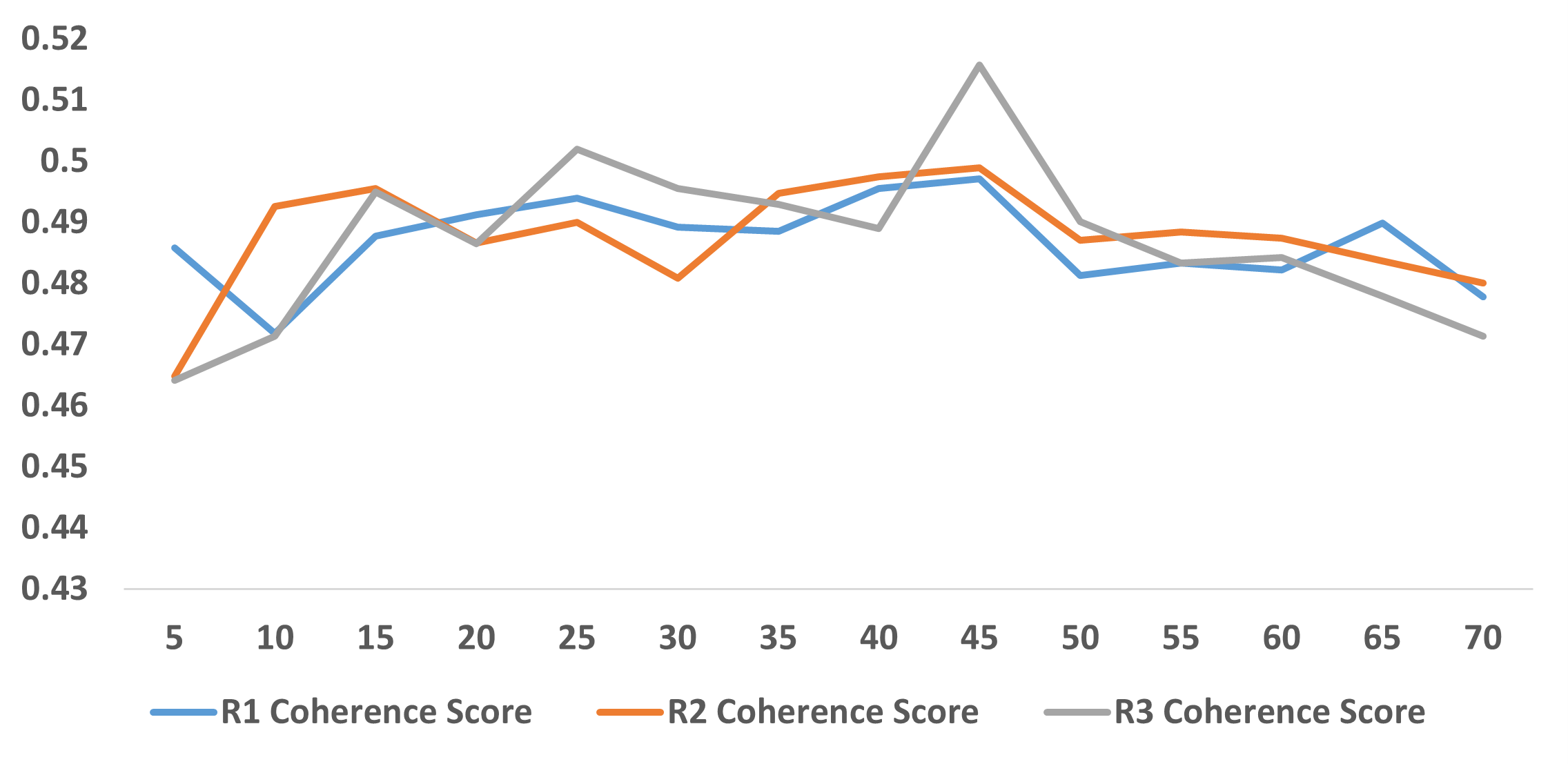}
\caption{The different coherence values for varied $K$ on different runs.}
\label{fig:coherence_scores}
\end{figure}
}

\subsection{The Prevalence \& Evolution of Top Ten LCSD platforms}\label{sec:evolutionTopLCSDPlatforms}
\begin{figure}[t]
\centering
\includegraphics[scale=0.5]{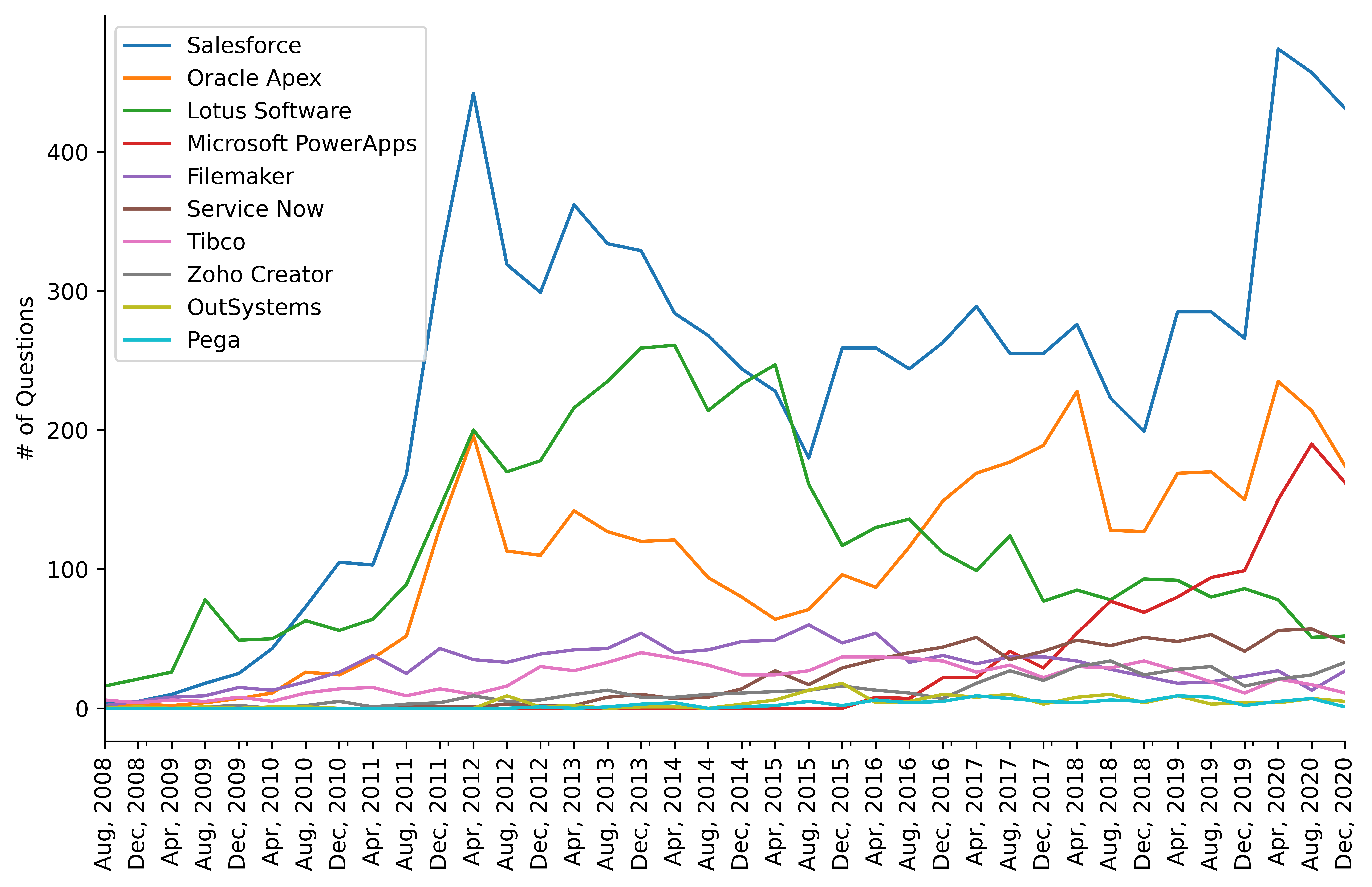}
\caption{\rev{The evolution of top ten LCSD platforms discussions over time.}}
\label{fig:top_platform_evolution_over_time}
\end{figure}

Our analysis of the evolution of topic categories (see \sec\ref{rq:evolve}) shows that there is an overall increase \rev{in} the number of new questions across the topics in SO. Our SO dataset is created by taking into account the LCSD platforms. In Figure~\ref{fig:top_platform_evolution_over_time}, we show how the 10 LCSD platforms evolve in our SO dataset over the past decade based on the number of new questions.  Salesforce\cite{salesforce} is the biggest and one of the oldest LCSD \rev{platforms} (released in 1999) in our dataset with around 30\% of all questions followed by Lotus Software\cite{lotus}, Oracle Apex~\cite{oracle_apex}, Microsoft powerapps\cite{powerapps}. Among these platforms, IBM Lotus Software was quite popular during the 2014s and gradually lost its popularity, and IBM finally sold it in 2018. Salesforce platform has been the most popular platform in terms of SO discussions since 2012. Our graph shows that these other three platforms, especially Microsoft Powerapps, \rev{are} gaining lots of attention during the pandemic, i.e., early 2020. 

\begin{figure}[t]
\centering
\includegraphics[scale=0.43]{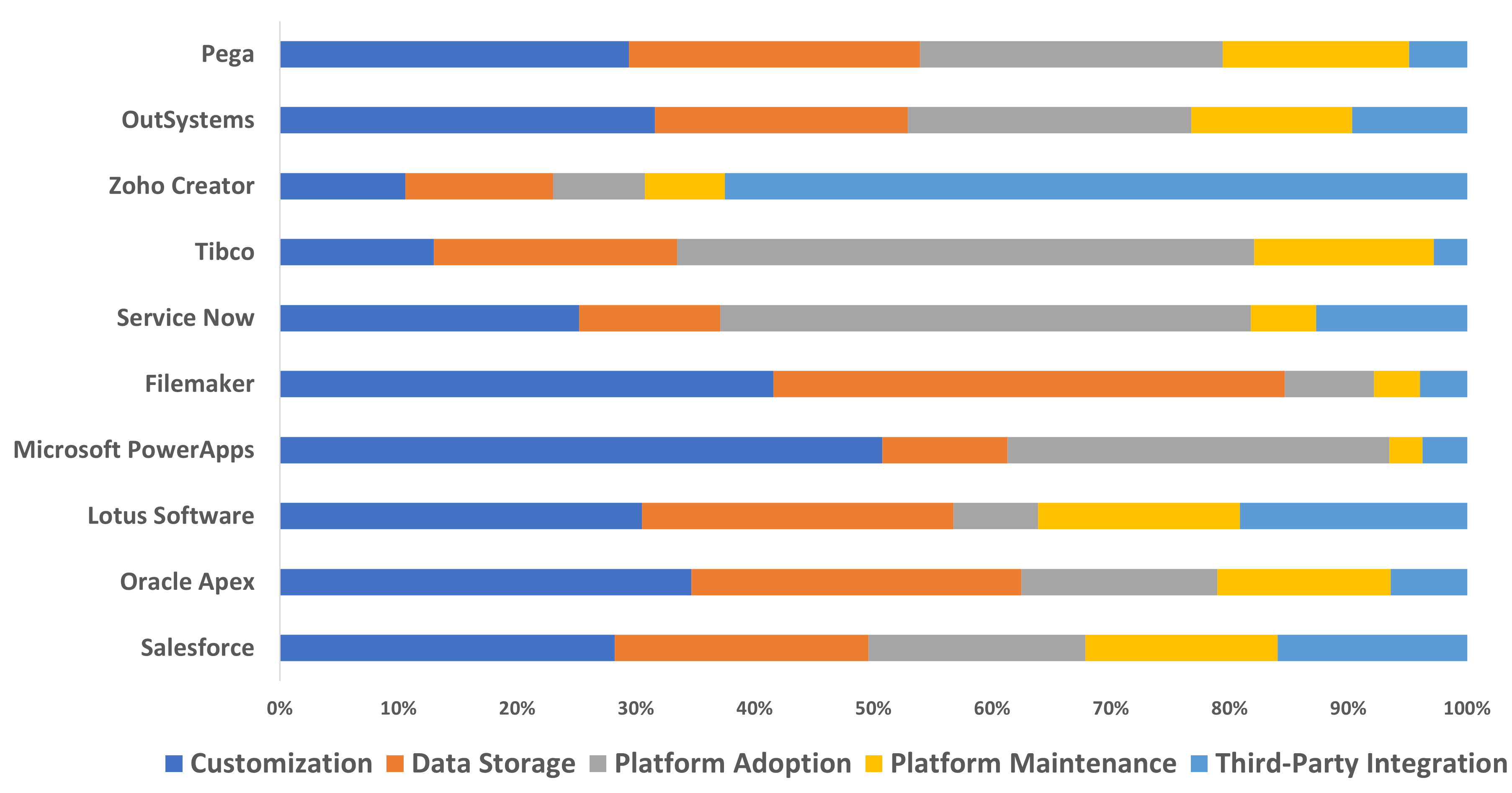}
\caption{The distribution of topic categories across top ten LCSD platforms.}
\label{fig:platform_topic_cat}
\end{figure}

We provide more context for these platforms in Figure~\ref{fig:platform_topic_cat} by illustrating the distribution of our observed five topic categories across the top ten LCSD platforms. We can see that Powerapps have the most number of queries in the Customization and Platform Adoption category. This happens because Powerapps is a relatively new LCSD platform (released in 2016) and it is gaining more and more attention from the community, and thus there are more queries such as business logic implementation (i.e., \dq{61685582}), connect Powerapps to a database in \dq{61611950}, user permission (i.e., \dq{61838119}). We can also see that older platforms such as Salesforce and Oracle Apex have more queries regarding Platform Maintenance, Third-Party Integration \rev{topic category}. Practitioners ask many different questions regarding these platforms such as deployment-related  (e.g., ``Deploying a salesforce.com flex app without visualforce'' in \dq{6614226}), third-party API integration (e.g., ``Google Map Integrated with Salesforce is Blank'' in \dq{9028682}), maintenance deployment (e.g., ``Salesforce deployment error because of test class failure'' in \dq{9171945}), interactive report in \dq{9700660}, customization with JSON \dq{9833992}, ``what is dashboard?'' in \dq{10911269} and Oracle Apex how to use platform in \dq{9438695}. Platform Adoption is a prevalent topic category in the Powerapps, ServiceNow, and Tibco platforms. We also notice that the Data Storage category is quite popular in Filemaker and Lotus Software. Interestingly we see that Zoho Creator LCSD platform around 60\% questions belong to Third-party API integration (especially email configuration \dq{48865565}). This data sheds light on the popular discussion topics of more recent and earlier LCSD platforms.
\begin{figure}[t]
\centering
\includegraphics[scale=0.60]{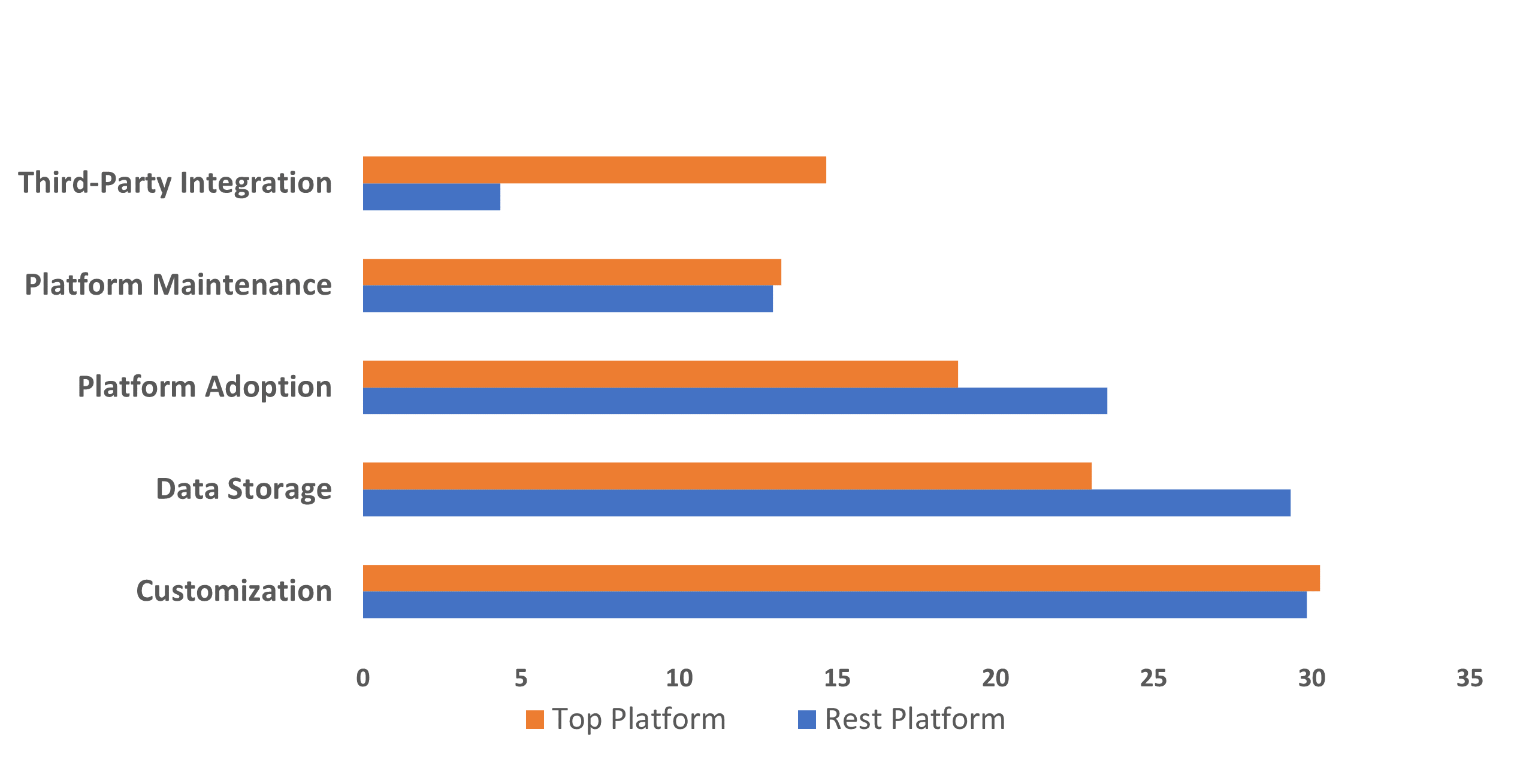}
\caption{The percentage of questions distributed across five topic categories for the top ten LCSD platforms versus \rev{the} rest of the platforms.}
\label{fig:top10vsrest}
\end{figure}

\rev{\subsection{The Case of ``aggregating" data across multiple LCSD platforms.}
In this study, we analysed 38 LCSD platforms and these platforms have distinct characteristics and challenges. Our goal is to offer researchers and practitioners a comprehensive overview of the LCSD domain as a whole, as opposed to focusing on a single LCSD platform. Hence, we integrated the data from all of these platforms. For instance,Fig.~\ref{fig:top_platform_evolution_over_time} \rev{demonstrates} that some of the most popular platforms such as Salesforce, Oracle Apex, \rev{and} Microsoft Powerapps have more questions in SO than other LCSD platforms. Fig.~\ref{fig:platform_topic_cat} demonstrates that questions across these platforms over different topic categories differs slightly. However, Fig.~\ref{fig:top10vsrest} shows that questions for Application Customization and Platform Maintenance topic category for top ten platforms vs others remain about the same at around 30\% and 13\% respectively. Popular platforms have more questions related to Third Party API integration (15\%)  than others (4\%). \rev{The} top ten platforms have relatively \rev{fewer} questions (15\%) in Data Storage (23\% vs 29\%) and Platform Adoption (19\% vs 24\%) Category compared to other platforms. Overall, we found that the observed topics are found across all the platforms that we studied. Given the popularity of some platforms over others, it is understandable that those platforms are discussed are more and as such some platforms can have more coverage (in terms of number of questions) in a topic over other platforms. However, the prevalence of all platforms across each topic shows that the topics are generally well-represented across the platforms.}

\begin{sidewaysfigure}
\vspace{12cm}
\hspace*{-2.5cm}%
\includegraphics[scale=1.2]{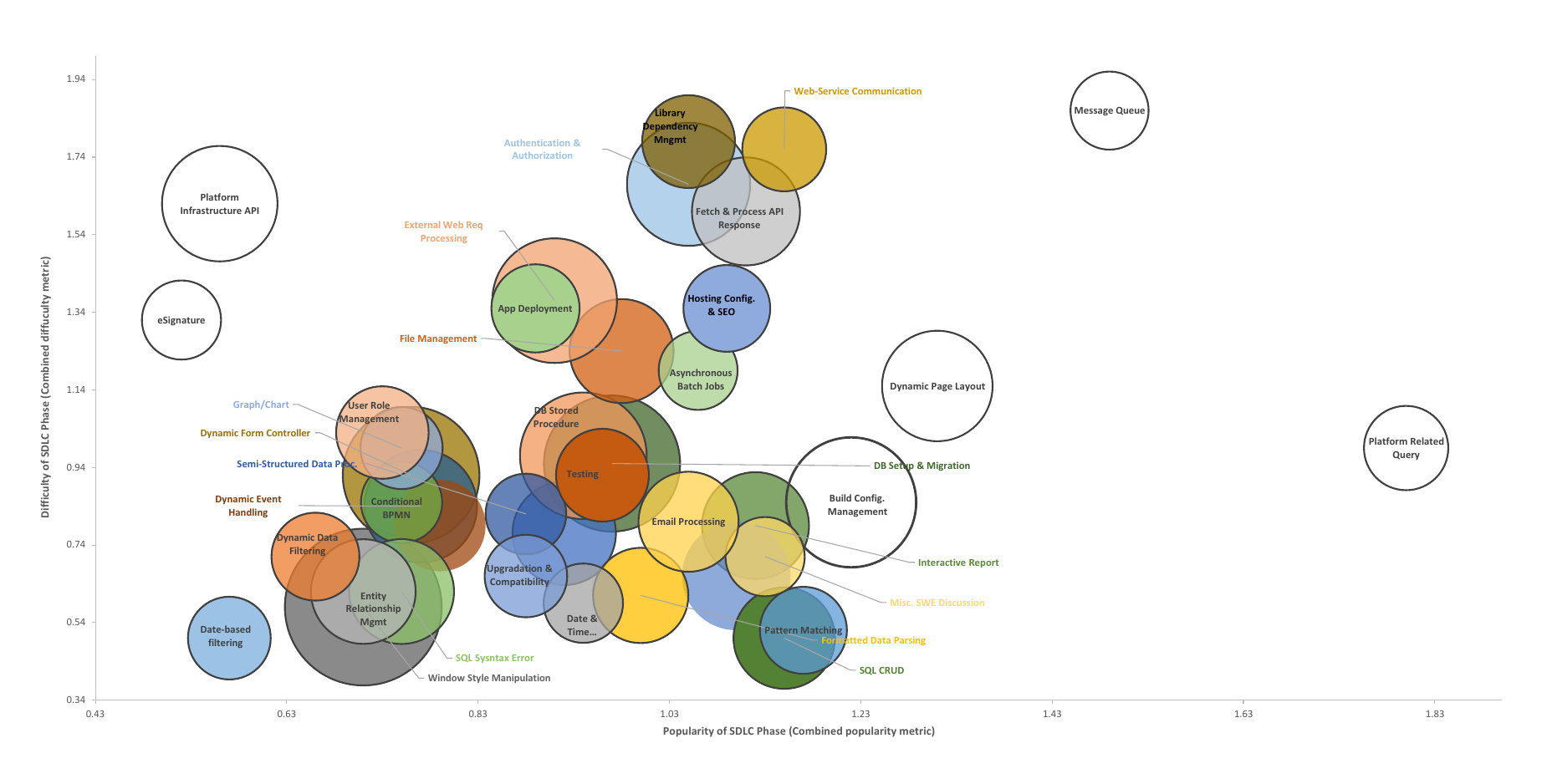}
\caption{The popularity vs. difficulty bubble chart for 40 \rev{LCSD-related} topics.}
\label{fig:40_topics_bubble_chart}
\end{sidewaysfigure}
\begin{table}[!htp]\centering
\caption{The summary of our core findings and recommendations}\label{tab:summary_findings}
\resizebox{1\columnwidth}{!}{%
\begin{tabular}{lp{1.5cm}p{4.8cm}p{5.4cm}}\toprule
\textbf{RQ} &\textbf{Theme} &\textbf{Core Findings} &\textbf{Recommendations} \\\midrule

1 & Platform Customization & We identified 40 LCSD topics that fall into five categories. Customization of platform features continues to be the most discussed challenge, referring to the difficulty developers face when seeking to adapt a given feature to a particular situation. 
& The difficulty with customization can be observed when the LCSD platforms do not offer anyway to change the default interface/functionality. The platforms may offer coding interface where individual components (e.g., a button) can be customized by writing code. \\ \cmidrule{2-4}

 & API Integration & Around 12\% of the topics discuss challenges related to the integration of third party apps into an LCSD platform like the integration of REST services. 
& The LCSD platforms may introduce specialized and programmable interfaces to support the call of a REST service and to be able to process the service output into a format that can be usable within the platform. \\ \midrule

2 & Platform Adoption & Topics related to customization and platform adoption are being discussed in recent years, because developers are increasingly asking for more documentation and other supporting features as the platforms are being adopted in real-world scenarios. 
& \multirow{ 2}{*}{1} { While the official documentation resources can be further improved by the LCSD vendor, automated tools can be investigated to generate documentation by learning of existing adoption of the platforms.} \\ \midrule

3 & How to Use & More than 50\% of the questions are How-type, showing the needs for better learning resources. & The official documentation of implementation can be further enhanced by processing the crowd-shared knowledge of the usage discussions of the platforms (e.g., from Stack Overflow)\\ \cmidrule{2-4}
 & Server Setup & What and Why-type questions are most dominant for server setup related. 
& LCSD platform should provide provide better visualizer and debugger for the practitioners to improve troubleshooting.   \\ \midrule

4 & Development Effort & Implementation is the most dominant SDLC phase (65\%). Interestingly we also around one third of the are related to Application Design (17\%) and Requirement Analysis \& Planning (9\%) 
& LCSD platform providers should take necessary steps to provide better community support in SO to address these challenges. Practitioners should also be aware LCSD platforms can improve the development of traditional software development team but they are not yet panacea for all problems.  \\ \midrule

5 & Deployment vs Maintenance & Questions related to both the deployment-SDLC and Maintenance-SDLC as the most popular and hardest to get accepted answers. 
& Platform providers should provide better level of abstraction for cloud management, application deployment and monitoring. Educators can provide better resources to learn about cloud platforms. \\ \cmidrule{2-4}

 & Messaging & The topic message queue from platform adoption is the most difficult to get an accepted answer. This topic contains discussion about the adoption of a general micro-service architecture within LCSD platforms 
& LCSD platforms can improve the adoption of micro-service architecture and the communication between different the microservices with better message queuing  \\

\bottomrule
\end{tabular}%
}
\label{tab:summaryOfFindings}
\end{table}

\subsection{Implications}\label{sec:implications}
In \tbl\ref{tab:summaryOfFindings}, we summarize the core findings of our study and provide recommendations for each findings. The findings from our study can guide the following three stakeholders: \begin{inparaenum}[(1)]
        \item LCSD platform Providers to improve the documentation, deployment, and maintenance support,
        \item LCSD Practitioners/Developers to gain a better understanding of the trade-offs between rapid development and customization constraints,
        \item Community of LCSD Researchers \& Educators to have a deeper understanding of the significant challenges facing the broader research area to make software development more accessible.
    \end{inparaenum} We discuss the implications below.


\rev{In this empirical study, we infer implications and recommendations based on our observation of practitioners' discussion in SO. So further validation from the developers' survey can provide more insight. However, the diversity of the low-code platforms and topics makes it non-trivial to design a proper survey with representative sample of LCSD practitioners. Therefore, the findings can be used to design multiple LCSD related surveys focusing on different low-code topics and platforms.}

\begin{figure}[t]
\centering
\includegraphics[scale=0.50]{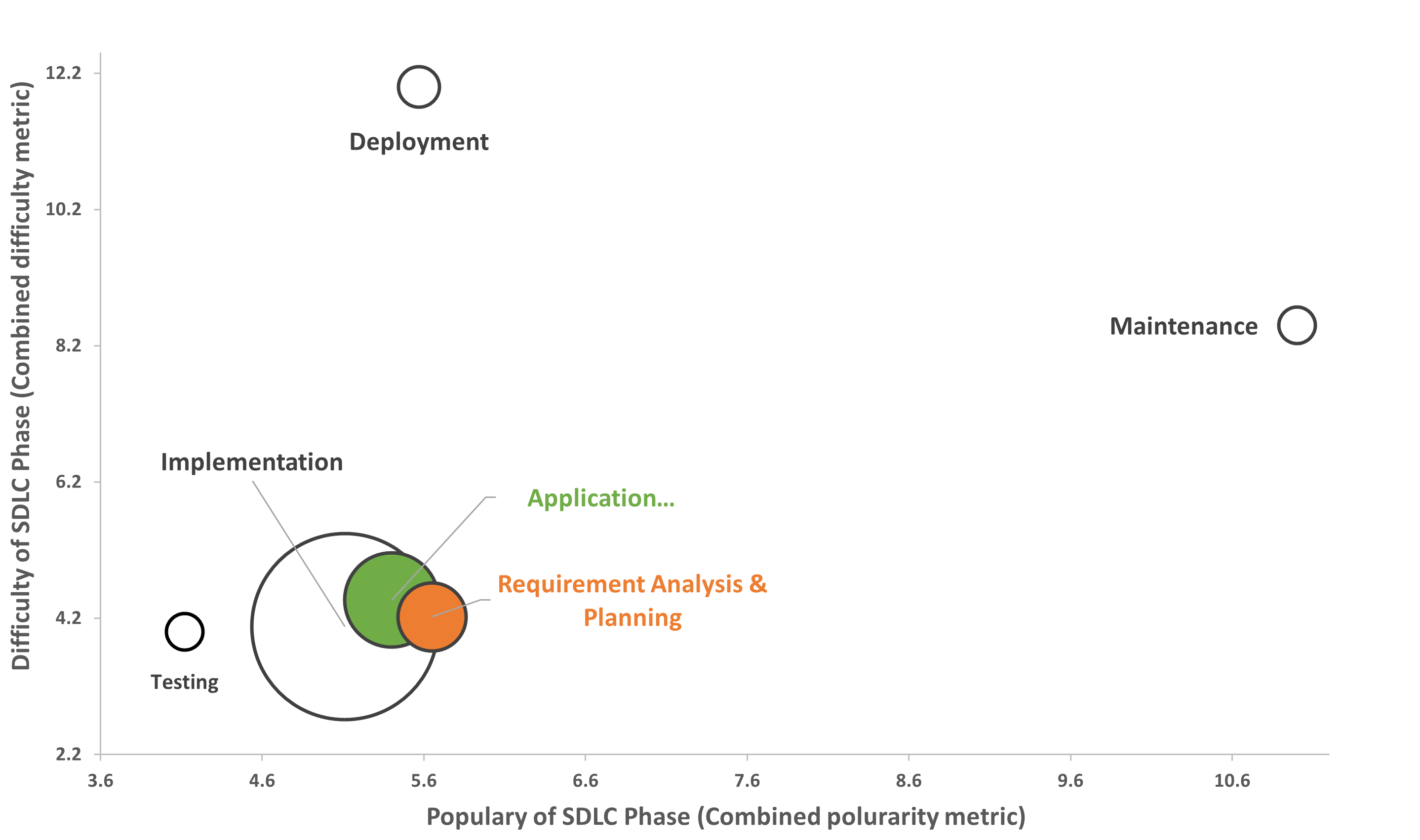}
\caption{The popularity vs. difficulty bubble chart for low-code software development life 
cycle phases}
\label{fig:SDLC_pop_diff}
\end{figure}

\begin{figure}[t]
\centering
\includegraphics[scale=0.50]{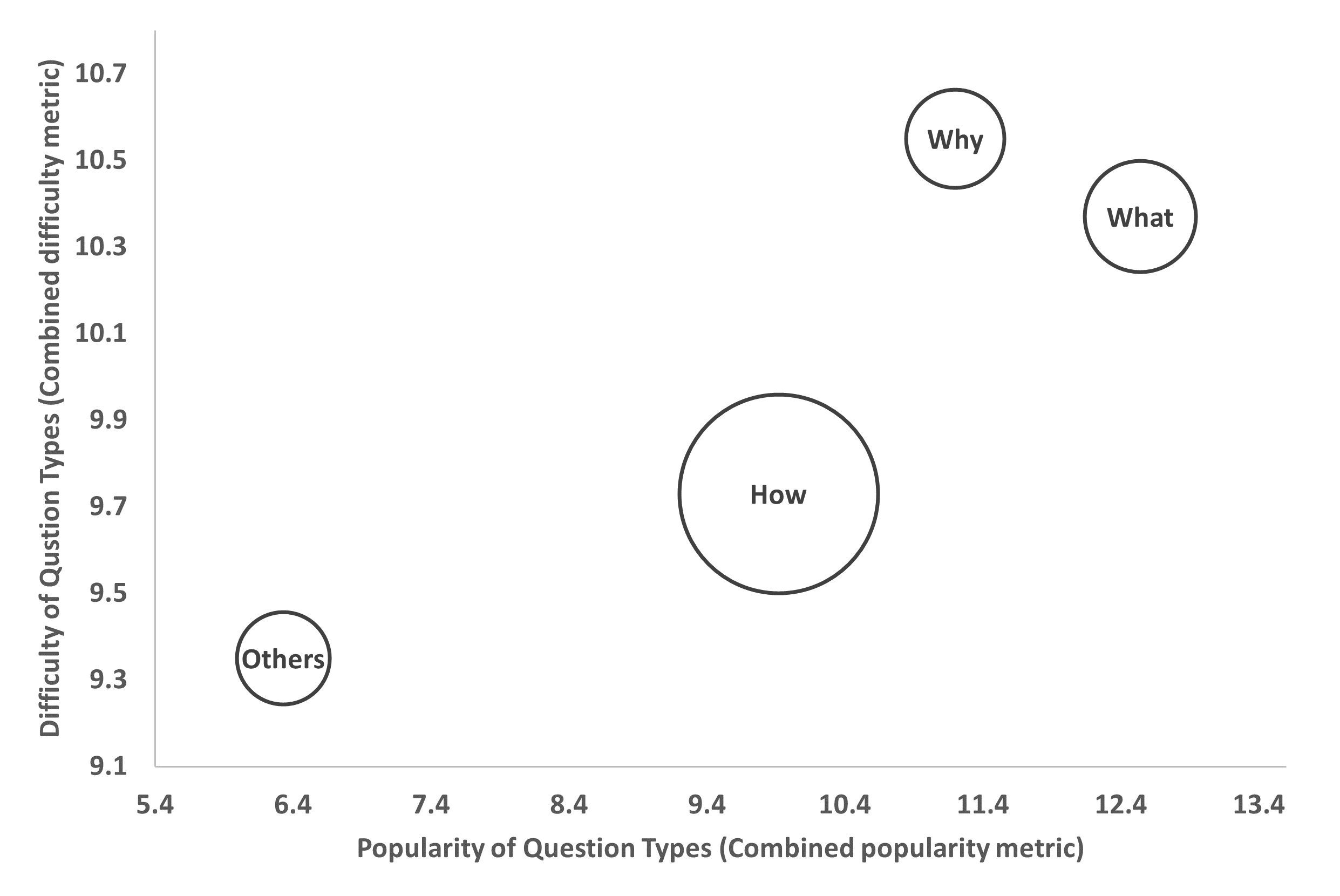}
\caption{The popularity vs. difficulty of different types of LCSD-related questions}
\label{fig:bubble_diff_pop_per_type}
\end{figure}

\bf{\ul{ LCSD Platform Vendors.}}
In order to better understand the issues of LCSD, we present a bubble chart with difficulty and popularity of different aspects of LCSD such as Topic Category in Figure~\ref{fig:bubble_diff_pop_per_category}, Types of questions in Figure~\ref{fig:bubble_diff_pop_per_type} and agile SDLC phases in Figure~\ref{fig:SDLC_pop_diff}. These findings coupled with the evolution of LCSD platforms (\ref{fig:top_platform_evolution_over_time}) and discussions (\ref{fig:topic_absolute_impact}) shows that Customization and Data Storage related queries are more prevalent, with the majority of these queries occurring during Implementation agile SDLC stage. However, one of our interesting findings is \textit{Platform Adoption} related queries are increasing in popularity. LCSD practitioners find LCSD platform infrastructure and server configuration-related quires tough and popular during the Deployment and Maintenance phase. The top five most challenging topics belong to Platform Adoption and Maintenance topic category. 

Many new practitioners make queries regarding LCSD platforms, learning resources, basic application and UI customization, and how to get started with this new emerging technology. Figure~\ref{fig:40_topics_bubble_chart} shows that \textit{Platform Related Query} topic is the most popular among LCSD practitioners. We find that \textit{Documentation} related queries are both top-rated and challenging. Our findings also suggest that many practitioners still face challenges during testing, especially with third-party testing tools like JUnit (in \dq{9811992}) and troubleshooting. Consequently, many of the questions on this topic remain unanswered. It reveals that to ensure smooth adoption of the LCSD platforms, providers should provide better and \rev{more} effective documentation and learning resources to reduce entry-level barriers and smooth out the learning curve. 

\textbf{\ul{ LCSD Practitioners/Developers.}}
Gartner~\cite{gartner} estimates that by 2022, more than half of the \rev{organizations} will adapt LCSD to some \rev{extent}. Additionally, our analysis reveals a rising trend for LCSD approaches, particularly during Covid-19 pandemic (Fig.~\ref{fig:all_questions_absolute_impact}). We can also see that new LCSD platforms such as Microsoft Powerapps are gaining many developers' attention. LCSD platform enables practitioners with diverse experience to contribute to the development process even without a software development background. However, our finding shows that practitioners find debugging, application accessibility, and documentation challenges. Hence, the practitioners should take the necessary steps to understand the tradeoffs of LCSD platforms' features deeply. The project manager should adopt specific strategies to customize, debug, and test the application. For example, many practitioners struggle with general Third-Party API integration and database design and query. We find that DevOps-related tasks such as CI/CD, Server configuration, and \rev{monitoring-related} queries are most challenging to the practitioners. So, a well-functioning LCSD team should allocate time and resources to them. It provides valuable insights for project managers to manage resources better (i.e., human resources and development time).

Figure~\ref{fig:SDLC_pop_diff} shows that Maintenance is the most popular development phase, followed by Deployment, and Testing is the least popular SDLC phase. Similarly, the figure also \rev{shows} that questions asked in deployment phase are the most difficult followed by Maintenance. Implementation, Requirement analysis and planning, Application design phase \rev{are} in the middle range in terms of popularity and difficulty spectrum. Thus, our analysis indicates that LCSD practitioners face more broad and complex application maintenance and deployment-related challenges, on which LCSD platform vendors should concentrate their efforts. This finding can influence the \rev{decision-making} process of LCSD developers and practitioners like prioritizing their efforts during the design, development, and deployment of software that uses LCSD platforms. For example, if sufficient support or tools are not available for scalable usage and deployment of an LCSD platform, developers may look for alternatives that have better deployment and maintenance \rev{support}.

One fundamental shortcoming of LCSD platforms is that their abstraction and feature limitations can make customization and debugging extremely difficult.
Additionally, managed cloud platforms make data management and deployability more challenging~\cite{sahay2020supporting, luo2021characteristics}. The findings in this study help to present some strengths and limitations of \rev{the} overall LCSD paradigm, which complements the findings of other studies~\cite{sahay2020supporting, alsaadi2021factors, luo2021characteristics, waszkowski2019low-automating, adrian2020app, ness2019potential}. The analysis could assist LCSD teams in selecting the appropriate LCSD platforms, which is critical for future success.

\textbf{\ul{ LCSD Researchers \& Educators.}}
The findings of this study have many implications for researchers and educators of LCSD platforms and the border research community to improve the software development process.
We discover that What-type and How-type questions are popular among LCSD practitioners. They also find them challenging because of adequate usable documentation. Thus, practitioners ask questions about certain limits or how to implement certain features, and in the accepted answer, some other user simply points to the official documentation page (e.g., ``Domino Data Service API Documentation'' in \dq{59739877} and \dq{5806293}). \rev{Many of the challenges faced by low-code petitioners are similar to traditional software developers.  So, researchers from border software engineering domain can contribute to improving aspects such as improving documentation~\cite{bayer2006view, khan2021automatic, bhat2006overcoming}, improving API description usage~\cite{uddin2015api, uddin2021automatic} and make it more accessible to general practitioners.} In the Customization and Data Storage topic category, we find practitioners asking help in generic programming queries, database design, and file management. So, \rev{research} on those topics will also help the adoption of LCSD. Some LCSD platforms provide great in-build support for unit and functional testing. However, we find around \textit{2.1\% of questions belong to Testing topic}. Most of these LCSD platforms heavily rely on cloud computing, and thus research improvement of server configuration and library management, i.e., DevOps~\cite{zhu2016devops} in general, will aid in better platforms. On the other hand, educators can focus their efforts on making the learning resources on Automatic testing, Server config. and DevOps \rev{practices} such as CI/CD more accessible to the citizen developers. 

Figure~\ref{fig:bubble_diff_pop_per_type}  shows What-type of posts are most popular, followed by Why-type, How-type, and Others-type. Additionally, it demonstrates that the most challenging question type is Why-type, followed by What-type, How-type, and others. So, although How-type questions are most dominant, the details types (i.e., Why-type, What-type) of questions are more popular and challenging. This analysis implies that LCSD practitioners have a harder time finding detailed information regarding different platform features. As a result, LCSD platform providers should improve their documentation.
Intuitively, How-type questions can be answered with better documentation for LCSD platforms. Given the official API documentation can often \rev{be} incomplete and obsolete~\cite{Uddin-HowAPIDocumentationFails-IEEESW2015,Khan-DocSmell-SANER2021} and given that our research shows that LCSD developers use SO to ask questions about various topics, LCSD researchers can develop techniques and tools to automatically improve the documentation of LCSD platforms by analyzing SO questions and answers. Indeed, existing research efforts show that solutions posted in SO can be used to improve API documentation by supporting diverse development tasks and programming languages~\cite{Chakraborty-NewLangSupportSO-IST2021,Uddin-OpinerAPIUsageScenario-IST2020,Uddin-OpinerAPIUsageScenario-TOSEM2020,Uddin-OpinerReviewAlgo-ASE2017,Uddin-OpinionSurvey-TSE2019,Uddin-OpinionValue-TSE2019,Uddin-OpinerReviewToolDemo-ASE2017}. 

We find that the LCSD paradigm's challenges can be different from traditional software development~\cite{sahay2020supporting}. Simultaneously, researchers can study how to
provide better tools for practitioners to customize the application.
Security is an open research opportunity for such platforms as a security
vulnerability in such platforms or frameworks could compromise millions of applications and users~\cite{lin2020software}. Researchers can develop better testing approaches to ensure faster development and dependability. Educators can also benefit from the results presented in Table~\ref{tab:topicPopularity}, \ref{tab:topicPopularity} and Figure~\ref{fig:40_topics_bubble_chart} to prioritize their focus on different \rev{topics} such as \textit{Library Dependency Mngmt,  Web-Service Communication, Asynchronous Batch Jobs, Testing,  Dynamic Form Controller}.

\section{Threats to Validity}\label{subsec:validity}


\nd\bf{Internal validity} threats, in our study, relate to the authors' bias while conducting the analysis as we have manually labeled the topics. We mitigate the bias in our manual labeling of topics, types of questions, and  LCSD phases by consulting the labels among multiple authors and \rev{resolving} any conflicts via discussion. Four of the authors actively participated in the labelling process. The first author reviewed the final labels and refined the labels by consulting with the second author.

\nd\bf{Construct Validity} threats
relate to the errors that may occur in data collection, like identifying relevant LCSD tags. To mitigate this, we created our initial list of tags, as stated in Section \ref{sec:methodology}, by analyzing the posts in SO related to the leading LCSD platforms. Then we expanded our tag list using state-of-art approach~\cite{bagherzadeh2019going,abdellatif2020challenges,ahmed2018concurrency, rosen2016mobile}. Another potential threat is the topic modeling technique, where we choose $K$ = 45 as the optimal number of topics for our dataset $B$. This optimal number of topics has a direct impact on the output of LDA. We experimented with different values of $K$ following related works~\cite{abdellatif2020challenges, bagherzadeh2019going}. We used the coherence score and manual examination to find $K$'s optimal value that gives us the most relevant and generalized low-code related topics~\cite{alamin2021empirical, iot21, abdellatif2020challenges}. 

\nd\bf{External Validity} threats relate to the generalizability of our findings. Our study is based on data from developers' \rev{discussions} on SO. However, there are other forums LCSD developers may use to discuss. We only considered questions and accepted answers in our topic modeling. We also had the option of choosing the best answer. In SO, the accepted answer and best answer may be different. Accepted answer is the one approved by the questioner while the best answer is voted by all the viewers. \rev{as discussed in Section~\ref{sub-sec:accepted_answer_negative_score} it is quite difficult to detect if an answer is relevant to the question or not. Thus We chose the accepted answer in this study because we believe that the questioner is the best judge of whether the answer solves the problem or not. Even without the unaccepted answers, our dataset contains around 38K posts (27K questions + 11K accepted answers). This also conforms with previous works~\cite{iot21, abdellatif2020challenges,alamin2021empirical, yang2016query}. Some novice practitioners post duplicate questions, assign incorrect tags, and provide inadequate descriptions, which receives an overall negative score from the community. To ensure that topics contain relevant discussion of high quality, we only use posts with non-negative scores.} Nevertheless, we believe using SO's data provides us with generalizability because SO is a widely used Q\&A platform for developers. However, we also believe this study can be complemented by including the best answers to the questions in SO, as we discussed earlier, including discussions from other forums, surveying, and interviewing low-code developers.


\section{Related Work} \label{sec:related_work}

We previously published a paper at the MSR 2021 based on an empirical study of LCSD topics in SO (see \cite{alamin2021empirical}). We compared the findings of this paper against our previous paper in \sec\ref{sec:intro}. Other related work can broadly be divided into two categories: SE (Software Engineering) research on/using \begin{inparaenum}[(1)]
\item low code software development (\sec\ref{sec:rel-lcsd}),  and
\item topic modeling (\sec\ref{sec:rel-se-res-topic}).
\end{inparaenum} 

\subsection{Research on Low Code Software Development and Methodologies}\label{sec:rel-lcsd}
LCSD is a new technology, with only a handful of research papers published in this field. Some research has been conducted on the potential applications of this developing technology in various software applications~\cite{lowcodeapp} or for automating business process in manufacturing~\cite{waszkowski2019low-automating}, healthcare~\cite{ness2019potential, woo2020rise}, Digital transformation~\cite{phalake2021low}, Industrial engineering education\cite{adrian2020app}, IoT systems using LCSD~\cite{ihirwe2020low}. 
Sipio et al.~\cite{di2020democratizing} present the benefits and future potential of LCSD by sharing their experience of building a custom recommendation system in the LCSD platform. Kourouklidis et al.~\cite{kourouklidis2020towards} discuss
the low-code solution to monitor the machine learning model's performance. Sahay
et al.~\cite{sahay2020supporting} survey LCDP and compare different LCDPs based on
their helpful features and functionalities. Khorram et al.~\cite{lowcodetesting} analyse commercial LCSD platforms and present a
list of features and testing challenges. Zhuang et al.~\cite{zhuang2021easyfl} created a low-code platform called EasyFL where researchers and educators can easily build systems for privacy-preserving distributed learning method. Ihirwe et al.~\cite{lowcodeIot} analyse 16 LCSD platforms and identifies what IoT application-related features and services each platform provides. All of these studies compare a single LCSD platform and its support and limitations for various sorts of applications~\cite{alonso2020towards}, rather than taking a holistic picture of the difficulties that the broader community faces.

There are also some studies where researchers proposed different techniques to improve LCSD platform such as Overeem et al.~\cite{overeem2021proposing} on LCSD platform's impact analysis, Jacinto et al.~\cite{jacinto2020test} improve testing for LCSD platforms.

Additionally, there are some studies that describe the difficulties faced by LCSD practitioners. The main research methodology and objectives of these studies, however, are significantly different from this study. Yajing et al.~\cite{luo2021characteristics} analyse the LCSD platform's characteristics including programming languages used, major implementation units, supporting technologies, applications being developed, domains, etc., along with the benefits, limitations, and challenges by collecting relevant posts from SO and Reddit. In this study, we use tag-based approach to find relevant LCSD-related posts which is much more reliable than text-based searching. Furthermore, the SO related discussion used in this study is significantly larger and our research objective about LCSD platforms challenges are quite different. Timothy et al.~\cite{lethbridge2021low} discuss experiences with several low-code platforms and provide  recommendations focusing on low-code platforms enabling scaling, understandability, documentability, testability, vendor-independence, and the overall user experience for developers as end-users who do some development. Danial et al.~\cite{dahlberg2020developer} and ALSAADI et al.~\cite{alsaadi2021factors} Surveyed on factors hindering the widespread adaptation of LCSD by interviewing LCSD developers or conducting a survey. To the best of our knowledge, ours is the first empirical study of LCSD platforms based on developers' discussions from Stack Overflow, and hence our findings complement those of other studies.

\subsection{Topic Modeling in Software Engineering}\label{sec:rel-se-res-topic}
Our motivation to use topic modeling to understand  LCSD discussions stems from
existing research in software engineering that shows that topics generated from
textual contents can be a good approximation of the underlying
\it{themes}~\cite{Chen-SurveyTopicInSE-EMSE2016,Sun-SoftwareMaintenanceHistoryTopic-CIS2015,Sun-ExploreTopicModelSurvey-SNPD2016}.
Topic models are used recently to understand software
logging~\cite{Li-StudySoftwareLoggingUsingTopic-EMSE2018} and previously for
diverse other tasks, such as concept and feature
location~\cite{Cleary-ConceptLocationTopic-EMSE2009,Poshyvanyk-FeatureLocationTopic-TSE2007},
traceability linking (e.g., bug)~\cite{Rao-TraceabilityBugTopic-MSR2011,asuncion2010software},
to understand software and source code history
evolution~\cite{Hu-EvolutionDynamicTopic-SANER2015,Thomas-SoftwareEvolutionUsingTopic-SCP2014,Thomas-EvolutionSourceCodeHistoryTopic-MSR2011}, to facilitate code search by categorizing
software~\cite{Tian-SoftwareCategorizeTopic-MSR2009}, to refactor software code
base~\cite{Bavota-RefactoringTopic-TSE2014}, as well as to explain software
defect~\cite{Chen-SoftwareDefectTopic-MSR2012}, and various software maintenance
tasks~\cite{Sun-SoftwareMaintenanceTopic-IST2015,Sun-SoftwareMaintenanceHistoryTopic-CIS2015}.
The SO posts are subject to several studies on various aspects
of software development using topic modeling, such as what developers are
discussing in general~\cite{barua2014developers} or about a
particular aspect, e.g., concurrency~\cite{ahmed2018concurrency}, big
data~\cite{bagherzadeh2019going}, chatbot~\cite{abdellatif2020challenges}. 


\begin{table}[ht]
\centering
\caption{Comparing the popularity and difficulty metrics of different domains}
\resizebox{\columnwidth}{!}{%
\begin{tabular}{llrrrrrrr}\toprule
    \textbf{Type} & \textbf{Metrics} & {\textbf{LCSD}} & {\textbf{IoT}} & {\textbf{Big Data}} & {\textbf{Chatbot}} & {\textbf{Security}} & {\textbf{Mobile}}  & \bf{Concurrency}\\
    \midrule
    \multirow{5}[0]{*}{\textbf{P}} & \# Posts &     33,766 &     53,173  &          125,671  &          3,890  &          94,541  &          1,604,483    & 245,541\\
          & Avg View  &   1330.6 &     1,320.3  &           1,560.4  &          512.4  &        2,461.1  &               2,300.0  & 1,641\\
          & Avg Favorite & 1.2 & 1.5   & 1.9   & 1.6   & 3.8   & 2.8 & 0.8\\
          & Avg Score & 0.7 & 0.8   & 1.4   & 0.7   & 2.7   & 2.1 & 2.5 \\
    \midrule
    \multirow{2}[0]{*}{\textbf{D}} & \% W/o Acc. Ans & 41\% & 64\%    & 60.3\%  & 67.7\%  & 48.2\%  & 52\%  & 43.8\%\\
          & Med Hrs to Acc. & 5.7 & 2.9   & 3.3   & 14.8  & 0.9   & 0.7 & 0.7\\
     \bottomrule
    \end{tabular}%
    }
  \label{tab:popDiffCompLCSDvsRest}%
\end{table}%

\begin{table}[ht]
  \centering
  \caption{Comparative analysis of question types across different domains}
    \begin{tabular}{lrrrr}\toprule
    \textbf{Question Type} & \textbf{How} & \textbf{What} & \textbf{Why} & \textbf{Others} \\
    \midrule
    LCSD & 55.6\% & 17.9\% & 14\% & 12.5\% \\
    IoT   & 47.3\%  & 37.9\%  & 20\%    & 8.3\% \\
    Chatbot & 61.8\%  & 11.7\%  & 25.4\%  & 1.2\% \\
    
    \bottomrule
    \end{tabular}%
  \label{tab:qTypeIoTvsRest}%
\end{table}%

In particular, SO posts have been used in various studies where the researchers analysed topics for that particular domains. For instance, SO posts has been used to study developers challenges in IoT \cite{iot21}, big data \cite{bagherzadeh2019going}, chatbots \cite{abdellatif2020challenges} and so on. The distributions and the nature of these posts differs. As SO is arguably the most popular public forum for developers, the analysis of these domains' characteristics may help us identify the SO community better. Therefore, a systematic analysis of these domains is interesting. Following related studies\cite{iot21}, we use six metrics in this study: \begin{inparaenum}[(1)]
        \item Total \#posts,
        \item Avg views,
        \item Avg favorite,
        \item Avg score,
        \item Percentage of questions without accepted answers,
        \item Median hours to get accepted answers per domain.
    \end{inparaenum}
The first four metrics are popularity metrics and the last two are difficulty metrics.

In this study, we do not replicate the findings of the original study in our dataset. Rather we only report the findings from the original study. So following related work~\cite{iot21}, we compared our LCSD-related discussions with other five domains: IoT~\cite{iot21},big data\cite{bagherzadeh2019going}, security~\cite{yang2016security}, mobile apps~\cite{Rosen-MobileDeveloperSO-ESE2015}, chatbots~\cite{abdellatif2020challenges} and concurrency~\cite{ahmed2018concurrency}.

Table~\ref{tab:popDiffCompLCSDvsRest} provides an overview of the seven metrics. We can see that it has a greater number of SO posts than chatbot domains but fewer than the other five domains. There are two other studies on Blockchain~\cite{wan2019discussed} and deep learning~\cite{han2020programmers} where the total number of posts are 32,375 and 25,887, respectively. However, these two studies did not report the other metrics, so they are excluded from the Table. Although the LCSD-related discussion may have fewer posts than these other domains, as discussed in RQ3, this number is increasing rapidly.

We can also observe that the LCSD domain shows similarities with IoT, Concurrency domain in terms of Avg. View count. Security and Mobile domain seem most popular in terms of Avg. Favourite count and LCSD rank lower in this metric. LCSD domains most resemble with IoT domain in terms of Avg. View, Avg. Favourite and Avg. Score. In terms of difficulty metrics percentage of posts without accepted answers, LCSD domain ranks lowest, which is good. However, it takes much longer to get accepted answers( 0.7 hours for Mobile, 2.9 for IoT). Only the chatbot domain requires more time to get accepted answers (i.e., 5.3 hours in LCSD vs 14.8 hours in chatbot).

We further discuss the LCSD, IoT, and Chatbot domains with the distribution of different types of questions in Table~\ref{tab:qTypeIoTvsRest}. We find that LCSD domain is in the middle of IoT and Chatbot domains in terms of How-type (57\%) compared to IoT (47\%) and chatbot (62\%). This signifies that LCSD domain practitioners ask more about implementation-related questions. In the Why-type question percentage of LCSD-related domains is lowest (14\%) compared with IoT (20\%) and chatbot (25\%). This suggests that practitioners in the LCSD domain enquire about relatively modest troubleshooting issues. For What-type, we notice that the IoT domain dominates with 38\% of questions, compared to the LCSD domain's 12\%. Practitioners of LCSD are less inquisitive about domain architecture and technologies compared to IoT domain. As a result of these analyses, we can see that LCSD domain practitioners exhibit several traits that distinguish them from practitioners in other domains, which LCSD vendors and educators should take into account.

\section{Conclusions} \label{sec:conclusion}
Low Code Software Development (LCSD) is a novel paradigm for developing software applications utilizing visual programming with minimum hand-coding. We present an empirical study that provides insights into the types of discussions low-code developers discuss in Stack Overflow (SO). We find 40 low-code topics in our dataset of 33.7K SO posts (question + accepted answers). We \rev{collected} these posts based on 64 SO tags belonging to the popular 38  LCSD platforms. We categorize them into five high-level groups, namely Application Customization (30\% Questions, 11 Topics),  Data Storage (25\% Questions, 9 Topics), Platform Adoption (20\% Questions, 9 Topics),  Platform Maintenance (14\% Questions, 6 Topics), and Third-Party Integration (12\% Questions, 5 Topics). We find that the Platform Adoption \rev{topic category} has gained popularity recently. Platform Related Query and Message Queue \rev{topics} from this category are the most popular. On the other hand, We also find that practitioners find Platform Adoption and Maintenance related \rev{topics} most challenging. How-type questions are the most common, but our research reveals that practitioners find what-type and why-type questions more difficult. Despite extensive support for testing, deployment, and maintenance, our analysis shows that server configuration, data migration, and system module upgrading-related queries are widespread and complex to LCSD practitioners. Despite significant testing, deployment, and maintenance support, our analysis finds numerous and complex queries regarding server configuration, data migration, and system module updating. Our analysis finds that better tutorial-based documentation can help solve many of these problems. We also find that during the Covid-19 pandemic, LCSD platforms were very popular with developers, especially when it came to dynamic form-based applications. We hope that all of these findings will help various  LCSD stakeholders (e.g., LCSD platform vendors, practitioners, SE researchers) to take necessary actions to address the various  LCSD challenges. Since the growth indicates that this technology is likely to be widely adopted by various companies for their internal and customer-facing applications, platform providers should address the prevailing developers' challenges. Our future work will focus on \begin{inparaenum}[(1)]
\item getting developers' feedback on our study findings based on surveys and developer interviews, and 
\item developing tools and techniques to automatically address the challenges in the LCSD platforms that we observed.
\end{inparaenum}



\begin{small}
\balance
\bibliographystyle{abbrv}
\bibliography{bibliography}

\begin{thebibliography}{100}

\bibitem{abdellatif2020challenges}
A.~Abdellatif, D.~Costa, K.~Badran, R.~Abdalkareem, and E.~Shihab.
\newblock Challenges in chatbot development: A study of stack overflow posts.
\newblock In {\em Proceedings of the 17th International Conference on Mining
  Software Repositories}, MSR '20, page 174–185, New York, NY, USA, 2020.
  Association for Computing Machinery.

\bibitem{adrian2020app}
B.~Adrian, S.~Hinrichsen, and A.~Nikolenko.
\newblock App development via low-code programming as part of modern industrial
  engineering education.
\newblock In {\em International Conference on Applied Human Factors and
  Ergonomics}, pages 45--51. Springer, 2020.

\bibitem{agrawal2018wrong}
A.~Agrawal, W.~Fu, and T.~Menzies.
\newblock What is wrong with topic modeling? and how to fix it using
  search-based software engineering.
\newblock {\em Information and Software Technology}, 98:74--88, 2018.

\bibitem{ahmed2018concurrency}
S.~Ahmed and M.~Bagherzadeh.
\newblock What do concurrency developers ask about? a large-scale study using
  stack overflow.
\newblock In {\em Proceedings of the 12th ACM/IEEE International Symposium on
  Empirical Software Engineering and Measurement}, ESEM '18, New York, NY, USA,
  2018. Association for Computing Machinery.

\bibitem{akiki2020eud}
P.~A. Akiki, P.~A. Akiki, A.~K. Bandara, and Y.~Yu.
\newblock Eud-mars: End-user development of model-driven adaptive robotics
  software systems.
\newblock {\em Science of Computer Programming}, 200:102534, 2020.

\bibitem{alamin2021empirical}
M.~A.~A. Alamin, S.~Malakar, G.~Uddin, S.~Afroz, T.~B. Haider, and A.~Iqbal.
\newblock An empirical study of developer discussions on low-code software
  development challenges.
\newblock In {\em 2021 IEEE/ACM 18th International Conference on Mining
  Software Repositories (MSR)}, pages 46--57. IEEE, 2021.

\bibitem{alonso2020towards}
A.~N. Alonso, J.~Abreu, D.~Nunes, A.~Vieira, L.~Santos, T.~Soares, and
  J.~Pereira.
\newblock Towards a polyglot data access layer for a low-code application
  development platform.
\newblock {\em arXiv preprint arXiv:2004.13495}, 2020.

\bibitem{alsaadi2021factors}
H.~A. ALSAADI, D.~T. RADAIN, M.~M. ALZAHRANI, W.~F. ALSHAMMARI, D.~ALAHMADI,
  and B.~FAKIEH.
\newblock Factors that affect the utilization of low-code development
  platforms: survey study.
\newblock {\em Romanian Journal of Information Technology and Automatic
  Control}, 31(3):123--140, 2021.

\bibitem{amplifystudio}
{AWS Amplify Studio overview}.
\newblock {Available: \url{https://aws.amazon.com/amplify/studio/}}.
\newblock [Online; accessed 5-January-2022].

\bibitem{oracle_apex}
{Oracle Apex Platform}.
\newblock {Available: \url{https://apex.oracle.com/}}.
\newblock [Online; accessed 5-January-2022].

\bibitem{appengine}
{App Engine: a fully managed, serverless platform for developing and hosting
  web applications at scale.}
\newblock {Available: \url{https://cloud.google.com/appengine/docs}}.
\newblock [Online; accessed 13-December-2021].

\bibitem{appian}
{Appian platform overview}.
\newblock {Available: \url{https://www.appian.com/}}.
\newblock [Online; accessed 5-January-2022].

\bibitem{arun2010finding}
R.~Arun, V.~Suresh, C.~V. Madhavan, and M.~N. Murthy.
\newblock On finding the natural number of topics with latent dirichlet
  allocation: Some observations.
\newblock In {\em Pacific-Asia conference on knowledge discovery and data
  mining}, pages 391--402. Springer, 2010.

\bibitem{asaduzzaman2013answering}
M.~Asaduzzaman, A.~S. Mashiyat, C.~K. Roy, and K.~A. Schneider.
\newblock Answering questions about unanswered questions of stack overflow.
\newblock In {\em 2013 10th Working Conference on Mining Software Repositories
  (MSR)}, pages 97--100. IEEE, 2013.

\bibitem{asuncion2010software}
H.~U. Asuncion, A.~U. Asuncion, and R.~N. Taylor.
\newblock Software traceability with topic modeling.
\newblock In {\em 2010 ACM/IEEE 32nd International Conference on Software
  Engineering}, volume~1, pages 95--104. IEEE, 2010.

\bibitem{bagherzadeh2019going}
M.~Bagherzadeh and R.~Khatchadourian.
\newblock Going big: A large-scale study on what big data developers ask.
\newblock In {\em Proceedings of the 2019 27th ACM Joint Meeting on European
  Software Engineering Conference and Symposium on the Foundations of Software
  Engineering}, ESEC/FSE 2019, pages 432--442, New York, NY, USA, 2019. ACM.

\bibitem{bajaj2014mining}
K.~Bajaj, K.~Pattabiraman, and A.~Mesbah.
\newblock Mining questions asked by web developers.
\newblock In {\em Proceedings of the 11th Working Conference on Mining Software
  Repositories}, pages 112--121, 2014.

\bibitem{bandeira2019we}
A.~Bandeira, C.~A. Medeiros, M.~Paixao, and P.~H. Maia.
\newblock We need to talk about microservices: an analysis from the discussions
  on stackoverflow.
\newblock In {\em 2019 IEEE/ACM 16th International Conference on Mining
  Software Repositories (MSR)}, pages 255--259. IEEE, 2019.

\bibitem{barua2014developers}
A.~Barua, S.~W. Thomas, and A.~E. Hassan.
\newblock What are developers talking about? an analysis of topics and trends
  in stack overflow.
\newblock {\em Empirical Software Engineering}, 19(3):619--654, 2014.

\bibitem{basciani2014mdeforge}
F.~Basciani, L.~Iovino, A.~Pierantonio, et~al.
\newblock Mdeforge: an extensible web-based modeling platform.
\newblock In {\em 2nd International Workshop on Model-Driven Engineering on and
  for the Cloud, CloudMDE 2014, Co-located with the 17th International
  Conference on Model Driven Engineering Languages and Systems, MoDELS 2014},
  volume 1242, pages 66--75. CEUR-WS, 2014.

\bibitem{basil1975iterative}
V.~R. Basil and A.~J. Turner.
\newblock Iterative enhancement: A practical technique for software
  development.
\newblock {\em IEEE Transactions on Software Engineering}, (4):390--396, 1975.

\bibitem{Bavota-RefactoringTopic-TSE2014}
G.~Bavota, R.~Oliveto, M.~Gethers, D.~Poshyvanyk, and A.~D. Lucia.
\newblock Methodbook: Recommending move method refactorings via relational
  topic models.
\newblock {\em IEEE Transactions on Software Engineering}, 40(7):671--694,
  2014.

\bibitem{bayer2006view}
J.~Bayer and D.~Muthig.
\newblock A view-based approach for improving software documentation practices.
\newblock In {\em 13th Annual IEEE International Symposium and Workshop on
  Engineering of Computer-Based Systems (ECBS'06)}, pages 10--pp. IEEE, 2006.

\bibitem{beck2001manifesto}
K.~Beck, M.~Beedle, A.~Van~Bennekum, A.~Cockburn, W.~Cunningham, M.~Fowler,
  J.~Grenning, J.~Highsmith, A.~Hunt, R.~Jeffries, et~al.
\newblock Manifesto for agile software development.
\newblock 2001.

\bibitem{beynon1999rapid}
P.~Beynon-Davies, C.~Carne, H.~Mackay, and D.~Tudhope.
\newblock Rapid application development (rad): an empirical review.
\newblock {\em European Journal of Information Systems}, 8(3):211--223, 1999.

\bibitem{bhat2006overcoming}
J.~M. Bhat, M.~Gupta, and S.~N. Murthy.
\newblock Overcoming requirements engineering challenges: Lessons from offshore
  outsourcing.
\newblock {\em IEEE software}, 23(5):38--44, 2006.

\bibitem{blei2003latent}
D.~M. Blei, A.~Y. Ng, and M.~I. Jordan.
\newblock Latent dirichlet allocation.
\newblock {\em Journal of Machine Learning Research}, 3(4-5):993--1022, 2003.

\bibitem{botterweck2006model}
G.~Botterweck.
\newblock A model-driven approach to the engineering of multiple user
  interfaces.
\newblock In {\em International Conference on Model Driven Engineering
  Languages and Systems}, pages 106--115. Springer, 2006.

\bibitem{brambilla2017modelmdse}
M.~Brambilla, J.~Cabot, and M.~Wimmer.
\newblock Model-driven software engineering in practice.
\newblock {\em Synthesis lectures on software engineering}, 3(1):1--207, 2017.

\bibitem{brambilla2017model}
M.~Brambilla, E.~Umuhoza, and R.~Acerbis.
\newblock Model-driven development of user interfaces for iot systems via
  domain-specific components and patterns.
\newblock {\em Journal of Internet Services and Applications}, 8(1):1--21,
  2017.

\bibitem{burnett1995visual}
M.~M. Burnett and D.~W. McIntyre.
\newblock Visual programming.
\newblock {\em COMPUTER-LOS ALAMITOS-}, 28:14--14, 1995.

\bibitem{Chakraborty-NewLangSupportSO-IST2021}
P.~Chakraborty, R.~Shahriyar, A.~Iqbal, and G.~Uddin.
\newblock How do developers discuss and support new programming languages in
  technical q\&a site? an empirical study of go, swift, and rust in stack
  overflow.
\newblock {\em Information and Software Technology (IST)}, page~19, 2021.

\bibitem{Chen-SoftwareDefectTopic-MSR2012}
T.-H. Chen, S.~W. Thomas, M.~Nagappan, and A.~E. Hassan.
\newblock Explaining software defects using topic models.
\newblock In {\em 9th working conference on mining software repositories},
  pages 189--198, 2012.

\bibitem{Chen-SurveyTopicInSE-EMSE2016}
T.-H.~P. Chen, S.~W. Thomas, and A.~E. Hassan.
\newblock A survey on the use of topic models when mining software
  repositories.
\newblock {\em Empirical Software Engineering}, 21(5):1843--1919, 2016.

\bibitem{Cleary-ConceptLocationTopic-EMSE2009}
B.~Cleary, C.~Exton, J.~Buckley, and M.~English.
\newblock An empirical analysis of information retrieval based concept location
  techniques in software comprehension.
\newblock {\em Empirical Software Engineering}, 14:93--130, 2009.

\bibitem{costabile2007visual}
M.~F. Costabile, D.~Fogli, P.~Mussio, and A.~Piccinno.
\newblock Visual interactive systems for end-user development: a model-based
  design methodology.
\newblock {\em IEEE transactions on systems, man, and cybernetics-part a:
  systems and humans}, 37(6):1029--1046, 2007.

\bibitem{dahlberg2020developer}
D.~Dahlberg.
\newblock Developer experience of a low-code platform: An exploratory study,
  2020.

\bibitem{de2014labeling}
A.~De~Lucia, M.~Di~Penta, R.~Oliveto, A.~Panichella, and S.~Panichella.
\newblock Labeling source code with information retrieval methods: an empirical
  study.
\newblock {\em Empirical Software Engineering}, 19(5):1383--1420, 2014.

\bibitem{di2020democratizing}
C.~Di~Sipio, D.~Di~Ruscio, and P.~T. Nguyen.
\newblock Democratizing the development of recommender systems by means of
  low-code platforms.
\newblock In {\em Proceedings of the 23rd ACM/IEEE International Conference on
  Model Driven Engineering Languages and Systems: Companion Proceedings}, pages
  1--9, 2020.

\bibitem{SOdump}
S.~Exchange.
\newblock { Stack exchange data dump }.
\newblock {Available: \url{https://archive.org/details/stackexchange}}, 2020.
\newblock [Online; accessed 5-January-2022].

\bibitem{fincher2005making}
S.~Fincher and J.~Tenenberg.
\newblock Making sense of card sorting data.
\newblock {\em Expert Systems}, 22(3):89--93, 2005.

\bibitem{fischer2004meta}
G.~Fischer, E.~Giaccardi, Y.~Ye, A.~G. Sutcliffe, and N.~Mehandjiev.
\newblock Meta-design: a manifesto for end-user development.
\newblock {\em Communications of the ACM}, 47(9):33--37, 2004.

\bibitem{fors2016design}
N.~Fors.
\newblock {\em The Design and Implementation of Bloqqi-A Feature-Based Diagram
  Programming Language}.
\newblock PhD thesis, Lund University, 2016.

\bibitem{lowcodeapp}
M.~Fryling.
\newblock Low code app development.
\newblock {\em J. Comput. Sci. Coll.}, 34(6):119, Apr. 2019.

\bibitem{gartner}
{Enterprise Low-Code Application Platforms (LCAP) Reviews and Ratings}.
\newblock {Available:
  \url{https://www.gartner.com/reviews/market/enterprise-low-code-application-platform}}.
\newblock [Online; accessed 5-January-2022].

\bibitem{googleAppSheet}
{AppSheet, Low-code application development}.
\newblock {Available: \url{https://www.appsheet.com}}.
\newblock [Online; accessed 13-December-2021].

\bibitem{googleappmaker}
{Google App Maker platform overview}.
\newblock {Available: \url{https://developers.google.com/appmaker}}.
\newblock [Online; accessed 5-January-2022].

\bibitem{google-disc}
{Google App Maker will be shut down on January 19, 2021}.
\newblock
  \url{https://workspaceupdates.googleblog.com/2020/01/app-maker-update.html}.
\newblock [Online; accessed 5-January-2022].

\bibitem{halbert1984programming}
D.~C. Halbert.
\newblock {\em Programming by example}.
\newblock PhD thesis, University of California, Berkeley, 1984.

\bibitem{han2020programmers}
J.~Han, E.~Shihab, Z.~Wan, S.~Deng, and X.~Xia.
\newblock What do programmers discuss about deep learning frameworks.
\newblock {\em Empirical Software Engineering}, 25(4):2694--2747, 2020.

\bibitem{honeycode}
{Amazon Honeycode platform overview}.
\newblock {Available: \url{https://www.honeycode.aws/}}.
\newblock [Online; accessed 5-January-2022].

\bibitem{Hu-EvolutionDynamicTopic-SANER2015}
J.~Hu, X.~Sun, D.~Lo, and B.~Li.
\newblock Modeling the evolution of development topics using dynamic topic
  models.
\newblock In {\em IEEE 22nd International Conference on Software Analysis,
  Evolution, and Reengineering}, pages 3--12, 2015.

\bibitem{ihirwe2020low}
F.~Ihirwe, D.~Di~Ruscio, S.~Mazzini, P.~Pierini, and A.~Pierantonio.
\newblock Low-code engineering for internet of things: A state of research.
\newblock In {\em Proceedings of the 23rd ACM/IEEE International Conference on
  Model Driven Engineering Languages and Systems: Companion Proceedings}, pages
  1--8, 2020.

\bibitem{lowcodeIot}
F.~Ihirwe, D.~Di~Ruscio, S.~Mazzini, P.~Pierini, and A.~Pierantonio.
\newblock Low-code engineering for internet of things: A state of research.
\newblock In {\em Proceedings of the 23rd ACM/IEEE International Conference on
  Model Driven Engineering Languages and Systems: Companion Proceedings},
  MODELS '20, New York, NY, USA, 2020. Association for Computing Machinery.

\bibitem{jacinto2020test}
A.~Jacinto, M.~Louren{\c{c}}o, and C.~Ferreira.
\newblock Test mocks for low-code applications built with outsystems.
\newblock In {\em Proceedings of the 23rd ACM/IEEE International Conference on
  Model Driven Engineering Languages and Systems: Companion Proceedings}, pages
  1--5, 2020.

\bibitem{Kendall-TauMetric-Biometrica1938}
M.~G. Kendall.
\newblock A new measure of rank correlation.
\newblock {\em Biometrika}, 30(1):81--93, 1938.

\bibitem{khan2021automatic}
J.~Y. Khan, M.~T.~I. Khondaker, G.~Uddin, and A.~Iqbal.
\newblock Automatic detection of five api documentation smells:
  Practitioners’ perspectives.
\newblock In {\em 2021 IEEE International Conference on Software Analysis,
  Evolution and Reengineering (SANER)}, pages 318--329. IEEE, 2021.

\bibitem{Khan-DocSmell-SANER2021}
J.~Y. Khan, M.~T.~I. Khondaker, G.~Uddin, and A.~Iqbal.
\newblock Automatic detection of five api documentation smells:
  Practitioners’ perspectives.
\newblock In {\em IEEE International Conference on Software Analysis, Evolution
  and Reengineering (SANER)}, page~12, 2021.

\bibitem{lowcodetesting}
F.~Khorram, J.-M. Mottu, and G.~Suny\'{e}.
\newblock Challenges \& opportunities in low-code testing.
\newblock In {\em Proceedings of the 23rd ACM/IEEE International Conference on
  Model Driven Engineering Languages and Systems: Companion Proceedings},
  MODELS '20, New York, NY, USA, 2020. Association for Computing Machinery.

\bibitem{kourouklidis2020towards}
P.~Kourouklidis, D.~Kolovos, N.~Matragkas, and J.~Noppen.
\newblock Towards a low-code solution for monitoring machine learning model
  performance.
\newblock In {\em Proceedings of the 23rd ACM/IEEE International Conference on
  Model Driven Engineering Languages and Systems: Companion Proceedings}, pages
  1--8, 2020.

\bibitem{kruskal1957historical}
W.~H. Kruskal.
\newblock Historical notes on the wilcoxon unpaired two-sample test.
\newblock {\em Journal of the American Statistical Association},
  52(279):356--360, 1957.

\bibitem{lethbridge2021low}
T.~C. Lethbridge.
\newblock Low-code is often high-code, so we must design low-code platforms to
  enable proper software engineering.
\newblock In {\em International Symposium on Leveraging Applications of Formal
  Methods}, pages 202--212. Springer, 2021.

\bibitem{Li-StudySoftwareLoggingUsingTopic-EMSE2018}
H.~Li, T.-H.~P. Chen, W.~Shang, and A.~E. Hassan.
\newblock Studying software logging using topic models.
\newblock {\em Empirical Software Engineering}, 23:2655–2694, 2018.

\bibitem{lin2020software}
G.~Lin, S.~Wen, Q.-L. Han, J.~Zhang, and Y.~Xiang.
\newblock Software vulnerability detection using deep neural networks: a
  survey.
\newblock {\em Proceedings of the IEEE}, 108(10):1825--1848, 2020.

\bibitem{linares2013exploratory}
M.~Linares-V{\'a}squez, B.~Dit, and D.~Poshyvanyk.
\newblock An exploratory analysis of mobile development issues using stack
  overflow.
\newblock In {\em 2013 10th Working Conference on Mining Software Repositories
  (MSR)}, pages 93--96. IEEE, 2013.

\bibitem{loper2002nltk}
E.~Loper and S.~Bird.
\newblock Nltk: the natural language toolkit.
\newblock {\em arXiv preprint cs/0205028}, 2002.

\bibitem{lotus}
{IBM Lotus software}.
\newblock {Available: \url{https://help.hcltechsw.com/}}.
\newblock [Online; accessed 5-January-2022].

\bibitem{lowcodewiki}
{Low-code development platform }.
\newblock {Available:
  \url{https://en.wikipedia.org/wiki/Low-code_development_platform}}.
\newblock [Online; accessed 5-January-2022].

\bibitem{luo2021characteristics}
Y.~Luo, P.~Liang, C.~Wang, M.~Shahin, and J.~Zhan.
\newblock Characteristics and challenges of low-code development: The
  practitioners' perspective.
\newblock In {\em Proceedings of the 15th ACM/IEEE International Symposium on
  Empirical Software Engineering and Measurement (ESEM)}, pages 1--11, 2021.

\bibitem{mccallum2002mallet}
A.~K. McCallum.
\newblock Mallet: A machine learning for language toolkit.
\newblock {\em http://mallet. cs. umass. edu}, 2002.

\bibitem{mchugh2012interrater}
M.~L. McHugh.
\newblock Interrater reliability: the kappa statistic.
\newblock {\em Biochemia medica}, 22(3):276--282, 2012.

\bibitem{mendix}
{Mendix platform overview}.
\newblock {Available: \url{https://www.mendix.com/}}.
\newblock [Online; accessed 5-January-2022].

\bibitem{mernik2005and}
M.~Mernik, J.~Heering, and A.~M. Sloane.
\newblock When and how to develop domain-specific languages.
\newblock {\em ACM computing surveys (CSUR)}, 37(4):316--344, 2005.

\bibitem{microsoftpowerfx}
{Microsoft Power FX}.
\newblock {Available:
  \url{https://docs.microsoft.com/en-us/power-platform/power-fx/overview}}.
\newblock [Online; accessed 13-December-2021].

\bibitem{myers2006invited}
B.~A. Myers, A.~J. Ko, and M.~M. Burnett.
\newblock Invited research overview: end-user programming.
\newblock In {\em CHI'06 extended abstracts on Human factors in computing
  systems}, pages 75--80, 2006.

\bibitem{ness2019potential}
C.~Ness and M.~E. Hansen.
\newblock Potential of low-code in the healthcare sector: an exploratory study
  of the potential of low-code development in the healthcare sector in norway.
\newblock Master's thesis, 2019.

\bibitem{oneblink}
{OneBlink platform overview}.
\newblock {Available: \url{https://www.oneblink.io/}}.
\newblock [Online; accessed 5-January-2022].

\bibitem{overeem2021proposing}
M.~Overeem and S.~Jansen.
\newblock Proposing a framework for impact analysis for low-code development
  platforms.
\newblock In {\em 2021 ACM/IEEE International Conference on Model Driven
  Engineering Languages and Systems Companion (MODELS-C)}, pages 88--97. IEEE,
  2021.

\bibitem{website:stackoverflow}
S.~Overflow.
\newblock {\em Stack Overflow Questions}.
\newblock \url{https://stackoverflow.com/questions/}, 2020.
\newblock Last accessed on 14 November 2020.

\bibitem{pandemic-low-code}
{Programming Gains Speed As Developers Turn to Low-Code During the Pandemic}.
\newblock {Available:
  \url{https://www.designnews.com/automation/programming-gains-speed-developers-turn-low-code-during-pandemic}}.
\newblock [Online; accessed 5-August-2022].

\bibitem{Pane-MoreNatureEUSE-Springer2006}
J.~Pane and B.~Myers.
\newblock {\em More Natural Programming Languages and Environments}, pages
  31--50.
\newblock Springer, 10 2006.

\bibitem{paterno2013end}
F.~Patern{\`o}.
\newblock End user development: Survey of an emerging field for empowering
  people.
\newblock {\em International Scholarly Research Notices}, 2013, 2013.

\bibitem{pcmag}
{The Best Low-Code Development Platforms}.
\newblock {Available:
  \url{https://www.pcmag.com/picks/the-best-low-code-development-platforms}}.
\newblock [Online; accessed 5-January-2022].

\bibitem{phalake2021low}
V.~S. Phalake and S.~D. Joshi.
\newblock Low code development platform for digital transformation.
\newblock In {\em Information and Communication Technology for Competitive
  Strategies (ICTCS 2020)}, pages 689--697. Springer, 2021.

\bibitem{pleuss2013model}
A.~Pleuss, S.~Wollny, and G.~Botterweck.
\newblock Model-driven development and evolution of customized user interfaces.
\newblock In {\em Proceedings of the 5th ACM SIGCHI symposium on Engineering
  interactive computing systems}, pages 13--22, 2013.

\bibitem{ponzanelli2014improving}
L.~Ponzanelli, A.~Mocci, A.~Bacchelli, M.~Lanza, and D.~Fullerton.
\newblock Improving low quality stack overflow post detection.
\newblock In {\em 2014 IEEE international conference on software maintenance
  and evolution}, pages 541--544. IEEE, 2014.

\bibitem{Poshyvanyk-FeatureLocationTopic-TSE2007}
D.~Poshyvanyk, Y.-G. Guéhéneuc, A.~Marcus, G.~Antoniol, and V.~T. Rajlich.
\newblock Feature location using probabilistic ranking of methods based on
  execution scenarios and information retrieval.
\newblock {\em IEEE Transactions on Software Engineering}, 33(6):420--432,
  2007.

\bibitem{powerapps}
{Microsoft power apps platform overview}.
\newblock {Available: \url{https://powerapps.microsoft.com/en-us/}}.
\newblock [Online; accessed 5-January-2022].

\bibitem{quickbase}
{Quickbase platform overview}.
\newblock {Available:
  \url{https://www.quickbase.com/product/product-overview}}.
\newblock [Online; accessed 5-January-2022].

\bibitem{ramasubramanian2013effective}
C.~Ramasubramanian and R.~Ramya.
\newblock Effective pre-processing activities in text mining using improved
  porter’s stemming algorithm.
\newblock {\em International Journal of Advanced Research in Computer and
  Communication Engineering}, 2(12):4536--4538, 2013.

\bibitem{Rao-TraceabilityBugTopic-MSR2011}
S.~Rao and A.~C. Kak.
\newblock Retrieval from software libraries for bug localization: a comparative
  study of generic and composite text models.
\newblock In {\em 8th Working Conference on Mining Software Repositories}, page
  43–52, 2011.

\bibitem{rehurek2010software}
R.~Rehurek and P.~Sojka.
\newblock Software framework for topic modelling with large corpora.
\newblock In {\em In Proceedings of the LREC 2010 Workshop on New Challenges
  for NLP Frameworks}. Citeseer, 2010.

\bibitem{ren2019discovering}
X.~Ren, Z.~Xing, X.~Xia, G.~Li, and J.~Sun.
\newblock Discovering, explaining and summarizing controversial discussions in
  community q\&a sites.
\newblock In {\em 2019 34th IEEE/ACM International Conference on Automated
  Software Engineering (ASE)}, pages 151--162. IEEE, 2019.

\bibitem{resnick2009scratch}
M.~Resnick, J.~Maloney, A.~Monroy-Hern{\'a}ndez, N.~Rusk, E.~Eastmond,
  K.~Brennan, A.~Millner, E.~Rosenbaum, J.~Silver, B.~Silverman, et~al.
\newblock Scratch: programming for all.
\newblock {\em Communications of the ACM}, 52(11):60--67, 2009.

\bibitem{Robillard-APIProperty-IEEETSE2012}
M.~P. Robillard, E.~Bodden, D.~Kawrykow, M.~Mezini, and T.~Ratchford.
\newblock Automated {API} property inference techniques.
\newblock {\em IEEE Transactions on Software Engineering}, page~28, 2012.

\bibitem{roder2015exploring}
M.~R{\"o}der, A.~Both, and A.~Hinneburg.
\newblock Exploring the space of topic coherence measures.
\newblock In {\em Proceedings of the eighth ACM international conference on Web
  search and data mining}, pages 399--408, 2015.

\bibitem{Rosen-MobileDeveloperSO-ESE2015}
C.~Rosen and E.~Shihab.
\newblock What are mobile developers asking about? a large scale study using
  stack overflow.
\newblock {\em Empirical Software Engineering}, page~33, 2015.

\bibitem{rosen2016mobile}
C.~Rosen and E.~Shihab.
\newblock What are mobile developers asking about? a large scale study using
  stack overflow.
\newblock {\em Empirical Software Engineering}, 21(3):1192--1223, 2016.

\bibitem{rymer2019forrester}
J.~R. Rymer, R.~Koplowitz, and S.~A. Leaders.
\newblock The forrester wave(tm) low-code development platforms for ad\&d
  professionals, q1 2019.
\newblock 2019.

\bibitem{sahay2020supporting}
A.~Sahay, A.~Indamutsa, D.~Di~Ruscio, and A.~Pierantonio.
\newblock Supporting the understanding and comparison of low-code development
  platforms.
\newblock In {\em 2020 46th Euromicro Conference on Software Engineering and
  Advanced Applications (SEAA)}, pages 171--178. IEEE, 2020.

\bibitem{salesforce}
{Salesforce platform overview}.
\newblock {Available: \url{https://www.salesforce.com/in/?ir=1}}.
\newblock [Online; accessed 5-January-2022].

\bibitem{sinha2010human}
G.~Sinha, R.~Shahi, and M.~Shankar.
\newblock Human computer interaction.
\newblock In {\em 2010 3rd International Conference on Emerging Trends in
  Engineering and Technology}, pages 1--4. IEEE, 2010.

\bibitem{Sun-SoftwareMaintenanceTopic-IST2015}
X.~Sun, B.~Li, H.~Leung, B.~Li, and Y.~Li.
\newblock Msr4sm: Using topic models to effectively mining software
  repositories for software maintenance tasks.
\newblock {\em Information and Software Technology}, 66:671--694, 2015.

\bibitem{Sun-SoftwareMaintenanceHistoryTopic-CIS2015}
X.~Sun, B.~Li, Y.~Li, and Y.~Chen.
\newblock What information in software historical repositories do we need to
  support software maintenance tasks? an approach based on topic model.
\newblock {\em Computer and Information Science}, pages 22--37, 2015.

\bibitem{Sun-ExploreTopicModelSurvey-SNPD2016}
X.~Sun, X.~Liu, B.~Li, Y.~Duan, H.~Yang, and J.~Hu.
\newblock Exploring topic models in software engineering data analysis: A
  survey.
\newblock In {\em 17th IEEE/ACIS International Conference on Software
  Engineering, Artificial Intelligence, Networking and Parallel/Distributed
  Computing}, pages 357--362, 2016.

\bibitem{Thomas-EvolutionSourceCodeHistoryTopic-MSR2011}
S.~W. Thomas, B.~Adams, A.~E. Hassan, and D.~Blostein.
\newblock Modeling the evolution of topics in source code histories.
\newblock In {\em 8th working conference on mining software repositories},
  pages 173--182, 2011.

\bibitem{Thomas-SoftwareEvolutionUsingTopic-SCP2014}
S.~W. Thomas, B.~Adams, A.~E. Hassan, and D.~Blostein.
\newblock Studying software evolution using topic models.
\newblock {\em Science of Computer Programming}, 80(B):457--479, 2014.

\bibitem{Tian-SoftwareCategorizeTopic-MSR2009}
K.~Tian, M.~Revelle, and D.~Poshyvanyk.
\newblock Using latent dirichlet allocation for automatic categorization of
  software.
\newblock In {\em 6th international working conference on mining software
  repositories}, pages 163--166, 2009.

\bibitem{torres2018demand}
C.~Torres.
\newblock Demand for programmers hits full boil as us job market simmers.
\newblock {\em Bloomberg. Com}, 2018.

\bibitem{total_low_code}
{How many Low-Code/No-Code platforms are out there?}
\newblock {Available:
  \url{https://www.spreadsheetweb.com/how-many-low-code-no-code-platforms-are-out-there/}}.
\newblock [Online; accessed 5-August-2022].

\bibitem{treude2011programmers}
C.~Treude, O.~Barzilay, and M.-A. Storey.
\newblock How do programmers ask and answer questions on the web?(nier track).
\newblock In {\em Proceedings of the 33rd international conference on software
  engineering}, pages 804--807, 2011.

\bibitem{Uddin-OpinionSurvey-TSE2019}
G.~Uddin, O.~Baysal, L.~Guerroj, and F.~Khomh.
\newblock Understanding how and why developers seek and analyze api related
  opinions.
\newblock {\em IEEE Transactions on Software Engineering}, page~40, 2019.

\bibitem{uddin2017automatic}
G.~Uddin and F.~Khomh.
\newblock Automatic summarization of api reviews.
\newblock In {\em 2017 32nd IEEE/ACM International Conference on Automated
  Software Engineering (ASE)}, pages 159--170. IEEE, 2017.

\bibitem{Uddin-OpinerReviewAlgo-ASE2017}
G.~Uddin and F.~Khomh.
\newblock Automatic summarization of {API} reviews.
\newblock In {\em Proc. 32nd IEEE/ACM International Conference on Automated
  Software Engineering}, page~12, 2017.

\bibitem{Uddin-OpinerReviewToolDemo-ASE2017}
G.~Uddin and F.~Khomh.
\newblock Opiner: A search and summarization engine for {API} reviews.
\newblock In {\em Proc. 32nd IEEE/ACM International Conference on Automated
  Software Engineering}, page~6, 2017.

\bibitem{Uddin-OpinionValue-TSE2019}
G.~Uddin and F.~Khomh.
\newblock Automatic opinion mining from {API} reviews from stack overflow.
\newblock {\em IEEE Transactions on Software Engineering}, page~35, 2019.

\bibitem{Uddin-OpinerAPIUsageScenario-TOSEM2020}
G.~Uddin, F.~Khomh, and C.~K. Roy.
\newblock Automatic api usage scenario documentation from technical q\&a sites.
\newblock {\em ACM Transactions on Software Engineering and Methodology},
  page~43, 2020.

\bibitem{Uddin-OpinerAPIUsageScenario-IST2020}
G.~Uddin, F.~Khomh, and C.~K. Roy.
\newblock Automatic mining of api usage scenarios from stack overflow.
\newblock {\em Information and Software Technology (IST)}, page~16, 2020.

\bibitem{uddin2021automatic}
G.~Uddin, F.~Khomh, and C.~K. Roy.
\newblock Automatic api usage scenario documentation from technical q\&a sites.
\newblock {\em ACM Transactions on Software Engineering and Methodology
  (TOSEM)}, 30(3):1--45, 2021.

\bibitem{uddin2015api}
G.~Uddin and M.~P. Robillard.
\newblock How api documentation fails.
\newblock {\em Ieee software}, 32(4):68--75, 2015.

\bibitem{Uddin-HowAPIDocumentationFails-IEEESW2015}
G.~Uddin and M.~P. Robillard.
\newblock How api documentation fails.
\newblock {\em IEEE Softawre}, 32(4):76--83, 2015.

\bibitem{iot21}
G.~Uddin, F.~Sabir, Y.-G. Guéhéneuc, O.~Alam, and F.~Khomh.
\newblock An empirical study of iot topics in iot developer discussions on
  stack overflow.
\newblock {\em Empirical Software Engineering}, 26, 11 2021.

\bibitem{ur2014practical}
B.~Ur, E.~McManus, M.~Pak Yong~Ho, and M.~L. Littman.
\newblock Practical trigger-action programming in the smart home.
\newblock In {\em Proceedings of the SIGCHI Conference on Human Factors in
  Computing Systems}, pages 803--812, 2014.

\bibitem{van2000domain}
A.~Van~Deursen, P.~Klint, and J.~Visser.
\newblock Domain-specific languages: An annotated bibliography.
\newblock {\em ACM Sigplan Notices}, 35(6):26--36, 2000.

\bibitem{vincent2019magic}
P.~Vincent, K.~Lijima, M.~Driver, J.~Wong, and Y.~Natis.
\newblock Magic quadrant for enterprise low-code application platforms.
\newblock {\em Retrieved December}, 18:2019, 2019.

\bibitem{vinyl}
{Vinyl platform overview}.
\newblock {Available: \url{https://zudy.com/}}.
\newblock [Online; accessed 5-January-2022].

\bibitem{wan2019discussed}
Z.~Wan, X.~Xia, and A.~E. Hassan.
\newblock What is discussed about blockchain? a case study on the use of
  balanced lda and the reference architecture of a domain to capture online
  discussions about blockchain platforms across the stack exchange communities.
\newblock {\em IEEE Transactions on Software Engineering}, 2019.

\bibitem{waszkowski2019low-automating}
R.~Waszkowski.
\newblock Low-code platform for automating business processes in manufacturing.
\newblock {\em IFAC-PapersOnLine}, 52:376--381, 01 2019.

\bibitem{wolber2011app}
D.~Wolber.
\newblock App inventor and real-world motivation.
\newblock In {\em Proceedings of the 42nd ACM technical symposium on Computer
  science education}, pages 601--606, 2011.

\bibitem{wong2019low}
J.~Wong, M.~Driver, and P.~Vincent.
\newblock Low-code development technologies evaluation guide, 2019.

\bibitem{woo2020rise}
M.~Woo.
\newblock The rise of no/low code software development—no experience needed?
\newblock {\em Engineering (Beijing, China)}, 6(9):960, 2020.

\bibitem{yang2016query}
D.~Yang, A.~Hussain, and C.~V. Lopes.
\newblock From query to usable code: an analysis of stack overflow code
  snippets.
\newblock In {\em 2016 IEEE/ACM 13th Working Conference on Mining Software
  Repositories (MSR)}, pages 391--401. IEEE, 2016.

\bibitem{yang2016security}
X.-L. Yang, D.~Lo, X.~Xia, Z.-Y. Wan, and J.-L. Sun.
\newblock What security questions do developers ask? a large-scale study of
  stack overflow posts.
\newblock {\em Journal of Computer Science and Technology}, 31(5):910--924,
  2016.

\bibitem{zhu2016devops}
L.~Zhu, L.~Bass, and G.~Champlin-Scharff.
\newblock Devops and its practices.
\newblock {\em IEEE Software}, 33(3):32--34, 2016.

\bibitem{zhuang2021easyfl}
W.~Zhuang, X.~Gan, Y.~Wen, and S.~Zhang.
\newblock Easyfl: A low-code federated learning platform for dummies.
\newblock {\em arXiv preprint arXiv:2105.07603}, 2021.

\bibitem{zohocreator}
{Zoho Creator platform overview}.
\newblock {Available: \url{https://www.zoho.com/creator/}}.
\newblock [Online; accessed 5-January-2022].

\end{thebibliography}
\end{small}
\end{document}